%% file: review.tex
\documentclass[rmp,letterpaper,twocolumn,floatfix]{revtex4-1}
\usepackage[hyperfootnotes=false,linktocpage]{hyperref}
\usepackage{amssymb}
\usepackage{amsmath}
\usepackage{graphicx}
\usepackage{color}
\def\x{{\bf x}}
\def\y{{\bf y}}
\def\sp{\text{sign prob.}}

\input{defs}

\begin{document}
\author{Emanuel Gull}
\affiliation{Department of Physics, Columbia University, New York, NY 10027, USA}
\author{Alexander I. Lichtenstein}1
\affiliation{Institute of Theoretical Physics, University of Hamburg, 20355 Hamburg, Germany}
\author{Andrew J. Millis} 
\affiliation{Department of Physics, Columbia University, New York, NY 10027, USA}
\author{Alexey N. Rubtsov}
\affiliation{Department of Physics, Moscow State University,
119992 Moscow, Russia }
\author{Matthias Troyer} 
\affiliation{Theoretische Physik, ETH Zurich, 8093 Zurich, Switzerland}
\author{Philipp Werner}
\affiliation{Theoretische Physik, ETH Zurich, 8093 Zurich,
Switzerland}

\title{Continuous-time Monte Carlo methods for quantum impurity  models}
\begin{abstract}
Quantum impurity models describe an atom or molecule embedded in a host material with which it can exchange electrons. They are basic to nanoscience as representations of quantum dots and molecular conductors and play an increasingly important role in the theory of ``correlated electron'' materials as auxiliary problems whose solution gives the ``dynamical mean field'' approximation to the self energy and local correlation functions. These applications require a method of solution which provides access to both high and low energy scales and is effective for wide classes of physically realistic models. The continuous-time quantum Monte Carlo algorithms reviewed in this article meet this challenge.  We present derivations and descriptions of the algorithms in enough detail to allow other workers to write their own implementations, discuss the strengths and weaknesses of the methods, summarize the problems to which the new methods have been successfully applied and outline prospects for future applications.
\end{abstract}
\maketitle
\tableofcontents
\input{introduction}

\input{diagmc}
\input{weakcoupling}
\input{hybridization}

\input{infiniteU}

\input{phonons}

\input{nonequilibrium}

\input{technical}

\input{applicationsdmft}

\input{applicationsnano}
\input{applicationsnonequilibrium}

\input{prospects}

\bibliography{refs_shortened}

\end{document}

%% file: defs.tex
\newcommand{\nup}{n_{\uparrow}}
\newcommand{\ndown}{n_{\downarrow}}
\newcommand{\nupt}{\nup(\tau)}
\newcommand{\ndownt}{\ndown(\tau)}

\newcommand{\nupta}{\nup(\tau_1)}

\newcommand{\ndownta}{\ndown(\tau_1)}

\newcommand{\nuptb}{\nup(\tau_2)}

\newcommand{\ndowntb}{\ndown(\tau_2)}
\newcommand{\nuptk}{\nup(\tau_k)}
\newcommand{\ndowntk}{\ndown(\tau_k)}
\newcommand{\Tr}{\text{Tr}}

\newcommand{\Hhyb}{H_\text{hyb}}
\newcommand{\Hhybtwid}{\tilde{H}_\text{hyb}}

\newcommand{\vk}{{\mathbf{k}}}
\newcommand{\N}{{\mathbf{N}}}
\newcommand{\G}{{\mathbf{G}}}
\newcommand{\D}{{\mathbf{D}}}
\newcommand{\M}{{\mathbf{M}}}
\newcommand{\fatP}{{\mathbf{P}}}

\def\stau{\{ s_{i},\tau_{i}, x_i\}}
\def\staup{\{ s_{i}',\tau_{i}', x_i'\}}

%% file: introduction.tex
\section{Introduction}
\subsection{Overview}

This article aims to provide a comprehensive overview of recent developments which have made continuous-time quantum Monte Carlo (CT-QMC) approaches the method of choice for the solution of broad classes of quantum impurity models.   We present derivations and descriptions of the algorithms in enough detail to allow other workers to  write their own codes, and give a general introduction to diagrammatic Monte Carlo methods on which these algorithms are based. We discuss the strengths and weaknesses of the methods, and their range of applicability.  We summarize the problems to  which the new methods have been successfully applied, and outline prospects for future applications.  We hope that readers will come away from the review with an appreciation of the power and flexibility of the techniques and with the knowledge needed to apply them to new generations of problems in nanoscience, correlated electron physics, nonequilibrium systems and other areas. But before entering into specifics it is worth asking: `what are  quantum impurity models?' and also `why study them with continuous time -methods?'

\subsection{Quantum impurity models: definitions and examples }

Quantum impurity models were introduced  to describe the properties of a nominally magnetic transition metal   ion embedded in a non-magnetic host metal.  A magnetic transition metal atom such  as Fe and Co  has a partly filled $d$ shell, and the intra-$d$ Coulomb interactions act to organize the electrons in the $d$-shell into a high-spin local moment configuration. Hopping from the $d$ shell to the metal or vice versa favors non-magnetic configurations and thus competes with the local interactions.  In 1961 P. W. Anderson \cite{Anderson61}, following important earlier work of \textcite{Friedel51,Friedel56},  wrote down a mathematical model (now referred to as the Anderson Impurity Model) which encodes this competition. Anderson's concept has proven enormously fruitful, with implications extending far beyond its original context of impurity magnetism.  Quantum impurity models are basic to nanoscience as representations of quantum dots and molecular conductors \cite{Hanson07} and have been used to understand the adsorption of atoms onto surfaces \cite{Brako81,Langreth91}. They are of theoretical interest as solvable examples of nontrivial quantum field theories \cite{Wilson75,Affleck08}  and in recent years have played an increasingly  important role in condensed matter physics  as auxiliary problems whose solution gives the  ``dynamical mean field" (DMFT) approximation to the properties of correlated electron materials such as high temperature copper-oxide and pnictide superconductors  \cite{Georges96,Held06,Kotliar06}.   

A quantum impurity model (see e.g. \cite{MahanChapter4}) may be represented as a Hamiltonian with three basic terms: $H_\text{loc}$ which describes the ``impurity'': a system with a finite (typically small) number of degrees of freedom, $H_\text{bath}$ which describes the noninteracting but infinite (continuous spectrum) system to which the impurity is coupled, and $H_\text{hyb}$ which gives the coupling between the impurity and bath. Thus
\begin{equation}
H_\text{QI}=H_\text{loc}+H_\text{bath}+H_\text{hyb}.
\label{HQI}
\end{equation}
The physics represented by $H_\text{QI}$ is in general nontrivial because $\left[H_\text{loc},H_\text{hyb}\right]\neq 0$ (in physical terms,  coupling to the bath mixes the impurity eigenstates).   

In the situation of primary physical interest $H_\text{loc}$ may be represented in terms of a set of single-particle fermion states labeled by quantum numbers $a=1,\ldots,N$ (including both spatial and spin degrees of freedom) and created by operators $d^\dagger_a$ as
\begin{eqnarray}
H_\text{loc}&=&H_\text{loc}^0+H_\text{loc}^I,\label{hloc}
 \label{Hloc} \\
H_\text{loc}^0&=&\sum_{ab} E^{ab}d^\dagger_a d_b,
 \label{Hloc0}\\
H_\text{loc}^I&=&\sum_{pqrs}I^{pqrs}d^\dagger_p d^\dagger_q d_r d_s+\ldots. \label{HlocI}
\end{eqnarray}
The $ab$ components of the matrix  ${\mathbf E}$ describe the bare level structure, $I$ parametrizes  electron electron interactions and the ellipsis denotes terms with $6$ or more fermion operators. 

$H_\text{bath}$ may be thought of as describing bands of itinerant electrons, each labeled by a one-dimensional momentum  coordinate $k$ or band energy $\varepsilon_k$ and an index (spin and orbital) $\alpha$. One usually writes
\begin{equation}
H_\text{bath}=\sum_{k\alpha} \varepsilon_{k\alpha}c^\dagger_{k\alpha}c_{k\alpha}.
\end{equation}
The most commonly used form of the mixing term is characterized by a hybridization matrix ${\bf V}$
\begin{equation}
H_\text{hyb}=\sum_{k\alpha b}{V}_k^{\alpha b}c^\dagger_{k\alpha}d_b +\text{H.c.}, 
\label{Hmix}
\end{equation}
although exchange couplings of the form 
\begin{equation}
H_\text{hyb}^\text{exchange}=\sum_{k_1k_2abcd}J_{k_1k_2}^{abcd}c^\dagger_{k_1a}c_{k_2b}d^\dagger_cd_d
\label{Hexchange}
\end{equation}
also arise, most famously in the ``Kondo problem'' of a spin exchange-coupled to a bath of conduction electrons \cite{Kondo64}.

Coupling of impurity models to oscillators (representing for example phonons in a solid)  has also been considered. A discussion in the CT-QMC context is presented in Section \ref{phononsec}.

It is sometimes convenient to represent the partition function $Z$ of the  impurity model as an imaginary time path integral \cite{NegeleOrland}. In this representation it is easy to formally eliminate the bath degrees of freedom (a technique pioneered by \textcite{Feynman63}), obtaining an action  which for Hamiltonians involving a hybridization of the form of Eq.~(\ref{Hmix}) is
\begin{align}
Z =&\int \mathcal{D}[d^\dagger,d]e^{-S},\\
\label{SIAM}
S =&\sum_{ab}\iint_0^\beta d\tau d\tau{'}d_a^\dagger(\tau) \Big[\left( \partial_\tau+E^{ab}\right)\delta(\tau-\tau{'}) \\ \nonumber
&+\Delta^{ab}(\tau-\tau{'})\Big]d_b(\tau{'}) + \int_0^\beta d\tau H_\text{loc}^I (\tau).
\end{align}
In this formulation the {\em hybridization function}  
\begin{align}\label{hybfun}
\Delta^{ab}(i\omega_n)=\sum_{k\alpha} {V^*}_k^{a\alpha}\left(i\omega_n-\varepsilon_{k\alpha}\right)^{-1}{V}_k^{\alpha b}
\end{align}
compactly encapsulates those aspects of the bath that are relevant to the impurity model physics. It will play a crucial role in our subsequent discussions. 
It is also often useful to define the noninteracting impurity model Green's function $\mathcal{G}^0$ via
\begin{align}
\mathcal{G}^0= -(\partial_\tau + \mathbf{E} + \mathbf{\Delta})^{-1}.
\end{align}

The paradigmatic quantum impurity model is the single-impurity single-orbital Anderson model \cite{Anderson61}. In this model,  $H_\text{loc}$ describes a single orbital, so the label $a$ is spin up or down, $E^{ab}$ is (in the absence of magnetic fields) just a level energy  $\varepsilon_0$, and the  interaction term collapses to $Un_\uparrow n_\downarrow$. Thus
\begin{eqnarray}
H_\text{AIM}&=&\sum_\sigma \varepsilon_0d^\dagger_\sigma d_\sigma +Un_\uparrow n_\downarrow 
\label{AIM} \\ \nonumber
& +&\sum_{k\sigma}\Big(V_kc^\dagger_{k\sigma}d_\sigma+H.c.\Big) +\sum_{k\sigma}\varepsilon_k c^\dagger_{k\sigma}c_{k\sigma}.
\end{eqnarray}

Impurity models with more degrees of freedom  are assuming increasing importance. More degrees of freedom means a richer variety of physical phenomena, implying a more complicated structure for the interactions.   For example, in transition metal oxide materials with partially filled $d$ shells or in compounds involving rare earth or actinide atoms with partially filled $f$ shells, the interactions express not only the energy cost of multiply occupying the atom but also the Hund's rule physics that states of maximal spin and orbital angular momentum are preferred. Thus the interaction Hamiltonian describing the energetics of different configurations of electrons in the $d$ orbitals which play an important role in the physics of transition metal oxides with cubic perovskite structures is normally written in the ``Slater-Kanamori (SK)'' form \cite{Mizokawa95,Imada98}: 

\begin{align}
H_\text{loc}^I&=H_\text{SK}\equiv U\sum_an_{a\uparrow}n_{a\downarrow}+(U-2J)\sum_{a\neq b}n_{a\uparrow}n_{b\downarrow} 
\nonumber \\
&+(U-3J)\sum_{a> b,\sigma}n_{a\sigma}n_{b{\sigma}} \nonumber \\ 
&-J\sum_{a\neq b}\left(d^\dagger_{a\uparrow}d^\dagger_{a\downarrow}d_{b\uparrow}d_{b\downarrow} +d^\dagger_{a\uparrow}d^\dagger_{b\downarrow}d_{b\uparrow}d_{a\downarrow}\right).
\label{HSK}
\end{align}

In nanoscience applications the impurity typically represents the highest occupied  (HOMO) and lowest unoccupied (LUMO) molecular orbitals and the interactions are computed from Coulomb matrix elements involving these orbitals. In ``cluster'' dynamical mean field applications the  ``impurity''  is thought of as a (typically small) number of sites with the $E^{ab}$ representing an intersite hopping Hamiltonian, thus in a two-site approximation to the Hubbard model
\begin{align}
H_\text{loc}&=H_\text{cl}=\sum_\sigma \varepsilon_0\left(d^\dagger_{1\sigma}d_{1\sigma}+d^\dagger_{2\sigma}d_{2\sigma}\right) \\
&+\sum_\sigma t\left(d^\dagger_{1\sigma}d_{2\sigma}+d^\dagger_{2\sigma}d_{1\sigma}\right)
+U\left(n_{1\uparrow}n_{1\downarrow}+n_{2\uparrow}n_{2\downarrow}\right).\nonumber
\end{align}

Solving the quantum impurity model means computing the correlation functions of the $d$ operators. Of these the most important is the $d$ Green function ($T_\tau$ denotes time-ordering).
\begin{equation}
G_d^{ab}(\tau)=-\left\langle T_\tau d_a(\tau)d_b^\dagger(0)\right\rangle.
\label{Gddef}
\end{equation} 
In the absence of interactions,  $G_d^{ab}(i\omega_n)=\mathcal{G}_d^{0,ab}(i\omega_n)\equiv[\left(i\omega_n-{\mathbf E}-{\mathbf \Delta} \right)^{-1}]_{ab}$. The effect of interactions may be parametrized by the self energy $\mathbf{\Sigma}(i\omega_n)=\left(\boldsymbol{\mathcal{G}}^0\right)^{-1}-\mathbf{G}^{-1}$. 

Solving the quantum impurity model  is conceptually and algorithmically challenging. As Eq.~(\ref{SIAM}) demonstrates, a  quantum impurity model is  a quantum field theory in $0$ space $+$ $1$ time dimension. While $0+1$ dimensional quantum field theories are easier to solve than higher dimensional ones, they are still (in the general case) nontrivial. Only in a few cases are exact solutions known, and while in many more cases the form of the  ``universal'' low energy behavior has been determined, the dynamical mean field and nanoscience applications require information about behavior beyond the universal limit, as well as quantitative information about the parameters describing the universal limit.   A further complication is that   impurity models  typically involve several energy scales, including an interaction scale, often high, a hybridization scale, typically intermediate, and one or more dynamically generated  energy scales,  which in many cases are very low relative to the basic interaction and hybridization scales. A robust method which works for general models over a range of energy scales is required.

\subsection{State of the art prior to continuous time QMC} 

Quantum impurity models have been of long-standing interest and a wide range of approximate techniques have been developed to solve them,  including perturbative expansions in coupling constant \cite{Yosida75} and in flavor degeneracy \cite{Read83,Coleman84}, perturbative \cite{Anderson70,Solyom74} and functional \cite{Hedden04} renormalization group, as well as ``$X$-operator'' techniques  \cite{Hubbard64,Keiter70,Gunnarsson83} and  different formulations of  auxiliary (``slave'') particle methods  \cite{Abrikosov64,Barnes76,Read83,Coleman84,Florens04}. An important subclass of analytical methods is based on the resummation to all orders of a particular subset of diagrams. In the impurity model context the most important of these are the non-crossing approximation (NCA) \cite{Bickers87} and its generalizations \cite{Pruschke89,Haule01}. The terminology refers to the structure of diagrams in an expansion in the hybridization. In this expansion contractions of bath operators are represented as lines. If one uses a time-ordered perturbation theory diagrams may be classified by the number of  times that lead operator lines cross and all diagrams with zero or one crossing may be analytically summed by solving an integral equation. While uncontrolled, these approximations capture many aspects of the physics of impurity models and  can be formulated on the real frequency axis. They have therefore been used as inexpensive solvers for impurity models and to study nonequilibrium phenomena in nano-contacts \cite{Meir94}.   Very recently, techniques closely related to those described here have been used to formulate a numerically exact solution based on an expansion around the NCA \cite{Gull10_bold}. It is found that the approximations are accurate in the Mott insulating phase but do not capture the important diagrams in the metallic phase of the Anderson impurity model. Powerful field-theoretical and Bethe-ansatz-based analytical methods have been developed to classify and compute exactly the universal low energy behavior \cite{Wilson75,AndreiRMP,Affleck08} of broad classes of quantum impurity models. However,  the dynamical mean field and nanoscience applications require an approach that works for wide classes of physically relevant impurity models and gives access to physics beyond the universal limit. Thus, while analytical methods provide very valuable insights, they  do not provide the comprehensive  solutions, valid over a wide range of frequencies,  that are needed for modern applications.

Starting from work of \textcite{Wilson75} and subsequently of  \textcite{White92}, (see \cite{Schollwoeck05} for a review),  an important set of numerical methods has been developed based on intelligently chosen truncations of the Hilbert space of the many-body problem in question.  The ``numerical renormalization group'' (NRG) methods are based on  iterative diagonalization using a logarithmic discretization of the energy spectrum of the  lead states and are reviewed for example in \cite{Bulla08b} while the ``density matrix renormalization group'' (DMRG) techniques involve an isolation of the relevant low-lying states. These methods are complementary to the methods discussed here: CT-QMC methods are most naturally formulated in imaginary time and very efficiently handle a wide range of energy scales and relatively general classes of models, but require analytical continuation to obtain real-time information and  have difficulty resolving subtle low energy features such as the fine structure of quantum criticality. On the other hand, the NRG and DMRG methods can be formulated directly on the real frequency axis or in real time and are particularly powerful in resolving ground states and low-lying levels but encounter difficulties in providing information over a wide range of frequencies and the difficulties increase rapidly as one moves beyond the simple Anderson/Hubbard models.  Both DMRG \cite{Hallberg06,Nishimoto06} and NRG \cite{Bulla08b} methods have been implemented as ``solvers'' for the quantum impurity models of dynamical mean field theory.   But except for the single-orbital Anderson model, where NRG methods have proven to be useful, especially in situations where a precise understanding of the very low energy behavior is crucial, NRG and DMRG solvers are not in widespread and general use in the DMFT community. 

A more widely applied class of techniques is based on the ``exact diagonalization'' (ED) idea introduced in the early days of dynamical mean field theory by \textcite{Caffarel94}. These authors approximated the continuum of bath energies  and values of the hybridization by a small number of variationally chosen eigenstates and hybridization functions. $H_{QI}$ then becomes a finite system, which is exactly diagonalized, leading to a $G_d$ characterized by a delta function spectrum. The cost scales exponentially with the number of sites considered. The largest systems which are typically studied contain on the order of 15 sites with one non-degenerate orbital on each site. Thus, in the single-impurity Anderson model, Eq.~(\ref{AIM}),  the continuum of bath states $c_{k\sigma}$  may be approximated by $7$ or $8$ ($\times 2$ for spin) orbitals while for say a three orbital model only two or three bath orbitals per impurity state can be accommodated. With the development of more modern algorithms and computers, enough bath sites can be included that for the single-orbital Anderson impurity model the temperature dependence can be computed, the convergence of results with bath size can be studied \cite{Capone07}, and systematic comparisons to other methods can be made  \cite{Werner06,Comanac08}. Recently, results on small clusters \cite{Civelli05,Kyung06b,Kancharla08,Liebsch09} and single-impurity, multiorbital  models  \cite{Liebsch05,Liebsch08} have also been obtained, although here the number of bath sites per orbital is limited and the convergence with bath site number cannot yet be addressed rigorously \cite{Koch08}.

Quantum Monte Carlo techniques provide a general method for solving quantum field theories, and prior to the development of CT-QMC methods the principal impurity solver was the Hirsch-Fye quantum Monte Carlo method \cite{Hirsch86}. This method is based on writing an imaginary-time functional integral, discretizing the interval $[0, \beta)$ into $M$ equally spaced ``time-slices'' $\Delta \tau=\beta/M$ and then on each time-slice $i$ applying a discrete Hubbard-Stratonovich transformation which for  the single-orbital Anderson model is 
\begin{align}
e^{-\Delta \tau U\left(n_\uparrow n_\downarrow-\frac{n_\uparrow+n_\downarrow}{2}\right)}&=\frac{1}{2}\sum_{s_i=\pm 1}e^{\lambda s_i\left(n_\uparrow-n_\downarrow\right)},\\
\lambda&=\text{arcosh}\left[\exp\left(\frac{1}{2}\Delta\tau U\right)\right].
\end{align} 
For a fixed choice of Ising variables $\{s_i\}$ the problem thus becomes a noninteracting fermion model in a time dependent $z$-oriented magnetic field $h(\tau_i)=s_i$ which may be formally solved, so one is left with the problem of sampling the trace over the $2^M$ dimensional space of the $s_i$.

The Hirsch-Fye method was for almost two decades the method of choice, but ultimately three difficulties limit its power. The first is that it  requires an equally spaced time discretization. (A linear-in-$\beta$ method for impurity models \cite{Khatami09}, while fast, similarly requires discretization both of the bath and the imaginary time  axis.)  The second is that at large interactions and low temperatures equilibration may become an issue.  While techniques have been developed to ameliorate these problems \cite{Blumer08,Gorelik09} and new update techniques have been proposed \cite{Alvarez08,Nukala09}, the difficulties of managing the discretization and equilibration issues  within Hirsch-Fye  are real. It appears that the CT-QMC methods discussed here are now preferred by most practitioners. The third, and most fundamental difficulty is that for interactions other than the simple one-orbital Hubbard model the Hubbard-Stratonovich fields required to decouple the interactions proliferate and may have to be chosen complex so sampling the space of auxiliary fields becomes prohibitively difficult \cite{Sakai06}.

\subsection{Why continuous time?}

Imaginary-time path integral representations of quantum problems  such as Eq.~(\ref{SIAM}) are mathematically defined (see, e.g. \onlinecite{NegeleOrland} and references therein)  in terms of the result of a limiting process in which one rewrites the partition function, $Z=\exp(-\beta H)$, of a system described by a Hamiltonian $H$ at temperature $T=1/\beta$  by defining $\Delta \tau=\beta/N$, $\tau_j=j\Delta \tau$ as
\begin{align}\label{Trotter}
Z=e^{-(\tau_N-\tau_{N-1}) H} e^{-(\tau_{N-1}-\tau_{N-2}) H}\dots e^{-(\tau_{1}-\tau_{0}) H}.
\end{align}
The path integral is  defined by  inserting complete sets of states between every pair of exponentials and then taking the limit $\Delta \tau \rightarrow 0$. This mathematical definition motivates a numerical approach \cite{Suzuki76} in which one approximates the path integral by (i) retaining a non-zero $\Delta \tau$ and (ii) using a Monte Carlo method to estimate the sums over all intermediate states. The exact partition function is recovered after the twin steps of converging the Monte Carlo and extrapolating the results to $\Delta \tau=0$. While clever and efficient methods (for example, the Hirsch-Fye procedure \cite{Hirsch86} mentioned in the previous subsection) have been devised for performing the Monte Carlo, the time step extrapolation remains an issue. The difficulties are particularly severe for the quantum impurity problems of interest here because the basic object in the theory is the Green function, which drops rapidly as $\tau$ is increased from $0$ and has discontinuous derivatives at $\tau=0,\beta$ which need to be correctly evaluated (see e.g. Fig.~2 in \cite{Werner06}). The discretization errors are  large, and a very small $\Delta\tau$ and a precise extrapolation to $\Delta \tau=0$ are required to obtain accurate results. However, the low energy behavior of interest is carried by times $\tau \sim \beta/2$,  so that simulations on a homogeneous grid require many points.  Methods which do not involve an explicit time discretization) would therefore appear to be advantageous.

The basic idea behind all of the continuous time methods discussed in this review is to avoid the time discretization entirely by sampling the terms in a diagrammatic expansion, instead of sampling the configurations in a complete set of states.  One of the first important methods to do this is Handscomb's method \cite{Handscomb62,Handscomb64}. This method and  its generalization, the stochastic series expansion (SSE) algorithm \cite{Sandvik91} are based on  a Taylor expansion of the partition function in powers of $\beta H$ and have been successful for quantum magnets. However, they require that the spectrum of the Hamiltonian is bounded from above so applications to  boson problems require a truncation of the Hilbert space while applications to fermion problems are limited by a bad sign problem.

The continuous time methods in use now stem from work  of   \cite{Prokofev96} and \cite{Beard96}, who showed that simulations of  bosonic lattice models can be implemented simply and efficiently in continuous-time by a stochastic sampling of a diagrammatic perturbation theory for the partition function. The general scheme for treating diagrams with continuous variables of arbitrary nature -- diagrammatic Monte Carlo -- is formulated in \cite{Prokofev98,Prokofev98A}.  In these methods  the systematic errors associated with time discretization and the Suzuki-Trotter decomposition were eliminated. The gain in computational efficiency is so large that the problem of simulating unfrustrated bosonic lattice models can now be considered as solved, although special cases, for example bosons coupled to a gauge field (rotating atomic gases, charged bosons in a magnetic field) remain challenging. 

The success of CT-QMC methods for bosons stimulated  efforts to adapt the technique to fermionic problems \cite{Rombouts99}.  However, in contrast to standard (unfrustrated) bosonic systems, where diagrams all have the same signs, in fermionic models individual diagrams  may have positive or negative signs, so that the sampling of individual diagrams suffers from a severe sign problem. This  sign problem may be reduced by combining classes of diagrams analytically into determinants.  Unfortunately, Rombouts and collaborators found that  a prohibitively severe sign problem  remained in the parameter regimes relevant to strong correlation physics. This, and the fact that the lattice algorithm given in \cite{Rombouts99} was restricted to  density-density interactions, caused many researchers to abandon the approach -- except for  the special case of sign-problem-free models with an attractive interaction, where CT-QMC methods have successfully been used to investigate the BEC-to-BCS crossover in ultracold atomic gases \cite{Burovski06a,Burovski06,Burovski08}.

The increasing importance of impurity models has motivated a reexamination of CT-QMC methods. Impurity models  turn out to have a much less severe sign problem than the full lattice problem (indeed in some cases the sign problem is absent). The reduction in severity of the sign problem has allowed the development of flexible and powerful continuous-time quantum Monte Carlo impurity solvers, first in a weak-coupling formulation \cite{Rubtsov04, Rubtsov05}, soon thereafter in a complementary hybridization expansion formulation \cite{Werner06},  and more recently in an auxiliary field formulation \cite{Gull08_ctaux}. These methods have quickly been extended in many directions and applied to numerous dynamical mean field studies of model Hamiltonians.  They enabled accurate simulations of the Kondo lattice model \cite{Otsuki09}, the first quantitative studies of multi-orbital models with realistic rotationally invariant (non-diagonal)  interactions \cite{Rubtsov05,Werner06Kondo, Haule07,Werner07crystal, Werner08nfl, Chan09} and allowed much more efficient simulations of the multi-site clusters needed to study spatial correlation effects within dynamical mean field theory \cite{Haule07plaquette, Park08plaquette, Gull08_plaquette, Ferrero09,Gull09_8site,Werner098site,Sordi10,Mikelsons09B}. They have also enabled more realistic ``LDA+DMFT'' studies of materials \cite{Marianetti07}.

Continuous time quantum Monte Carlo methods can also be used to  efficiently compute four-point correlation functions, which are important for susceptibilities, phase boundaries, and in connection with recently developed extensions of  dynamical mean field  theory \cite{Toschi07,Rubtsov08,Slezak09,Kusunose06}. The methods have been applied to nanoscience topics including the properties of transition metal clusters on metal surfaces \cite{Savkin05,Gorelov07}. Previously inaccessible physics questions such as the quasiparticle dynamics and thermal crossovers in heavy fermion materials are being addressed \cite{Shim07Pu,Haule07c,Park08}  and applications to questions motivated by experiments on fermions in optical lattices have begun to appear \cite{Dao08,DeLeo08}. Extensions to nonequilibrium problems are now under development \cite{Muehlbacher08, Werner09, Schiro09, Werner10}.

While the new  CT-QMC methods have been transformative, opening wide classes of problems to systematic study, they have not solved the fermion sign problem.   As far as  is known,  sign problems are physical and unavoidable, at least in itinerant phases with unpaired fermions \cite{Troyer05} and indeed set the ultimate limits on the problems and parameter regimes which can be studied by the continuous time methods discussed here. Further discussion of sign problems  will be given in Sec.~\ref{signproblem} and in the context of the discussion of specific algorithms.

%% file: diagmc.tex
\section{Diagrammatic Monte Carlo in continuous time}\label{diagmcsec}
\subsection{Basic ideas} \label{ctqmcbasicideas}
The basic idea of the CT-QMC methods is very simple. One begins from a Hamiltonian $H=H_a+H_b$ which is split into two parts labeled by $a$ and $b$, writes the partition function $Z=e^{-\beta H}$ in the interaction representation with respect to $H_a$  and expands in powers of $H_b$, thus (${T_\tau}$ is the time ordering operator)
\begin{align}
Z=&\Tr\ {T_\tau} e^{-\beta H_a}\exp\left[-\int_0^\beta d\tau H_b(\tau)\right] \nonumber\\
=&\sum_k (-1)^k\int_0^\beta d\tau_1\ldots\int_{\tau_{k-1}}^\beta d\tau_k \nonumber\\
&\times\text{Tr}\big[e^{-\beta H_a}H_b(\tau_k)H_b(\tau_{k-1})\ldots H_b(\tau_1)\big].\label{Zseries1}
\end{align}

The trace evaluates to a number and diagrammatic Monte Carlo methods \cite{Prokofev98} enable a sampling over all orders $k$, all topologies of the paths/diagrams and all  times $\tau_1, \cdots, \tau_k$ in the same calculation.  Because the method is formulated in continuous time from the beginning, time discretization errors do not have to be controlled and the simulation can be arranged to ensure that the method focuses attention on the time regions which are most important to the process under study. Provided the spectrum of the perturbation term is bounded from above the contributions of  very large orders are exponentially suppressed by the  factor $\frac{1}{k!}$ originating from the expansion of an exponential. Thus the sampling process does not run off to infinite order and no truncation of the diagram order is needed. (Note that for bosonic operators a perturbation in the interaction would be divergent  since the spectrum cannot be bounded from above unless a cutoff in bosonic occupation number is introduced \cite{ItzyksonZuber}, so an expansion in the hybridization is usually employed.)

The method does not rely on an auxiliary field decomposition although it may be advantageously combined with one \cite{Gull08_ctaux}. Further, the method does not rely on a particular  partitioning into ``interacting'' and ``noninteracting'' parts; in principle the only requirement is that one may decompose the Hamiltonian in such a way that the time evolution associated with $H_a$ and the contractions of operators $H_b$ may easily be evaluated. In practice, the sign associated with interchanges of fermion operators means that the expansion must be arranged such that terms differing only in the contractions of fermion operators are combined; for example into determinants.

In the impurity model context four  types of expansion have been formulated, which we refer to as CT-HYB ($H_b=H_\text{hyb}$, Eq.~(\ref{Hmix})), CT-INT ($H_b=H^I_\text{loc}$, Eq.~(\ref{HlocI})), CT-AUX ($H_b=H^I_\text{loc}$ but with an additional auxiliary field decomposition) and CT-J (an expansion for Kondo-like problems with $H_b=H_\text{hyb}^\text{exchange}$, Eq.~(\ref{Hexchange})). 
The advantage of the hybridization expansion is that arbitrarily complicated impurity interactions can easily be treated; the disadvantage is that because $\left[H_\text{hyb},H_\text{loc}\right]\neq 0$ at least one of the operators is non-diagonal so the expansion generically requires the manipulation of 
matrix blocks whose size grows exponentially with the number of impurity orbitals. 
The present state of the art is that  $5$ spin-degenerate orbitals can be treated. Various truncation and approximation schemes provide limited access to larger problems but as the number of orbitals is increased the difficulties rapidly become insurmountable.  

CT-INT and CT-AUX are  variations of an  ``interaction expansion''. They are sometimes referred to as  ``weak coupling'' expansions, but this is a misnomer -- the expansion is in powers of the interaction but is not (in principle) restricted to small interactions. The series is always convergent for nonzero temperature and finite number of orbitals.  In  CT-INT and CT-AUX the scaling with number of impurity orbitals  is not exponential, so much larger systems can be treated.  However, the methods are most suited to Hubbard-like models with a single local density-density interaction. More complicated interactions typically  require multiple expansions in the several vertices and if the interactions do not commute (as is the case for the components of the spin exchange)  the difficulties increase.   

CT-J is  an expansion organized for Kondo-like models where the interaction vertex also creates particle-hole pairs in the conduction bands. 
It combines aspects of both the interaction and hybridization expansion. 

While all of the expansions are based on the same general idea, there are significant differences in the specifics of how the expansion is arranged,  the measurements are done and the errors are controlled. We therefore devote a separate section to each expansion.
 In the remainder of this section we provide an overview of general aspects of continuous time Monte Carlo methods.

\subsection{Monte Carlo basics: sampling, errors, Markov chains and the Metropolis algorithm}\label{markovandmetropolis}
In this subsection we recall some basic results pertaining to the Monte Carlo evaluation of high dimensional integrals. For reader unfamiliar with Monte Carlo, the books by \textcite{BookLandauBinder} and \textcite{BookKrauth} give an extensive introduction to the technique.

In the CT-QMC methods, as in many other classical or quantum many-body problems, one is faced with the issue of evaluating sums over phase spaces or configuration spaces which we denote generically by ${\mathcal{C}}$. ${\mathcal{C}}$ is typically of a very high dimension, so Monte Carlo techniques are the only practical methods of evaluation. A crucial quantity is the partition function, $Z$, which we will write formally as an integral over configurations $\x\in\mathcal{C}$ with weight $p(\x)$:
\begin{equation}
Z=\int_{\mathcal{C}} d\x p(\x).
\label{Zbasic}
\end{equation}
In a classical system $\x$ might be a point in phase space with a Boltzmann weight $p(\x)=\exp(-\beta E(\x))$, where $E(\x)$ is the energy of the configuration $\x$.  In the quantum problems described here $\x$ will represent a particular term in a diagrammatic partition function expansion.   

The expectation value of a quantity $A$ is given by the average, over the configuration space ${\mathcal C}$ with weight $p$, of a quantity ${\mathcal A}(\x)$: 
\begin{equation}
 \langle A \rangle_{p} = \frac{1}{Z}\int_{\mathcal{C}} d\x {\mathcal A}(\x)p(\x),
\label{MCObs}
\end{equation}
The auxiliary quantity ${\mathcal A}(\x)$  depends on the specific representation chosen in a particular algorithm. 

The average (\ref{MCObs}) can be estimated in a Monte Carlo procedure by selecting $M$ configurations $\x_i$ with a probability $p(\x)/Z$ and averaging the contributions ${\mathcal A}(\x_i)$:
\begin{equation}\label{MCObs2}
\langle A \rangle_p \approx \langle A \rangle_{MC} \equiv \frac{1}{M}\sum_{i=1}^M {\mathcal A}(\x_i).
\end{equation}

According to the central limit theorem, if the number of configurations is large enough the estimate (\ref{MCObs2}) will be normally distributed around the exact value $\langle A \rangle_p$ with variance
\begin{equation}
\left(\Delta A\right)^2\equiv \langle \left(A_{MC}-A_p\right)^2\rangle=\frac{{\rm Var} A}{M}.
\label{variance}
\end{equation}

It will sometimes be advantageous or necessary to sample configurations $\x_i$ with a distribution $\rho(\x)$ different  from $p(\x)$. The expectation value $\langle A \rangle_\rho$ in the ensemble then has to be reweighed:
\begin{eqnarray}\label{MCreweight}
\langle A \rangle &=& \frac{1}{Z} \int_\mathcal{C}d\x {\mathcal A}(\x) p(\x) \nonumber \\ &=& 
 \frac{ \int_\mathcal{C}d\x {\mathcal A}(\x) \frac{p(\x)}{\rho(\x)} \rho(\x) }{ \int_\mathcal{C} d\x\frac{p(\x)}{\rho(\x)} \rho(\x) }
\equiv \frac{\langle A \frac{p}{\rho} \rangle_\rho }{\langle \frac{p}{\rho}  \rangle_\rho}.
\label{reweighing}
\end{eqnarray}
To estimate this expectation value one needs to sample both the numerator and denominator and  collect averages of ${\mathcal A}(\x_i)p(\x_i)/\rho(\x_i)$ and $p(\x_i)/\rho(\x_i)$. Care must be taken in estimating the statistical errors of such ratios, since cross-correlations will make na\"ive error propagation unreliable.  A jackknife or bootstrap procedure (see, e.g. \cite{numrec92}) is needed.

Integrals with general distributions such as Eqs.~(\ref{Zbasic}), (\ref{reweighing}) 
are best sampled by generating configurations using a Markov process. A Markov process is fully characterized by a transition matrix $W_{\x\y}$ specifying the probability to go from state $\x$ to state $\y$ in one step of the Markov process. Normalization (conservation of probabilities) requires $\sum_\y W_{\x\y}=1$. Starting from an arbitrary distribution the Markov process will converge exponentially to a stationary distribution $p(\x)$ if two conditions are satisfied.
\begin{itemize}
\item {\it Ergodicity}: It has to be possible to reach any configuration $\x$
from any other configuration $\y$ in a finite number of Markov steps: for all $\x$ and $\y$ there exists an integer $N<\infty$ such that for all $n\ge N$ the probability  $(W^n)_{\x\y}\ne 0$.
\item {\it Balance}: Stationarity implies that the distribution $p(\x)$  fulfills the balance condition
\begin{equation}\label{bal1}
\int_{\mathcal C} d\x\,p(\x) W_{\x\y} = p(\y),\end{equation}
that is $p(\x)$ is a left eigenvector of the
transition matrix $W_{\x\y}$.
A sufficient but not necessary condition usually used instead of the balance condition is the {\it detailed balance} condition
\begin{align}\label{detbal1}
\frac{W_{\x\y}}{W_{\y\x}} = \frac{p(\y)}{p(\x)},
\end{align}
which we will use below.
\end{itemize}

The first, and still most widely used, algorithm that satisfies detailed balance is the Metropolis-Hastings algorithm \cite{metropolis1953,hastings1970}. There, an update from a configuration $\x$ to a new configuration $\y$  is proposed with a probability $W_{\x\y}^{\text{prop}}$ but accepted only  with probability $W_{\x\y}^{\text{acc}}$.  If the proposal is rejected  the old configuration $\x$ is used again.
The transition matrix is
\begin{align}
W_{\x\y} = W_{\x\y}^{\text{prop}}W_{\x\y}^{\text{acc}}
\end{align}
and the detailed balance condition (\ref{detbal1}) is satisfied by using the Metropolis-Hastings acceptance rate
\begin{equation}\label{metrogen}
 W_{\x\y}^{\text{acc}}= \min\left[ 1,  R_{\x\y} \right].
\end{equation}
with the acceptance ratio $R_{\x\y}$ given by
\begin{equation}\label{transitionrate}
 R_{\x\y}=\frac{p(\y) W_{\y\x}^{\text{prop}}}{p(\x) W_{\x\y}^{\text{prop}}}
\end{equation}
and $R_{\y\x}=1/R_{\x\y}$. To simplify the notation we will often quote just $R_{\x\y}$, and imply that $\min[ 1,  R_{\x\y} ]$ is the actual acceptance probability. Note that the acceptance ratio $R_{\x\y}$ includes both the weights and the proposal probabilities. In the following sections we will always specify both the proposal probabilities $W_{\x\y}^{\text{prop}}$ and the acceptance ratios $R_{\x\y}$.

\subsection{Diagrammatic Monte Carlo -- the sampling of path integrals and other diagrammatic expansions}
\label{SeriesMC}
The partition function Eq.~(\ref{Zseries1}) may be expressed as a sum of integrals originating from a diagrammatic expansion:
\begin{equation}
\label{ct_pf}
Z=\sum_{k=0}^\infty\sum_{\gamma\in\Gamma_k} \int_0^\beta d\tau_1 \ldots \int_{\tau_{k-1}}^\beta d\tau_k w(k,\gamma,\tau_1, \ldots,
\tau_k),
\end{equation}
which has the form of Eq.~(\ref{Zbasic}). The individual configurations are of the form
\begin{equation}
\x = (k,\gamma, (\tau_1,\ldots,\tau_k)),
\end{equation} 
where $k$ is the expansion or diagram order and  $\tau_1,\ldots,\tau_k\in[0,\beta)$ are the times of the $k$ vertices in the configuration. 
The parameter $\gamma\in\Gamma_k$ includes all discrete variables, such as the topology of the diagram 
and spin, orbital, lattice site, and auxiliary spin indices associated with the interaction vertices.

A configuration $\x$ has a weight
\begin{equation}\label{weightw}
p(\x) = w(k,\gamma,\tau_1, \ldots,\tau_k)d\tau_1 \cdots d\tau_k ,
\end{equation}
which we will assume to be non-negative for now. The case of negative weights is discussed in Sec. \ref{signproblem}. Although these weights are well-defined probability densities they involve infinitesimals $d\tau$, which one might worry could cause difficulties with proposal and acceptance probabilities in the random walk in configuration space. As  \cite{Prokofev96,Prokofev98,Prokofev98A,Beard96} showed, this is not the case.

\begin{figure}[tb]
\begin{center}
\includegraphics[width=0.95\columnwidth]{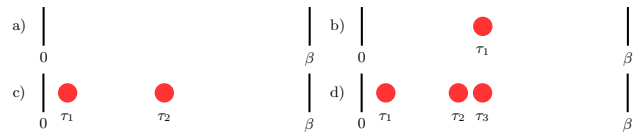}
\end{center}
\caption{Diagrammatic representation of configurations $\x=\{(k;\tau_1, \ldots \tau_k)\} \in\mathcal{C}$ showing examples with orders $k=0, 1, 2, 3$ and vertices (represented by dots) at times $\tau_1, \dots, \tau_3$.}
\label{fig:diagsimple}
\end{figure}

The various algorithms reviewed here differ in the representations, weights, and updates, as well as in the most convenient representation for the measurement of observables, but all express the partition function in the general form (\ref{ct_pf}).  To illustrate the Monte-Carlo sampling of such  continuous-time partition function expansions and in particular to demonstrate that the infinitesimal does not cause problems,  we consider the very simple partition function  
\begin{align}
Z = \sum_{k=0}^\infty\int_0^\beta d\tau_1 \int_0^\beta d\tau_2 \cdots \int_0^\beta d\tau_k \frac{w(k)}{k!},
\end{align}
which using time ordering can be rewritten as
\begin{align}\label{PFExp}
Z = \sum_{k=0}^\infty\int_0^\beta d\tau_1 \int_{\tau_1}^\beta d\tau_2  \cdots \int_{\tau_{k-1}}^\beta d\tau_k w(k).
\end{align}
The distribution describing the probability of a diagram of order $k$ with vertices at times $\{\tau_j\}$ is (here we make the times explicit)
\begin{equation}
p((k,\tau_1,\ldots,\tau_k))=w(k)\prod_{i=1}^kd\tau_i.
\label{pdef}
\end{equation}
In the following we will always assume time-ordering $\tau_1 \le \tau_2\le\ldots\le\tau_k$ and visualize the configurations using a diagrammatic representation as in Fig. \ref{fig:diagsimple}.

Transitions between configurations $\x$ and $\y$ are realized by updates. Updates in diagrammatic Monte Carlo codes typically
involve (i) updates that increase the order $k$ by inserting an additional vertex at a time $\tau$ and (ii) updates that decrease the order $k$ by removing a vertex $\tau_j$.
These insertion and removal updates are necessary to satisfy the ergodicity
requirement and are often sufficient:  we can reach any configuration from another one by removing all
the existing vertices and then inserting new ones. Additional updates keeping the order $k$ constant are typically not required for ergodicity but may speed up equilibration and improve the sampling efficiency. In some special circumstances, for example if all odd order diagrams have zero weight, updates which insert or remove multiple vertices are required. 

\begin{figure}[bt]
\begin{center}
\includegraphics[width=0.6\columnwidth]{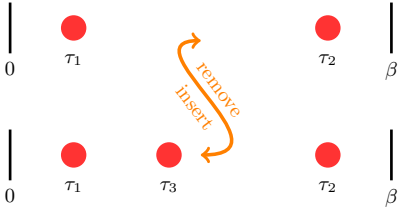}
\end{center}
\caption{An insertion update (top to bottom) inserting a vertex at time $\tau_3$ and the corresponding removal update (bottom to top), removing the vertex at time $\tau_3$.} \label{raiselower}
\end{figure}

In the following we will focus on the insertion and removal updates, illustrated in Fig. \ref{raiselower}.
For the insertion let us start from a configuration  $ (k,\vec{\tau})= (k,\tau_1,\ldots,\tau_k)$ of order $k$. We propose to insert a new vertex at a time $\tau$ uniformly chosen in the interval $[0,\beta)$, to obtain a new time-ordered configuration $(k+1,\vec{\tau}')= (k+1,\tau_1,\ldots,\tau,\ldots,\tau_{k})\equiv (k+1,\tau_1',\ldots,\tau_{k+1}')$.
The proposal rate for this insertion is given by the probability density
\begin{align}\label{Wpropup}
W^{\text{prop}}_{(k,\vec{\tau}),(k+1,\vec{\tau}')}=\frac{d\tau}{\beta}.
\end{align}

The reverse move is the removal of a randomly chosen vertex. The probability of removing a particular vertex to go back from $(k+1,\vec{\tau}')$ to $(k,\vec{\tau})$ is just one over the number of
available vertices:
\begin{align}\label{Wpropdn}
W^{\text{prop}}_{(k+1,\vec{\tau}'),(k,\vec{\tau})} = \frac{1}{k+1}.
\end{align}

To obtain the Metropolis acceptance rates we first calculate the acceptance ratio
\begin{align}
\label{acceptance1}
 &R_{(k,\vec{\tau}),(k+1,\vec{\tau}')}=\frac{p((k+1,\vec{\tau}'))}{p((k,\vec{\tau}))} \frac{W_{(k+1,\vec{\tau}'),(k,\vec{\tau})}^{\text{prop}}}{W_{(k,\vec{\tau}),(k+1,\vec{\tau}')}^{\text{prop}}} \\
 &= \frac{w(k+1)d\tau_1'\cdots d\tau_{k+1}'}{w(k)d\tau_1\cdots d\tau_{k}}\frac{1/(k+1)}{ d\tau/\beta} 
 = \frac{w(k+1)}{w(k)} \frac{\beta}{k+1}. \nonumber
\end{align}
Observe that all infinitesimals cancel: the additional infinitesimal in the weight $p((k+1,\vec{\tau}'))$ is canceled by the infinitesimal of the proposal rate for insertions.

Equation~\ref{acceptance1} implies that the acceptance rates $W^{\text{acc}}$ are well defined finite numbers given by
\begin{eqnarray}
W^{\text{acc}}_{(k,\vec{\tau}),(k+1,\vec{\tau}')}&=&\min\left[1, R_{(k,\vec{\tau}),(k+1,\vec{\tau}')}\right],\\
W^{\text{acc}}_{(k+1,\vec{\tau}'),(k,\vec{\tau})} &=&\min\left[1,1/ R_{(k,\vec{\tau}),(k+1,\vec{\tau}')}\right].
\end{eqnarray}

Acceptance rates for updates that preserve the order $k$, such as shifting some of the $\tau_i$ times or updating the discrete parameters $\gamma$ are straightforward to evaluate since there all infinitesimals cancel trivially.

\begin{figure}[tb] 
\begin{center}
\includegraphics[width=0.95\columnwidth]{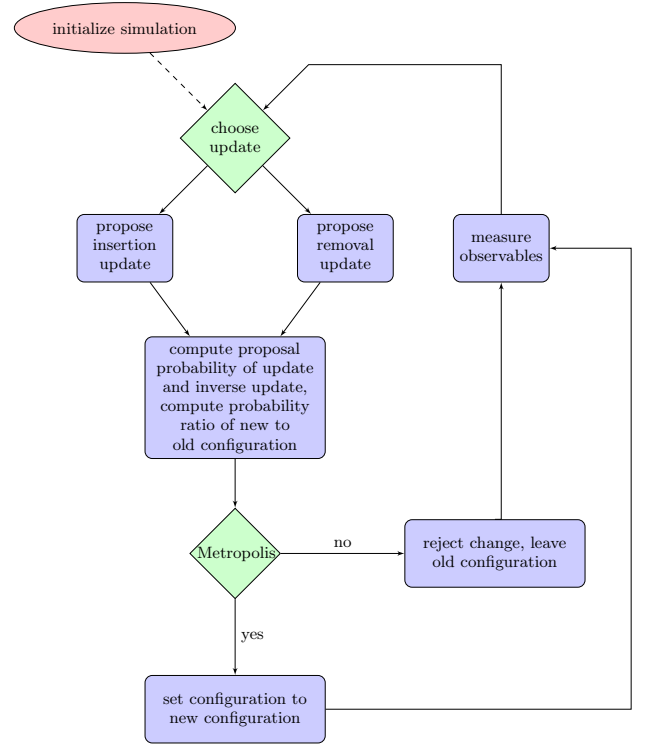}
\caption{Continuous-time Quantum Monte Carlo flow diagram.
}\label{flowdiagram}
\end{center}
\end{figure}

The general scheme of diagrammatic Monte Carlo algorithms is illustrated in
Fig.~\ref{flowdiagram}. One cannot stress often enough that measurements are performed again on the old configuration if the proposed update has been rejected.

\subsection{The negative sign problem} \label{signproblem}
Until now we have tacitly assumed that the expansion coefficients of our partition function expansion are always positive or zero. This has allowed us to interpret the weights as probability densities on the configuration space and the stochastic  sampling of these configurations in a Monte Carlo simulation. If the weights $p(\x)$ become negative, as is often the case in fermionic simulations due to the anti-commutation relations between fermionic operators, they can no longer be regarded as probabilities.
The common solution is to sample with respect to the absolute value of the weight $\rho(\x)=|p(\x))|$ and reweight the measurements according to Eq.~(\ref{reweighing}). The ratio $p(\x)/\rho(\x)$ is then just  $\text{sign}(p(x))$  $= p(\x)/ |p(\x)|$. This gives for the average ({\ref{MCObs})
\begin{equation}\label{signprob}
 \langle A \rangle = \frac{\langle A \cdot \text{sign}\rangle_{|p|}}{\langle \text{sign}\rangle_{|p|}},
\end{equation} 	
which can be evaluated by sampling numerator and denominator separately with respect to the positive weight  $|p(\x)|$.

While sampling with the absolute value and reweighing allows Monte Carlo simulations of systems with negative weights, it does not solve the ``sign problem''. Sampling Eq.~(\ref{signprob}) suffers from exponentially growing errors. To see this let us consider the average sign
\begin{equation}
\langle \text{sign}\rangle = \frac{\int_{\mathcal{C}} d\x\ \text{sign}(\x) |p(\x)|}{\int_{\mathcal{C}} d\x |p(\x)|} = \frac{Z}{Z_{|p|}},
\end{equation}
which is just the ratio of the partion function $Z$ and the partition function of a ``bosonic'' system with positive weights $|p(\x)|$. This ratio can be expressed through the difference $\Delta F$ in free energies of these two systems
\begin{equation}\label{frat}
\langle \text{sign}\rangle = \frac{Z}{Z_{|p|}} = \exp (-\beta \Delta F),
\end{equation}
and decreases exponentially as the temperature is lowered or the volume of the system increased.

The sign {\it problem}  is thus the accurate measurement of  this near-zero sign from individual measurements that are $+1$ or $-1$ , a {\it cancellation} problem. The variance of the sign is
\begin{equation}
{\rm Var}\;\text{sign} = \langle \text{sign}^2\rangle - \langle \text{sign}\rangle^2 = 1- \exp (-2\beta \Delta F)\approx 1
\end{equation}
and the relative error after $M$ measurements
\begin{equation}
\Delta\text{sign} = \frac{{\sqrt{{\rm Var}\;\text{sign} /M}}}{\langle \text{sign}\rangle} \approx \frac{\exp (\beta \Delta F)}{\sqrt{M}}
\label{signvariance}\end{equation}
grows exponentially with decreasing temperature and increasing system size.

The sign problem has been proven to be nondeterminstic polynomial (NP) hard, and hence 
in general no polynomial time solution is believed to exist  \cite{Troyer05}.  However, the severity of the sign problem (in the notation of Eq.~(\ref{signvariance}) the magnitude of the coefficient $\exp (\beta \Delta F)$) depends both on the model considered and on the representation chosen for the model. Impurity models tend to have less severe sign problems than comparable finite-sized lattice models (`turning off' the coupling to the bath often makes the sign problem worse). In special cases the sign problem is absent.
For example, Yoo and coworkers proved that there is no sign problem in Hirsch-Fye simulations of the single impurity single orbital  Anderson impurity model \cite{Yoo05}, and this proof can be easily extended to some multi-orbital models and  adapted to the continuous time algorithms presented in this review.

A trivial sign  problem arises if the operator $-H_b$ is negative and odd perturbation orders are allowed. A simple example is  the weak coupling  expansion of the repulsive (positive $U$) Hubbard model.  In this particular case the sign problem may be avoided by  a trick discussed in Section \ref{weakcoupling}. 

In the hybridization expansion a severe sign problem may occur if the hybridization  function and the bare level energy  do not commute $\left[\Delta^{ab},E^{ab}\right]\neq 0$ \cite{Wang10}. An apparently related difficulty occurs in  the weak coupling approach if the hybridization function and interaction are not diagonal in the orbital/spin occupation  number basis \cite{Gorelov09}.  In the larger systems dealt with in cluster dynamical mean field theory, fermion loops occur and produce a sign problem.  Because the sign problem is model and representation dependent, further discussion is postponed to the sections pertaining to specific algorithms.

%% file: weakcoupling.tex
\section{Interaction Expansion algorithm CT-INT} \label{weak_chapter}
The interaction expansion algorithm CT-INT was the first continuous-time impurity solver to be introduced \cite{Rubtsov04}. It  proceeds from Eq.~(\ref{Zseries1}), with $H_b$ taken to be the interaction part $H_\text{loc}^I$ of Eq.~(\ref{HlocI}), and $H_a = H_\text{bath}+H_\text{hyb}+H_\text{loc}^0$ (see Eqs.~(\ref{HQI}) and (\ref{hloc})). It has a better scaling  with system size than the hybridization algorithm and can treat more general interactions than CT-AUX. A ``trivial'' sign problem arises for repulsive interactions, where terms of the form $(-U)^k$ appear. Elimination of this sign problem is an important issue in the design of the algorithm.

\subsection{Partition function expansion}
\label{weakcoupling}
\begin{figure}[tb]
\begin{center}
\includegraphics[width=0.95\columnwidth]{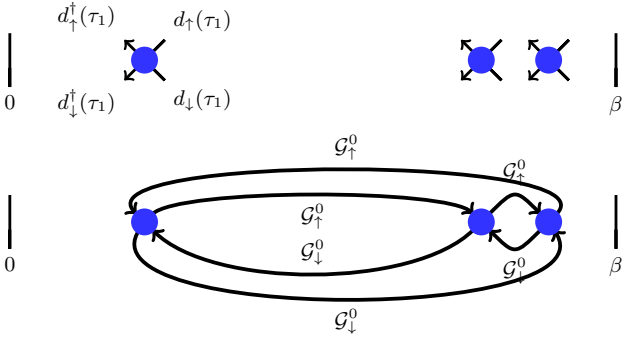}
\end{center}
\caption{Depiction of a third order term in the weak coupling expansion. Upper panel: Hubbard interaction vertices
denoted by circles. Each
$U n_\uparrow(\tau) n_\downarrow(\tau)$ - vertex has four operators. Lower panel: one possible contraction of the interaction vertices.
\label{weakvert}}
\end{figure}

We illustrate the method by considering the simplest model, the one orbital single site Anderson impurity model Eq.~(\ref{AIM}) which, for this expansion, is most conveniently formulated in terms of the action $S=S_0+S_U$ with
\begin{align}
S_0 &= -\sum_\sigma\iint_0^\beta d\tau d\tau'd_\sigma^\dagger(\tau)\mathcal{G}^0_\sigma(\tau-\tau')^{-1}d_\sigma(\tau'),\\
S_U &= U\int_0^\beta d\tau \nupt\ndownt,\label{S_U}
\end{align}
where $\mathcal{G}^0_\sigma = (i\omega_n - \epsilon_0 - \Delta_\sigma)^{-1}, $ and $\epsilon_0$ is the impurity energy level.
We consider more general models in Sec.~\ref{clusterr}.
The expansion of the partition function in powers of $U$ reads
\begin{align}
Z/Z_0 &= 1\\ \nonumber &+\frac{(-U)}{1!}\int_0^\beta d\tau_1 \langle n_\uparrow(\tau_1) n_\downarrow(\tau_1) \rangle_0
         \\ \nonumber &+\frac{(-U)^2}{2!}\iint_0^\beta d\tau_1 d\tau_2 \langle n_\uparrow(\tau_1) n_\downarrow(\tau_1) n_\uparrow(\tau_2) n_\downarrow(\tau_2) \rangle_0
         \\ \nonumber &+\cdots,
\end{align}
where the notation $\langle \ldots \rangle_0 = \frac{1}{Z_0} \int
\mathcal{D}[d^\dag,d] e^{-S_0} [\ldots]$ denotes an average in the
\emph{non-interacting} ensemble with quadratic action $S_0$ (see low order
terms in Fig.~\ref{weakvert}), and $Z_0 = \int
\mathcal{D}[d^\dag,d] e^{-S_0}$. Employing Wick's theorem
\cite{Wick50} we may express the expectation value in terms of
determinants of the non-interacting Green's function $-\langle
T d(\tau_i) d^\dagger(\tau_j)\rangle_0$ $=$ $\mathcal{G}^0(\tau_i-\tau_j)$:
\begin{align}\label{wick}
&\langle \nupta\ndownta \nuptb \ndowntb \cdots \nuptk \ndowntk \rangle_0 = \nonumber\\
&\hspace{50mm}\det \D_k^\uparrow \det \D_k^\downarrow, \\
&(\D_k^\sigma)_{ij} = \mathcal{G}^{0}_{\sigma}(\tau_i-\tau_j).
\end{align}
Summing the contractions into a determinant instead of sampling them individually reduces the size of the configuration space and avoids a sign problem coming from the fermionic exchange.

We thus arrive at the following series for the partition function:
\begin{equation}
\label{signedexpansion}
 Z/Z_0=\sum_{k=0}^{\infty}\frac{(-U)^k}{k!}\int_0^\beta d\tau_1 \ldots d\tau_k \left(\prod_\sigma
\det \D_k^\sigma \right).
\end{equation}
Two ``sign problems'' may potentially occur in this expansion: an ``intrinsic'' sign problem arising from fermion exchange because the determinants might become negative and a ``trivial''
sign problem, arising for $U>0$ from the $(-U)^k$ factor. 
The arguments of \textcite{Yoo05} prove that for the single impurity Anderson model each of the determinants is no-negative, so there is no intrinsic sign problem. This is not necessarily the case for the more general models considered in subsection \ref{clusterr}. The ``trivial'' sign problem arising for $U>0$ can be managed in several ways.
For the single band, single-impurity Anderson model 
\cite{Rubtsov03} 
showed that the replacement
$d^\dag_\downarrow \to \tilde{d}_\downarrow, d_\downarrow \to
\tilde{d}^\dag_\downarrow$ leads to following changes in the
parameters of the effective action:
\begin{align}
{{ \epsilon_{0\downarrow} \to -\epsilon_{0\downarrow},}\atop
{ \epsilon_{0\uparrow} \to  \epsilon_{0\uparrow}+U,}}\ \ \ 
{{\Delta_\downarrow (\tau) \to -\Delta_\downarrow(-\tau)}\atop
{ U\to -U.}}
\end{align}
The repulsive interaction becomes attractive and the ``trivial'' sign problem due to the interaction term vanishes.

This approach performs a particle-hole transformation on the down spins only such that up and down spins are treated inequivalently.
While the entire series formally maintains spin inversion symmetry (in the absence of a magnetic field), restoring it dynamically by Monte Carlo sampling is challenging in practice. It is better to avoid the symmetry breaking as follows.

First, observe that the transformed Hamiltonian can be viewed in the original variables as an expansion in $U n_\uparrow (n_\downarrow-1)$; this leads to a down-spin determinant with diagonal elements replaced by
$\mathcal{G}^0_\downarrow(0)-1$. The absence of a sign problem means that the down spin determinant must generate a minus sign that compensates the $(-U)$ factor. This approach may be generalized: expanding in powers of
\begin{align}\label{NonSymmetrized}
S_U &= U \int_0^\beta d\tau \left(\nupt -
\alpha_{\uparrow}\right)\left(\ndownt -\alpha_{\downarrow}\right),
\end{align}
with the corresponding change $\epsilon_{0\sigma}$ $\to$ $\epsilon_{0\sigma}-U \alpha_{-\sigma}$, $\mathcal{G}^0$ $\to$ $\tilde{\mathcal{G}}^0$ in $S_0$,
leads to 
\begin{align}\label{detalpha}
&\det \D_k^\sigma =  \Big\langle T_\tau[n_{\sigma}(\tau_1)-
\alpha_{\sigma}]\cdots[n_{\sigma}(\tau_k)-\alpha_{\sigma}]\Big\rangle_0
\nonumber
\\
&= \begin{vmatrix} \tilde{\mathcal{G}}^0_\sigma(0) - \alpha_{\sigma} &
\tilde{\mathcal{G}}^0_\sigma(\tau_1-\tau_2)& \cdots
&\tilde{\mathcal{G}}^0_\sigma(\tau_1-\tau_k)
\\
\tilde{\mathcal{G}}^0_\sigma(\tau_2-\tau_1) & \tilde{\mathcal{G}}^0_\sigma(0) - \alpha_{\sigma} & \ddots & \vdots \\
\vdots & \ddots & \ddots & \vdots & \\
\tilde{\mathcal{G}}^0_\sigma(\tau_k-\tau_1) & \cdots & \cdots &
\tilde{\mathcal{G}}^0_\sigma(0)
- \alpha_{\sigma}\\
\end{vmatrix}.
\end{align}
\textcite{Rubtsov03} showed that for $\alpha_\uparrow+\alpha_\downarrow=1, \alpha_\uparrow \alpha\downarrow \le 0$ the trivial sign problem is absent.

Finally, it is advantageous to avoid this explicit
symmetry breaking at the cost of introducing an
auxiliary field $s = \uparrow,\downarrow$ and expanding in powers
of
\begin{align}\label{Symmetrized}
S_U &= \frac{U}{2} \int_0^\beta d\tau \sum_{s_\tau} \left(\nupt -
\alpha_{s_\tau\uparrow}\right)\left(\ndownt -\alpha_{s_\tau
\downarrow}\right),
\end{align}
Expanding this action we get an additional
random variable $s_i=\uparrow,\downarrow$ at each vertex that
needs to be sampled over. In practice this does not introduce any
difficulties: all expressions remain the unchanged, apart from an an additional index $\alpha_{s_i \sigma}$ instead of
$\alpha_\sigma$ in the determinants of Eq.~(\ref{detalpha}).

In the actual calculation it is useful to take the parameter
$\alpha_{s\sigma}=0.5+\delta$ for $s=\sigma$ and
$\alpha_{s\sigma}=-\delta$ otherwise. In principle, $\delta$ can
be taken to be zero but setting it to a small positive value
$\delta\approx 0.01$ allows to avoid numerical instabilities due
to nearly-singular matrices.

An interaction expansion has also been derived  in \cite{Assaad07}  for retarded interactions such as
\begin{align}
S_\text{ret} = \sum_{ab}\int_0^\beta d\tau d\tau' {\cal O}^a(\tau) W^{ab}(\tau-\tau'){\cal O}^b(\tau'),
\end{align}
where ${\cal O}$ denotes a fermion bilinear.
This formalism will be discussed in Sec.~\ref{weakphonon}.

\subsection{Updates}
\begin{figure}[hbt]
\begin{center}
\includegraphics[width=0.95\columnwidth]{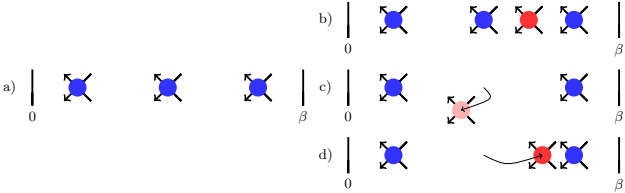}
\end{center}
\caption{Local updates for the CT-INT algorithm.
(a): starting configuration (b): Insertion of a vertex, (c): removal of a
vertex, (d): shift of a vertex in imaginary time.}
\label{weakmoves}
\end{figure}

The series
(\ref{signedexpansion}) and the corresponding one for (\ref{Symmetrized}) are of the type (\ref{ct_pf}),
and we can employ continuous-time sampling as described in Sec.~\ref{SeriesMC}.
We insert and remove interaction
vertices on the imaginary time axis, corresponding to the
terms $U (n_\uparrow(\tau)-\alpha_{s_\tau\uparrow}) (n_\downarrow(\tau)-\alpha_{s_\tau \downarrow}$) 
(see Figure \ref{weakmoves}).
Proposing a vertex insertion update with probability $d\tau/(2\beta)$ (for the imaginary time location and the orientation of the auxiliary spin $s_\tau$) and a removal update with probability $1/(k+1)$ we obtain
\begin{align} \label{RSIAM}
R= \frac{\beta U}{(k+1)} \prod_\sigma \frac{\det \D_{k+1}^\sigma} {\det \D_{k}^\sigma}.
\end{align}
Note that for the interaction defined in Eq.~(\ref{Symmetrized}) the prefactor $\frac{1}{2}$ is compensated by the factor $2$ in the ratio of proposal probabilities, which comes from the
two possible values of $s_\tau$, so that the acceptance ratio is the same as in the straightforward approach.

This update and its inverse are sufficient to be
ergodic. In evaluating the determinant ratios  the fast-update
technique described in Sec.~\ref{fastupdate} should be used, since
it allows to calculate the ratio $R$ in ${\rm O}(k^2)$ operations,
substantially faster than the na\"ive evaluation of determinants
with ${\rm O}(k^3)$ operations.

\subsection{Measurements}
Monte Carlo averages are calculated using Eq.~(\ref{MCObs}) and (\ref{MCObs2}), where the distribution $p$ of Eq.~(\ref{MCObs}) is given by the coefficients of Eq.~(\ref{signedexpansion}). In particular, the Green's function
\begin{align}\label{Gweakseries}
&G_\sigma(\tau-\tau')=-\frac{Z_0}{Z}\sum_{k=0}^\infty
\frac{(-U)^k}{k!}\int d\tau_1\ldots d\tau_k\nonumber\\
&\hspace{4mm}\times \Big\langle
T_\tau d_\sigma(\tau)d_\sigma^\dagger(\tau')
n_{1\uparrow}(\tau_1)n_{1\downarrow}(\tau_1)\cdots n_{k\downarrow}(\tau_k)\Big \rangle_0
\end{align}
is estimated by $G_{\tau_1\tau_1,...,\tau_k\tau_k}(\tau, \tau')$ (corresponding to $A(\x)$ in Eq.~(\ref{MCObs})):
\begin{align}
G_\sigma(\tau-\tau')&=\langle G_{\tau_1\tau_1,...,\tau_k\tau_k}(\tau, \tau')\rangle_\text{MC},\label{GFQMC}\\
G_{\tau_1\tau_1,...,\tau_k\tau_k}(\tau, \tau')&=-\frac{\langle
T_\tau d_\sigma(\tau)d^\dagger_\sigma(\tau') n_{1\sigma} n_{2\sigma}
\cdots n_{k\sigma}\rangle_0}{\langle n_{1\sigma} n_{2\sigma}
\cdots n_{k\sigma}\rangle_0}.\label{Gweakconf}
\end{align}
The $\langle \cdots \rangle_\text{MC}$ denotes a Monte Carlo average, while the $\langle \cdots \rangle_0$ denotes all possible Wick's contractions of one particular Monte Carlo configuration. The denominator is a determinant that cancels the Wick's contraction of a partition function configuration $p$, and the numerator determinant consists of a matrix with an additional
row $[\mathcal{G}^{0}_{\sigma}(\tau-\tau'),
\mathcal{G}^{0}_{\sigma}(\tau-\tau_1), \mathcal{G}^{0}_{\sigma}(\tau-\tau_2),
\dots, \mathcal{G}^{0}_{\sigma}(\tau-\tau_k)]$ and column
$(\mathcal{G}^{0}_{\sigma}(\tau-\tau'),
\mathcal{G}^{0}_{\sigma}(\tau_1-\tau'), \mathcal{G}^{0}_{\sigma}(\tau_2-\tau'),
\dots, \mathcal{G}^{0}_{\sigma}(\tau_k-\tau')]$.

The configuration $G_{\tau_1\tau_1,...\tau_k\tau_k}(\tau, \tau')$ in Eq.~(\ref{Gweakconf})
depends on two independent arguments $\tau, \tau'$, while the observable average Eq.~(\ref{Gweakseries})
is time-translation invariant. This symmetry of the effective action is
restored only after the averaging in Eq.~(\ref{GFQMC}).
It will be shown in Sec.~\ref{weakmeas} that it is best either to measure a quantity corresponding to $\Sigma G$ or to
perform a Fourier transform to Matsubara
frequencies analytically, so that the Green's function is calculated directly in the
frequency domain.

There is one particular observable estimate that can be obtained just from the properties
of the random walk itself, without any additional calculation: the average value of the perturbation operator.
One can see from a term-to-term comparison of the respective series that the average perturbation order $\langle k\rangle$
is proportional to the inverse temperature and the average value of the interaction operator,
\begin{equation}
\langle k\rangle_\text{MC}= \langle S_U\rangle.
\end{equation}
Therefore, the expecation value of the interaction operator $U n_\uparrow n_\downarrow$ is $\langle k\rangle_{MC}/\beta$.

\subsection{Generalization to clusters, multi-orbital problems and retarded interactions}\label{clusterr}
In the case of the Hubbard model on a cluster,  the only difference to the single orbital case is that creation and annihilation operators acquire an additional site index. We can absorb all quadratic hopping terms in $\mathcal{G}^0$ and
perform the interaction expansion in
\begin{align}
S_U = U \sum_i (n_{i\uparrow}-
\alpha_{i\uparrow})(n_{i\downarrow}-\alpha_{i\downarrow}),
\end{align}
where $i$ runs over the sites of the cluster.
The $\alpha_{i\sigma}$-terms are chosen as in the single site case; optionally with an auxiliary spin $s_i$ at each site.

The Green's functions $\mathcal{G}^0_{ij}(\tau_i - \tau_j)$ are site-dependent, but the
spin up and spin down contributions still factor into separate determinants:
\begin{align}
\frac{Z}{Z_0} = &\sum_{k=0}^\infty \int d\tau_1 \ldots d\tau_k
\!\!\!\mathop{\sum_{s_1 \cdots s_k= \pm 1}}_{i_1 \cdots i_k}
\!\!\!\frac{(-U)^k}{k!}\prod_\sigma \det \D_{k}^\sigma,
\end{align}
where $(\D_k^\sigma)_{ij} = \mathcal{G}_{ij,\sigma}^{0}(\tau_i -
\tau_j)-\delta_{ij}\alpha_{i\sigma}$. It follows immediately that
there is no sign problem in the half-filled case,
where the determinants of the up- and down matrices are identical. However,
away from half filling a sign problem occurs in general, see e.g.
Fig.~5 in Ref.~\cite{Gull08_ctaux}.

For the updates a generalization of Eq.~(\ref{RSIAM}) should be used, where
$\beta$ is replaced by the factor $\beta N_c$, with $N_c$ the number of sites in the
cluster.

In the general case of multiple orbital problems intrinsic sign problems typically occur, and management even of the trivial sign problem becomes more involved. The basic idea is to express the interaction $H_\text{loc}^I$ [Eq.~(\ref{hloc})] in action form as
\begin{align}
S_\text{loc} = \sum_{pqrs} \!\iint\! d\tau d\tau' I^{pqrs} (d_p^\dagger d_s -\alpha_{ps})(d_q^\dagger d_r -\alpha_{qr})
\end{align}
and then perform a multiple expansion in the interactions $I^{pqrs}.$ In multi-oribtal systems the number of terms proliferates; for $N$ orbitals there are of order $N^4$ terms, although in practice some of them vanish by symmetry. Denoting the tuple $(pqrs)$ at vertex $i$ by $\xi_i$ we have

\begin{align}
\frac{Z}{Z_0} &= \sum_{k=0}^\infty \sum_{\xi_1\ldots\xi_k}^{N_\xi}
\int d\tau_1\ldots d\tau_k \frac{(-1)^k
I_{\xi_1} \ldots  I_{\xi_k}}{k!} \langle v_{\xi_1}\cdots v_{\xi_k} \rangle_0,\label{wexp_mo}\nonumber\\
v_{\xi}&\equiv (d^\dag_{p_\xi} d_{s_\xi} -\alpha_{ps})(d^\dag_{q_\xi} d_{r_\xi} -\alpha_{qr}).
\end{align}
If we insert random (nonzero) matrix elements at random times in $[0,\beta)$,
the prefactor of the acceptance probability ratios in
Eq.~(\ref{RSIAM}), $\beta U/(k+1)$, is modified by a factor
$N_\xi$, becoming $\beta I N_\xi/(k+1)$.

Wick's theorem of Eq.(\ref{wexp_mo}) yields a determinant similar
to (\ref{detalpha}).
If the Green's function matrix $\mathcal{G}^0_{ij}$ for the
different orbitals is diagonal in the orbital indices, the
determinant factorizes into smaller-size determinants, each
However, in general there is no reason for the
determinant of $\D$ to have the same sign for all configurations.
The choice of $\alpha$-terms has an influence on the sign
statistics, and they need to be adjusted for each problem such
that the expansion is sign - free or at least has an average sign
that is as large as possible. How this is best done is still an
open question. An ansatz has been presented in
Ref.~\cite{Gorelov07}. The basic principle is to treat the
off-diagonal interaction terms with small but non-zero $\alpha$,
whereas the symmetrized form (\ref{Symmetrized}) is used for the
density-density part.

\section{Continuous-time auxiliary field algorithm CT-AUX}\label{ctaux_chapter}
A first continuous-time auxiliary field method for fermionic lattice models was developed by 
\textcite{Rombouts98,Rombouts99}, and applied to the nuclear Hamiltonian and small Hubbard lattices.
We present here a different formulation  \cite{Gull08_ctaux} that is also applicable to (cluster) impurity problems.
This continuous-time auxiliary field (CT-AUX) algorithm is based on an interaction expansion combined with an auxiliary field decomposition of the interaction vertices.
One may view CT-AUX as an ``optimal'' Hirsch-Fye algorithm, on a non-uniform time grid and with a varying number of ``time slices'' that are chosen automatically for given parameters.
The approach allows the combination of numerical techniques developed for the Hirsch-Fye algorithm (see e.g. Sec.~\ref{delayed}) with the advantages of a continuous-time method.
It was shown to be equivalent to the weak coupling algorithm in the case of the single-band Hubbard model \cite{Mikelsons09}.
Currently the CT-AUX impurity solver is the method of choice for large cluster simulations.

\subsection{Partition function expansion}
We present the derivation for the case of a cluster impurity problem with $N_c$ cluster sites. The generalization to multiorbital models with density density interactions is straightforward. The application to more general multiorbital models would involve techniques similar to \textcite{Sakai06,Sakai06b} and has not yet been attempted.
Starting from the partition function $Z = \Tr e^{-\beta (H_0 + H_U)}$ we add a non-zero constant $K$ to $H_U$:
\begin{align}
H_U &= U \sum_i^{N_c}\left(n_{i\uparrow}n_{i\downarrow}-\frac{n_{i\uparrow}+n_{i\downarrow}}{2}\right) -\frac{K}{\beta}, \label{HU}\\
H_0 &= H_\text{AIM}-H_U+K/\beta,
\end{align}
such that
\begin{align} \label{Zexp}
Z &= \Tr\left[ e^{-\beta H_0}T_\tau e^{\int d\tau \left(\frac{K}{\beta}-U\sum_i^{N_c}\left(n_{i\uparrow}n_{i\downarrow} - \frac{n_{i\uparrow}+n_{i\downarrow}}{2}\right) \right)}\right].
\end{align}
Expanding the exponential in powers of $H_U$ and  applying the auxiliary field decomposition \cite{Rombouts99}
\begin{align}\label{auxdec}
1 - \frac{\beta U}{K}\sum_i^{N_c}\left( n_{i\uparrow}n_{i\downarrow} - \frac{n_{i\uparrow}+n_{i\downarrow}}{2}\right) &= \frac{1}{2N_c}\sum_{i, s_i=\pm1}\!\!\!\!e^{\gamma s_i (n_{i\uparrow}-n_{i\downarrow})},\\
\cosh(\gamma) &=1+\frac{U\beta N_c}{2K},
\end{align}
we obtain
\begin{align} \label{Ztr}
&Z =  \sum_{k=0}^\infty \sum_{s_1, \cdots s_k\atop=\pm 1}\int_0^\beta\!\!\!\!d\tau_1 ...\int_{\tau_{k-1}}^\beta\!\!\!\!d\tau_k\left(\frac{K}{2\beta N_c}\right)^k Z_k(\{s_k, \tau_k, x_k\}),\\
&Z_{k}(\stau) \equiv \Tr \prod_{i=k}^{1} e^{ - \Delta \tau_{i} H_{0} } e^{s_{i} \gamma (n_{x_i\uparrow} - n_{x_i\downarrow})}\label{ZN},
\end{align}
with $\Delta \tau_{i} \equiv \tau_{i+1} - \tau_{i}$ for $i<k$ and $\Delta \tau_{k} \equiv \beta-\tau_{k}+\tau_{1}$.

Equation~(\ref{ZN}) is very similar to the equations for the BSS
\cite{BSS81} or Hirsch-Fye \cite{Hirsch86} algorithms (see
also \cite{Georges96}, appendix B1). Using the identity
$\Tr_{d_i^\dagger,d_i}\{e^{-\sum_{ij}d_i^\dagger
A_{ij}d_j}e^{-\sum_{ij}d_i^\dagger
B_{ij}d_j}e^{-\sum_{ij}d_i^\dagger C_{ij}d_j}\}$ $ =$ $ \det$
$\left(1 + e^{-\mathbf{A}}e^{-\mathbf{B}}e^{-\mathbf{C}}\right)$ and following the derivation
in \cite{Gull08_ctaux}, we obtain
\begin{eqnarray}
\frac{Z_{k}(\stau) }{Z_0} &=& \prod_{\sigma=\uparrow,\downarrow} \det \N_\sigma^{-1}(\stau),
\label{Zn_div_Z0}\\
\N^{-1}_{\sigma}(\stau) &\equiv& e^{V_{\sigma}^{\{s_{i}\}}} -\G_{0\sigma}^{ \{\tau_{i},x_i\}} \label{Dyson1}
\Big(e^{V_{\sigma}^{ \{s_{i}\} }} - 1 \Big),\hspace{5mm}
\\
e^{V_{\sigma}^{\{s_{i}\}}}&\equiv&\text{diag}\Big(e^{\gamma (-1)^{\sigma }s_1}, \ldots, e^{\gamma (-1)^{\sigma } s_k}\Big),\hspace{5mm}
\end{eqnarray}
with the notations $(-1)^{\uparrow}  \equiv 1$, $(-1)^{\downarrow}  \equiv -1$ and
$(\G_{0\sigma}^{ \{\tau_{i},x_i\}})_{i,j}=\mathcal{G}^0_{x_ix_j,\sigma}(\tau_i-\tau_j)$ for $i\neq j$,
$(\G_{0\sigma}^{ \{\tau_{i},x_i\}} )_{i,i}=\mathcal{G}^0_{x_ix_i,\sigma}(0^+)$ (we assume in this section that $\mathcal{G}^0_{x_ix_i,\sigma}(0^+)>0$).
As we handle a variable number of time slices at constantly shifting imaginary time locations, it is advantageous to formulate the
algorithm in terms of a matrix $\N_\sigma$, defined by $\G_\sigma = \N_\sigma \G_{0\sigma}$ instead of $\G.$
With Eq.~(\ref{Dyson1}) we express the weight of any (auxiliary spin, time, site) - configuration in terms of the bath Green's function $\mathcal{G}^0_\sigma,$
the constant $\gamma$ defined in Eq.~(\ref{auxdec}), and the determinant of two matrices $\N_\sigma$. The contribution of such a
configuration to the whole partition function is given by Eq.~(\ref{Zn_div_Z0}).

\def\sumSTau {\sum_{k\geq 0} \Big(\frac{K}{2\beta N_c}\Big)^k \sum_{s_i= \pm 1 \atop 1\leq i\leq k}
\int_0^\beta d\tau_1 \ldots  \int_{\tau_{k-1}}^\beta \!\!\!\! d\tau_k }
\begin{figure}[tb]
\begin{center}
\includegraphics[width=0.95\columnwidth]{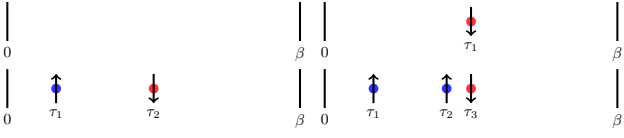}
\end{center}
\caption{Pictorial representation of configurations $\{k;(s_j, \tau_j)\} \in\mathcal{C}$ which are sampled by the CT-AUX algorithm. Diagrams for orders $0, 1, 2,$ and $3$. In this algorithm, an auxiliary
spin $s_j$ (represented here by the red and blue vertices and the direction of the arrows) needs to be sampled in addition to the imaginary time location $\tau_j$ of a vertex.
}
\end{figure}
\subsection{Updates}
In the CT-AUX-algorithm the partition function Eq.~(\ref{Ztr}) consists of a sum over all expansion orders $k$ up to infinity,
another discrete sum over auxiliary fields $s$ and sites $x$, and a $k$-dimensional time-ordered integral from zero to $\beta,$ so  we can employ the sampling scheme of chapter
\ref{SeriesMC}.

In addition to the imaginary time locations of the interaction vertices we also need to sample auxiliary spins $s_j$ associated with each
vertex. Thus, the configuration space $\mathcal{C}$ (Eq.~\ref{Zbasic}) is given by the set
\begin{align} \label{ctaux_confspace}
\mathcal{C}= &\{ \{\}, \{(s_1,\tau_1,x_1)\}, \{(s_1, \tau_1, x_1), (s_2,\tau_2, x_2)\},\\ \nonumber &\cdots, \{(s_1,\tau_1, x_1),\cdots,(s_k,\tau_k, x_k)\},\cdots\},
\end{align}
where the $s_j$ are auxiliary Ising spins that take values $\pm 1$, $k$ is the expansion order, $x_j$ denotes cluster sites and the $\tau_j$ are continuous variables between $0$ and $\beta$, which we assume to be time-ordered, i.e. $\tau_1 < \tau_2 < \cdots < \tau_k.$

Note that this representation is different from the one proposed in \cite{Rombouts99}, where the configuration
space consists of a number $N_\text{max}$ of  fixed ``slots'' at which interaction operators can be inserted into an operator chain (a ``fixed length'' representation).  This leads to additional combinatorial factors in the acceptance probabilities.

Although they are not sufficient for an ergodic sampling, we first consider spinflip updates at constant order which are fast to compute and very useful for reducing autocorrelation times:
\begin{eqnarray}&&((s_1,\tau_1,x_1),\cdots,(s_j,\tau_j,x_j),\cdots,(s_k,\tau_k,x_k)) \nonumber \\ &\rightarrow& ((s_1,\tau_1,x_1),\cdots,(-s_j,\tau_j,x_j),\cdots,(s_k,\tau_k,x_k)),\hspace{5mm}
\end{eqnarray}
the probability density ratios of the two configurations are computed from Eq.~(\ref{Zn_div_Z0}) as:
\begin{align}
R=\frac{p(\x')}{p(\x)} = \frac{\prod_\sigma \det \N_\sigma^{-1}(\staup)}{\prod_\sigma\det \N_\sigma^{-1}(\stau)}.
\end{align}
\begin{figure}[bt]
\begin{center}
\includegraphics[width=0.6\columnwidth]{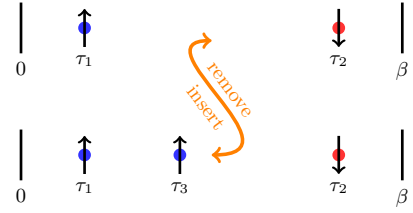}
\end{center}
\caption{An insertion update and its corresponding removal update within the CT-AUX algorithm.}\label{raiselower.fig}
\end{figure}
Vertex insertion updates from configuration $\x=\stau$ to configuration $\y=\staup$, on the other hand, have to be balanced by a removal updates (Fig.~\ref{raiselower.fig}).
The proposal probability accounts for choosing a random time between $0$ and $\beta$, a random site, and a random spin direction:
\begin{align}\label{Wpropupctaux}
W^{\text{prop}}_{\x\y} = \frac{1}{2N_c}\frac{d\tau}{\beta}.
\end{align}
The proposal probability of removing a spin, going from order $k+1$ to order $k$ consists of choosing one of $k+1$ spins:
\begin{align}\label{Wpropdnctaux}
W^{\text{prop}}_{\y\x} = \frac{1}{k+1}.
\end{align}
Therefore we obtain, following Eq.~(\ref{metrogen}):
\begin{align}\label{metro_ctaux}
R_{\x\y} = \frac{K}{k+1}\frac{\det \N_\uparrow(\y) \det \N_\downarrow(\y)}{\det \N_\uparrow(\x) \det \N_\downarrow(\x)}.
\end{align}
The efficient numerical computation of these expressions is discussed in Sec.~\ref{fastupdate} and \ref{spinflips}.

\subsection{Measurements}
\subsubsection{Measurement of the Green's function}\label{gfmeasctaux}
The main observable of interest is the Green's function $G_{pq,\sigma}(\tau,\tau')$ for cluster sites $p$ and $q$ and spin $\sigma$.
First, let us note that we are free to add two additional ``non-interacting''
spins $s=s'=0$ to Eq.~(\ref{ZN}) at any arbitrary time $\tau$ and $\tau'$ (we denote the corresponding matrices of size $n+2$ with a tilde).
$ZG_{pq,\sigma}(\tau,\tau')$ is then given by an expression similar to
Eq.~(\ref{ZN}), with an insertion of $d_\sigma(\tau)$ and $d^{\dagger}_\sigma(\tau')$ at
the corresponding times.
Using the same linear algebra as in Hirsch-Fye (Eq.~(118) of Ref.~\cite{Georges96}) we obtain
\begin{align}
\label{Gfinal}
G_{pq,\sigma} (\tau,\tau')
=&
\frac{1}{Z}
\displaystyle \sumSTau \nonumber\\
&\times Z_k (\stau)
\tilde G_{pq,\sigma} ^{\stau}(\tau,\tau'),
\end{align}
with
$\tilde G_{pq,\sigma} ^{\stau} =
\tilde N_{\sigma,pr}(\stau) \tilde G_{0,rq,\sigma}^{ \{\tau_{i}\}}.$
Since $s=s'=0$,
a block calculation yields
\begin{align}\label{GSreduc}
\tilde G_{pq,\sigma} ^{\stau}(\tau,\tau') &=  \mathcal{G}^0_{pq,\sigma} (\tau, \tau') \\ \nonumber
&+\sum_{l,m=1}^{k}
 \mathcal{G}^0_{px_l,\sigma} (\tau,\tau_{l})
M_{lm}
 \mathcal{G}^0_{x_mq,\sigma}(\tau_m,\tau').\\
M_{lm}&=[ (e^{V_{\sigma}^{\{s_{i}\}} } -1)\N_{\sigma}(\stau)]_{lm},\label{M_ij},
\end{align}
and $G_{pq,\sigma}(\tau,\tau')=\langle \tilde G_{pq,\sigma}(\tau,\tau')\rangle_\text{MC}$.
As in the CT-INT algorithm we may Fourier transform the above expression to obtain a measurement formula in frequency space:
\begin{align}\label{GFomega_ctaux}
\tilde G_{pq}(i\omega_n) &= \mathcal{G}_{pq}^0(i\omega_n) \\ \nonumber&- \sum_{lm}\frac{\mathcal{G}_{pl}^0(i\omega_n)\mathcal{G}_{mq}^0(i\omega_n)}{\beta} e^{i\omega_n \tau_l} M_{lm}e^{-i\omega_n \tau_m}.
\end{align}
By accumulating the Fourier coefficients directly, we avoid 
many of the discretization and related high frequency expansion problems (Sec.~\ref{weakmeas}).

A closer analysis of  Eq.~(\ref{GSreduc}) shows that it is possible and advantageous to measure $S=\M\mathcal{G}^0=\Sigma G$ directly,
as will be discussed in Section~\ref{efficient_measurements}.

\subsubsection{Role of the parameter $K$ -- potential energy}
Similar to the weak-coupling expansion parameter $\alpha$ of Sec.~\ref{weak_chapter}, the parameter $K$ of Eq.~(\ref{HU}) can be freely adjusted.
The average perturbation order $\langle k_{\text{ctaux}} \rangle$ is related to $K$, the potential energy and filling by
\begin{equation}
\langle k_{\text{ctaux}} \rangle_\text{MC} = K-\beta U \langle n_\uparrow n_\downarrow -(n_\uparrow+n_\downarrow)/2 \rangle,
\label{order_ctaux}
\end{equation}
and hence the perturbation order in the continuous-time auxiliary-field method grows linearly with $K$.

%% file: hybridization.tex
\section{Hybridization expansion solvers CT-HYB} \label{hyb_seg_chapter}
\subsection{The hybridization expansion representation}

A complementary approach to the CT-INT and CT-AUX solvers described in chapters \ref{weak_chapter} and \ref{ctaux_chapter} is the hybridization expansion algorithm (CT-HYB) developed  by Werner, Millis, Troyer, and collaborators \cite{Werner06,Werner06Kondo}. 
It proceeds from Eq.~(\ref{Zseries1}) with $H_b$ taken to be the hybridization term $H_\text{hyb}$ and $H_a = H_\text{bath}+H_\text{loc}$.
An advantage of this approach is that the average expansion order for a typical problem near the Mott transition is much smaller than in the interaction expansion methods and therefore lower temperatures are accessible \cite{Gull07}. General interactions can easily be treated as long as the local Hilbert space is not too large. The first paper, \cite{Werner06}, presented an algorithm and applications for the single impurity Anderson model. A generalization to multi-orbital models with complex interactions and the Kondo model  soon followed \cite{Werner06Kondo},  and this formalism was later extended by \textcite{Haule07} who introduced the ideas of basis truncation and sector statistics, and implemented the algorithm for models with off-diagonal hybridization functions. 

Since $\Hhyb$ $=$ 
$\sum_{pj}$ $(V_p^j c_{p}^\dagger d_j  + V_p^{j*}d_j^\dagger c_{p}) = \Hhybtwid$ 
$+$ $\Hhybtwid^\dagger$  contains two terms which create and annihilate electrons on the impurity, respectively, 
only even powers of the expansion and contributions with equal numbers of $\Hhybtwid$ and $\Hhybtwid^\dagger$ 
can yield a non-zero trace. The partition function therefore becomes 
\begin{align}
Z &= \sum_{k=0}^\infty \int_0^\beta d\tau_1 \ldots \int_{\tau_{k-1}}^\beta d\tau_k 
                       \int_0^\beta d\tau_1' \ldots  \int_{\tau_{k'-1}}^\beta d\tau_k' \\ \nonumber 
&\times\Tr\left[T_\tau e^{-\beta H_a} \Hhybtwid(\tau_k) \Hhybtwid^\dagger(\tau_k') \ldots \Hhybtwid(\tau_1) \Hhybtwid^\dagger(\tau_1')\right].
\end{align}
Inserting the $\Hhybtwid$ and $\Hhybtwid^\dagger$ operators explicitly yields
\begin{align}
Z &= \sum_{k=0}^\infty\int_0^\beta d\tau_1 \ldots \int_{\tau_{k-1}}^\beta d\tau_k 
                       \int_0^\beta d\tau_1' \ldots  \int_{\tau_{k-1}'}^\beta d\tau_k' \\ \nonumber
&\sum_{j_1, \cdots j_k \atop j_1', \cdots j_k'}\sum_{p_1, \cdots p_k \atop p_1',\cdots p_k'} V^{j_1}_{p_1}V^{j_1'*}_{p_1'} \cdots V^{j_k}_{p_k}V^{j_k'*}_{p_k'}  \\ \nonumber
&\times\Tr \Big[T_\tau e^{-\beta H_a} d_{j_k}(\tau_k)c^\dagger_{p_k}(\tau_k)c_{p_{k'}}(\tau_k')d_{j_k'}^\dagger(\tau_k')\\ &\nonumber\cdots d_{j_1}(\tau_1)c_{p_1}^\dagger(\tau_1) c_{p_1'}(\tau_1')d_{j_1'}^\dagger(\tau_1')\Big].
\end{align}
Separating the bath and impurity operators we obtain 
\begin{align}
Z &=\sum_{k=0}^\infty\int_0^\beta d\tau_1 \ldots \int_{\tau_{k-1}}^\beta d\tau_k 
                       \int_0^\beta d\tau_1' \ldots  \int_{\tau_{k-1}'}^\beta d\tau_k' \\ \nonumber
&\sum_{j_1, \cdots j_k \atop j_1', \cdots j_k'}\sum_{p_1, \cdots p_k \atop p_1',\cdots p_k'} V^{j_1}_{p_1}V^{j_1'*}_{p_1'} \cdots V^{j_k}_{p_k}V^{j_k'*}_{p_k'}  \nonumber\\ \nonumber
&\times\Tr_d\left[T_\tau e^{-\beta H_\text{loc}} d_{j_k}(\tau_k) d_{j_k'}^\dagger(\tau_k')\cdots d_{j_1}(\tau_1)d_{j_1'}^\dagger(\tau_1')\right] \\ \nonumber
&\times\Tr_c\left[T_\tau e^{-\beta H_\text{bath}} c^\dagger_{p_k}(\tau_k)c_{p_{k'}}(\tau_k')\cdots c_{p_1}^\dagger(\tau_1) c_{p_1'}(\tau_1')\right].
\end{align}
We can now integrate out the bath operators $c_p(\tau)$, since they are non-interacting and the time-evolution (given by $H_a$) no longer couples the impurity and the bath.  Defining the bath partition function
\begin{align}
Z_\text{bath} = \Tr e^{-\beta H_\text{bath}} = \prod_\sigma \prod_{p} (1+e^{-\beta \varepsilon_{p}}),
\end{align}
and the anti-periodic hybridization function $\mathbf{\Delta}$ (Eq.~(\ref{hybfun})),
\begin{align}
\Delta_{lm}(\tau) = \sum_p \frac{V^{l*}_p V^m_p}{e^{\varepsilon_p\beta}+1} \times \left\{ \begin{array}{ll}-e^{-\varepsilon_p(\tau-\beta)},& 0<\tau<\beta \\ e^{-\varepsilon_p\tau},&  -\beta<\tau<0\end{array}\right.,
\end{align}
we obtain the determinant 
\begin{eqnarray}
&&\frac{1}{Z_\text{bath}}\Tr_c \Big[T_\tau e^{-\beta H_\text{bath}} \sum_{p_1,\cdots p_k} \sum_{p_1',\cdots p_k'}  V^{j_1}_{p_1}V^{j_1'*}_{p_1'} \cdots V^{j_k}_{p_k}V^{j_k'*}_{p_k'} \nonumber \\
&&\times c^\dagger_{p_k}(\tau_k) c_{p_{k'}}(\tau_k')\cdots c_{p_1}^\dagger(\tau_1) c_{p_1'}(\tau_1')\Big] = \det \mathbf{\Delta},\label{fcdet}
\end{eqnarray}
for an arbitrary product of bath operators. Here, $\mathbf{\Delta}$ is a $k\times k$  matrix with elements 
$ \Delta_{lm} =  \Delta_{j_lj_m}(\tau_l - \tau_m)$. 
In practice, and in analogy to the algorithms in previous sections, it will be more convenient to handle the inverse of this matrix $\mathbf{\Delta}$, which we denote by $\mathbf{M}=\mathbf{\Delta}^{-1}$ (see Sec.~\ref{fastupdate}). 

The partition function expansion for the hybridization algorithm now reads (for time-ordered configurations)
\begin{eqnarray}
Z &=&  Z_\text{bath} \sum_k \iiint d\tau_1 \cdots d\tau_k' \sum_{j_1, \cdots j_k}\sum_{j_1', \cdots j_k'} \label{FullExpHyb}\\ 
&\times&\Tr_d\left[T_\tau e^{-\beta H_\text{loc}} d_{j_k}(\tau_k) d_{j_k'}^\dagger(\tau_k')\cdots d_{j_1}(\tau_1)d_{j_1'}^\dagger(\tau_1')\right] \det \mathbf{\Delta}.\nonumber
\end{eqnarray}
If the coupling to the bath is diagonal in the ``flavor'' (spin, site, orbital, \textellipsis) indices $j$, then $\mathbf{\Delta}$ is a block-diagonal matrix and Eq.~(\ref{FullExpHyb}) simplifies to
\begin{eqnarray}\label{SingleOrbitalExpHyb}
Z &=&  Z_\text{bath}\prod_j \sum_{k_j=0}^\infty \int_0^\beta d\tau^j_1 \ldots \int_{{\tau'}^j_{k_j-1}}^\beta d{\tau'}^j_{k_j} \\
&\times& \Tr_d\Big[T_\tau e^{-\beta H_\text{loc}} d_j(\tau^j_{k_j}) d_j^\dagger(\tau'^j_{k_j})\ldots d_j(\tau^j_{1})d_j^\dagger(\tau'^j_1)\Big] \det \mathbf{\Delta}_j.\nonumber
\end{eqnarray}

\subsection{Density - density interactions}\label{segment_section}

\begin{figure}[tb]
\begin{center}
\includegraphics[width=0.95\columnwidth]{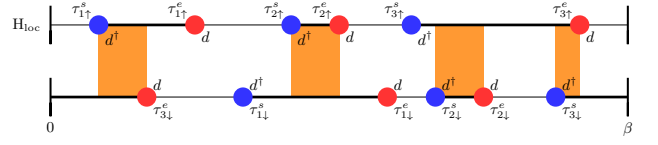}
\end{center}
\caption{Segment representation of term in hybridization expansion of single orbital Anderson model. Upper line: spin up orbital, lower line, spin down orbital: heavy line, orbital occupied; light line, orbital empty. For each orbital, length  of  black line (occupied orbitals) determines the chemical potential contribution to the weight  factor (\ref{localweight}). Shaded areas: regions where both up and down orbitals are filled, so the impurity is doubly occupied. The length of the shaded area enters into an overall weighting
factor for the potential energy (Hubbard $U$).}
\label{SegmentOverlap}
\end{figure}

We first consider (multi-orbital) models with density-density interactions.
In this case, the local Hamiltonian $H_{\text{loc}}$ commutes with the occupation number operator of each orbital.
We may therefore represent the time evolution of the impurity by collections of ``segments" which represent time intervals in which an electron of a given flavor resides on the impurity. An example of such a segment configuration for a single orbital model (two spin flavors) is shown in Fig.~\ref{SegmentOverlap}.

Since the local Hamiltonian is diagonal in the occupation number basis the contribution of the trace factor can be computed for each segment configuration. For a model with $n$ orbitals and a total length $L_j$ of segments in orbital $j$ and a total overlap $O_{ij}$ between segments of flavor $i$ and $j$ one obtains ($s$ is a sign depending on the operator sequence)
\begin{align}\label{localweight}
w_\text{loc}(\x) = \text{Tr}_d[\ldots] = s e^{\mu \sum_{j}^{n}L_j }e^{-\sum_{i<j}^{n}(U_{ij}O_{ij})},
\end{align}
except in the trivial case where there are no operators for certain flavors. In the latter case, several segment configurations, involving ``full" and ``empty" lines, contribute to the trace.

\subsection{Formulation for general interactions}\label{hyb_matrix_section}

\begin{figure}[tbh]
\includegraphics[width=0.95\columnwidth]{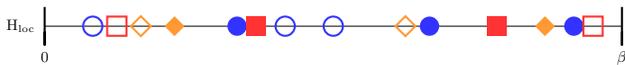}
\begin{center}
\caption{A typical term in the expansion (\ref{FullExpHyb}): three ``flavors'' (red, blue, yellow)  of fermionic creation and annihilation
operators (denoted by filled and empty triangles, squares, and circles) are placed at times between $0$ and
$\beta$. In the general case, orbital occupation is not conserved by local Hamiltonian so  two operators of the same type may
follow each other.}
\label{HybridizationGeneralConf.fig}
\end{center}
\end{figure}

If $H_\text{loc}$ is not diagonal in the occupation number basis defined by the $d^\dagger_\alpha$, a separation of flavors, as in the segment formalism, is no longer possible (see illustration in Fig.~\ref{HybridizationGeneralConf.fig}) and the calculation of 
$
w_\text{loc}(\x) = \text{Tr}_d \big[T_\tau e^{-\beta H_\text{loc}} \prod_{\alpha} 
d_\alpha(\tau_{k_\alpha}^\alpha)d^\dagger_\alpha(\tau_{k_\alpha}'^\alpha)
\ldots
d_\sigma(\tau_1^\alpha)d^\dagger_\alpha(\tau_1'^\alpha) 
\big]
$
becomes more involved. One strategy, proposed in
\cite{Werner06Kondo} is to represent the operators $d_\alpha$ and $d^\dagger_\alpha$ as matrices in the eigenbasis of $H_\text{loc}$  because in this representation the time evolution operators $e^{-H_\text{loc}\tau}$ become diagonal. 
The evaluation of the trace factor thus involves the multiplication of matrices whose size is equal to the size of the Hilbert space of $H_\text{loc}$. Since the dimension of the Hilbert space grows exponentially with the number of flavors, the calculation of the trace factor becomes the computational bottleneck of the simulation, and the matrix formalism is therefore restricted to a relatively small number of flavors ($\lesssim 10$). The technical part of evaluating these traces is described in detail in Sec.~\ref{tracesec}.

\textcite{Haule07} observed that  conserved quantum numbers may be exploited to facilitate the calculation of the trace. If the eigenstates of $H_\text{loc}$ are ordered according to conserved quantum numbers, the evaluation of the trace is reduced to block matrix multiplications (see Sec.~\ref{tracesec}) of the form
\begin{align}
w_\text{loc}(\x)=&\sum_{\text{contr.} m} \Tr_m\Big[\cdots \\ \nonumber \cdots &(O)_{m'',m'}(e^{-(\tau'-\tau) H_\text{loc}})_{m'}(O)_{m',m}(e^{-\tau H_\text{loc}})_m\Big],
\end{align}
where $O$ is either a creation or annihilation operator, $m$ denotes the index of the matrix block, and the sum runs over those sectors which are compatible with the operator sequence.  With this technique, 3-orbital models or 4-site clusters can be simulated efficiently \cite{Werner08nfl,Chan09,Haule07plaquette,Park08,Gull08_plaquette}. However, since the matrix blocks are dense and the largest blocks still grow exponentially with system size, the simulation of 5-orbital models becomes already quite expensive and the simulation of 7-orbital models with 5, 6 or 7 electrons is only feasible with current computer resources if the simulation is restricted to a few valence states and, within this subspace, the maximum  size of the blocks is  truncated (see \ref{truncation_sec}). Simulations based on such a truncated version of the matrix formalism were used in \cite{Shim07Pu,Marianetti08}. The Krylov method described in the next section avoids truncations to a large extent. 

\subsection{Krylov implementation}\label{Krylov}

An alternative  strategy to evaluate the trace in Eq.~(\ref{FullExpHyb}) was proposed in \cite{Laeuchli09} based on the observation that in the occupation number basis both  the $d_i^{(\dagger)}$-operator matrices and $H_\text{loc}$ are typically very sparse, so the $d_i^{(\dagger)}$-operators  can easily be applied to any given state while efficient Krylov-space methods can be used to evaluate the imaginary time evolution. This implementation involves only matrix-vector multiplications with sparse operators $d^{(\dagger)}$ and $H_\text{loc}$, and is thus doable even for systems for which the multiplication  of dense matrix blocks becomes prohibitively expensive. Furthermore, no explicit truncation of states of the local Hamiltonian is required, so that all excited states remain accessible at intermediate times $\tau$ in the trace. The outer trace may be approximated by a sum over the lowest energy states. If this is done  it is important to measure the various local observables at $\tau=\beta/2$ in order to be least affected by the truncation  of the trace at $\tau=0$ (and equivalently at $\tau=\beta$). 

The complexity of the Krylov algorithm is $O(N_ \text{dim} \times N_ \text{tr} \times N_\text{hyb} \times N_\text{iter} )$,
where $N_\text{dim}$ is the size of the impurity Hilbert space, $N_\text{tr}$ the number of states kept in the outer trace, $N_\text{hyb}$ the number of hybridization events, and $N_\text{iter}$ the number of Krylov iterations used for the calculation of the time evolution from one operator to the next. In Ref.~\cite{HochbruckLubich97} it has been shown rigorously that  these Krylov space algorithms converge rapidly as a function of $N_\text{iter}$, typically reaching convergence for very small iteration numbers $p\ll N_\text{dim}$, although the number of iterations depends on the  time interval $\tau$. In the worst case where all states in the trace are retained ($N_\text{tr}=N_\text{dim}$) and the complexity scales as $N_\text{dim}^2,$ where as in the  best case $N_\text{tr}=O(1)$ and the complexity is linear in the dimension of the Hilbert space. In comparison, the complexity of the approach described in Sec.~\ref{hyb_matrix_section} is cubic in $N_\text{dim}$. \textcite{Laeuchli09} showed that the Krylov approach with outer trace truncated to the lowest energy states becomes favorable  for models with more than 4 orbitals (or 4 sites). The systematic error resulting from the truncation of the outer trace becomes negligible at  temperatures below a few percent of the bandwidth. The Krylov-based hybridization expansion thus provides a method for the systematic investigation  of larger problems such as the dynamical mean field theory of  transition metal and actinide compounds.

\subsection{Updates} \label{hyb_sampling}
In order to sample Eq.~(\ref{FullExpHyb}) we perform a Monte Carlo simulation as described in Sec.~\ref{SeriesMC}.
We explain the sampling procedure for the formulation with density-density interactions. The two basic updates required for ergodicity are the insertion and the removal of a segment. 

\begin{figure}[t] 
  \begin{center}
  \includegraphics[width=0.6\columnwidth]{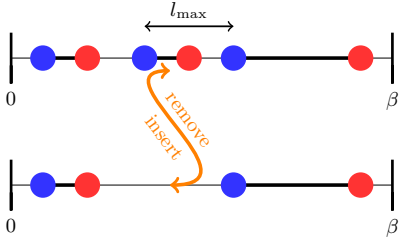}
  \end{center}
  \caption{An insertion update and its corresponding removal update within the hybridization algorithm.}\label{raiselowerh}
\end{figure}

Starting from a  configuration of segments $\x_k = \{(\tau_1^s, \tau_1^e), (\tau_2^s, \tau_2^e),\cdots,(\tau_k^s, \tau_k^e)\}$
we attempt to insert a new segment $s_{k+1}$ starting at $\tau^{s}$ to obtain a configuration $\y_{k+1}$. This move is rejected if $\tau^s$ lies on one of the existing segments, since we cannot create two identical fermions at the same site.
Otherwise, we choose a random time uniformly in the interval $[\tau^s,\tau^{s'})$ of length $l_\text{max}$ (Fig.~\ref{raiselowerh}), where $\tau^{s'}$ is 
the start of the next segment in $\x_k$. For the reverse move, the proposal probability 
is given by the probability of selecting that given segment for removal.

Therefore the proposal probabilities are
\begin{align}
W^{\text{prop}}_{\x\y} = \frac{d\tau^2}{\beta l_{\text{max}}}, \\
W^{\text{prop}}_{\y\x} = \frac{1}{k+1},
\end{align}
and the acceptance ratio becomes
\begin{equation}
\label{segacrat}
R_{\x\y}= \frac{p_\y W^{\text{prop}}_{\y\x}}{p_\x W^{\text{prop}}_{\x\y}} = 
\frac{\beta l_{\text{max}}}{k+1}\frac{w_\text{loc}(\y)\det \mathbf{\Delta}(\y) }{w_\text{loc}(\x) \det \mathbf{\Delta}(\x) }.
\end{equation}
\begin{figure}[tb]
\begin{center}
\includegraphics[width=0.95\columnwidth]{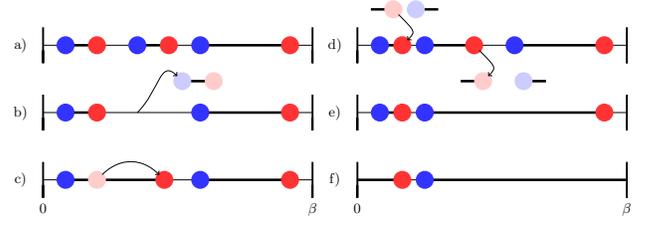}
\end{center}
\caption{
Updates of the hybridization algorithm as described in the text: 
(a) original configuration; (b) removal of a segment; (c) shift of an end point of a segment; (d) insertion of an antisegment; (e) removal of an antisegment; (f) removal of another antisegment such that the remaining segment "wraps" around $\beta$.}
\label{SegmentMoves}
\end{figure}

An important second update, equivalent to the insertion of a segment, is the insertion of an ``antisegment'': 
the insertion of a annihilator-creator pair istead of a creator-annihilator pair. The formulae for the acceptance ratio are the same as Eq.~(\ref{segacrat}).
Besides smaller autocorrelation times these updates cause the two zero-order contributions ``full occupation'' and ``no segment'' 
to be treated on equal footing.

Further updates, like the shift of a segment or the shift of one or both end points do not change the order
of the expansion, but help to reduce autocorrelation times. The shift moves are ``self - balancing'' (proposal probabilities for shift moves and their inverse are the same), so
\begin{align}
R_{\x\y} =\frac{w_\text{loc}(\y)\det \mathbf{\Delta}(\y) }{w_\text{loc}(\x)\det \mathbf{\Delta}(\x)}.
\end{align}

Global updates (Sec.~\ref{globalupdates}), e.g. the exchange of all segments of two orbitals, may be required to ensure ergodicity, {\it i.e.} that the random walk 
does not get trapped in one part of phase space.
Such updates require the configuration to be recomputed from scratch, and are in general of order $O(k^3)$. 
They are essential in calculations of magnetic phase boundaries \cite{Kunes09,Chan09,Poteryaev08}.

\subsection{Measurements} \label{segment_meas}

The CT-HYB algorithm generates configurations with the weight that they contribute to the partition function $Z$. 
To obtain expectation values of an observable we can either simulate the series of that observable (which, for the  Green's
function, corresponds to the ``worm'' algorithm described in Sec.~\ref{wormchapter}), or estimate the observable according to Eq.~(\ref{MCObs}).

\begin{figure}[bt]
\begin{center}
\includegraphics[width=0.95\columnwidth]{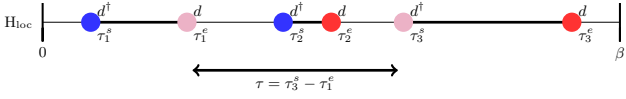}
\end{center}
\caption{Hybridization algorithm: Green's function configuration. 
A typical configuration for a Green's function, created by taking the partition function configuration of order $k=3$ and identifying the creation
operator at $\tau_3^s$ and the annihilation operator at $\tau_1^e$ as the Green's function operators to obtain a Green's function configuration corresponding to a partition function configuration at one order lower.
Red: creation and blue: annihilation operators of the partition function. Light purple: Green's function operators.}
\label{GreensConf}
\end{figure}

The single most important observable for quantum Monte Carlo impurity solvers is the finite temperature imaginary time Green's function
$G_{lm}(\tau) = -\langle T_\tau d_l(\tau)d_m^\dagger(0)\rangle$. The series for this observable is
\begin{align}\label{GFDiagramsHyb}
&G_{lm}(\tau_l, \tau_m) = - Z_\text{bath} \!\!\!\sum_{k,{j_1, \cdots j_k \atop j_1', \cdots j_k'}}\!\! \int d\tau_1...d\tau_k' \det \mathbf{\Delta}_k \nonumber 
\Tr_d\Big[T_\tau 
e^{-\beta H_\text{loc}} \\
&d_l(\tau_l) d_m^\dagger(\tau_m) d_{j_k}(\tau_k) d_{j_k'}^\dagger(\tau_k')
\ldots d_{j_1}(\tau_1)d_{j_1'}^\dagger(\tau_1')\Big].
\end{align}
This shows that Green's function configurations at expansion order $k$ are partition function configurations at expansion order $k$ with additional $d_l$ and $d_m^\dagger$ operators or,
alternatively, partition function operators at order $k+1$ with no hybridization line connecting to $d_l(\tau_l)$ and $d_m^\dagger(\tau_m)$.
In practice we obtain an estimator of $G_{lm}(\tau_l, \tau_m)$ by identifying two operators $d_l(\tau_l), d_m^\dagger(\tau_m)$ in a partition function configuration 
that are an imaginary time distance $\tau=\tau_l-\tau_m$ apart, and removing the hybridization line connecting them (see Fig.~\ref{GreensConf}). 
The insertion of local operators into a partition function configuration, as it is done in the interaction expansion formalism, is not ergodic in the hybridization expansion.

The size $(k-1)\times(k-1)$ hybridization matrix $\mathbf{\Delta}_{k-1}^{\tau_l,\tau_m}$ of all hybridization operators except for $d_l(\tau_l) $ and $d_m^\dagger(\tau_m)$ corresponds to $\mathbf{\Delta}$ with the column/row $s_l$ and $s_m$ corresponding to the operators $d_l$ and $d_m^\dagger$ removed, and the weight of a Green's function configuration $G_{lm}(\tau_l,\tau_m)$ is 
\begin{align}
\frac{p_{G_{lm}}}{Z} = \frac{\det \mathbf{\Delta}_{k-1}^{\tau_l, \tau_m}}{\det \mathbf{\Delta}}.
\end{align}
An expansion by minors or the inverse matrix formulas of Sec.~\ref{fastupdate} describe how such a determinant ratio is computed:
\begin{align}\label{hybgfmeas}
\frac{p_{G_{lm}}}{Z}=(\bf{\Delta})^{-1}_{s_m s_l}=M_{s_m s_l}.
\end{align}
We can bin this estimate into fine bins to obtain the Green's function estimator
\begin{align}
G_{lm}(\tau)&=\frac{1}{\beta}\left\langle \sum_{ij}^k M_{ji}\tilde{\delta}(\tau, \tau_m-\tau_l) \delta_{t(i)l}\delta_{t(j)m}\right\rangle_\text{MC}\!\!\!\!\!\!,\\
\tilde{\delta}(\tau, \tau')&=\left \{ {\delta(\tau-\tau'),\ \  \tau' > 0\atop -\delta(\tau-\tau'-\beta),\ \  \tau'<0,} \right.
\label{hybmeas}
\end{align}
with $t(i)$ denoting the orbital index of the operator at row / column i.
For a configuration at expansion order $k$ we obtain a total
of $k^2$ estimates for the Green's function -- or one for every creation-annihilation operator pair or every single element of the $(k \times k)$-matrix  $\mathbf{M}=\mathbf{\Delta}^{-1}$.
The measurement in Eq.~(\ref{hybmeas}) may suffer from bad statistics if very few hybridization lines are present ($k$ is small) in an orbital.
In this case, the Green function measurement should be based on the insertion of operators.

In the segment representation, very efficient estimators exist for the density, the double occupancy and the potential energy (and similarly for all observables that commute with the local Hamiltonian):
\begin{align}
E_\text{pot} &= \sum_{i>j}U_{ij} D_{ij},\\ 
D_{ij} &= \langle n_{i} n_{j} \rangle_\text{MC}.
\end{align}
The occupation $n_j$ of the $j$th flavor is estimated by the length $L_j$ (Eq.~\ref{localweight}) of all the segments: $n_j = \langle L_j/\beta\rangle.$
Double occupancies and interaction energies are obtained from the overlap $O_{ij}$ of segments as $D_{ij}=\langle O_{ij} \rangle/\beta$.
The system has a total magnetization of $S_{z} = (\langle L_{\uparrow}^\text{tot} - L_{\downarrow}^\text{tot}\rangle)/\beta$.
Overlaps and lengths of segments are computed at every Monte Carlo step, and thus these observables may be obtained with high accuracy at essentially no additional cost. 

Finally the average expansion order of the algorithm is an estimator for the kinetic energy \cite{Haule07}, similar to $E_\text{pot}$ in the case of the CT-INT and CT-AUX algorithms:
\begin{align}
E_\text{kin} = \frac{1}{\beta}\langle k \rangle_\text{MC}.
\end{align}

%% file: infiniteU.tex
\section{Infinite-$U$ method CT-J}\label{infinite-U}

\subsection{Overview}
In many cases the physics of interest is captured by low energy effective theories in which some (often high energy) degrees of freedom have been integrated out, leaving a model described by a restricted Hilbert space. A standard example is the ``Schrieffer-Wolf'' transformation which obtains the Kondo Hamiltonian (describing a single $S=1/2$ spin exchange-coupled to a conduction band) as the low energy theory of the single-impurity Anderson model in the regime where the charge fluctuations are suppressed and the impurity is occupied by only one electron.

The projection is conceptually  advantageous, because it allows one to focus on the important degrees of freedom. There is also a computational advantage: while the CT-HYB method  accomplishes the projection (because the simulation produces a large weight for the relevant states and a small weight for the states which are projected out), transitions between the states in the low energy manifold  require excursions to states 
with very small weight,
leading to large auto-correlation times. The direct study of a projected Hamiltonian avoids this problem. 

Otsuki and collaborators have presented a CT-QMC method for dealing with projected Hamiltonians \cite{Otsuki07}. Their papers focus on a particular class of ``Coqblin-Schrieffer'' (CS) or generalized Kondo models arising in the context of the physics of heavy fermion compounds. We follow their presentation here.
be applicable to a much wider range of downfolded models.

Coqblin-Schrieffer  models arise when an impurity spends most of its time in a state of definite charge, with occasional virtual fluctuations into different charge states. An example is the Anderson model, Eq.~(\ref{AIM}), in the large $U$, weak $V$ limit where, if the level energy is correctly tuned, at almost all times the impurity is occupied by one electron which may be of spin up or down. Fluctuations into a state with density $n=0$ or $n=2$ followed by a return to a state $n=1$ allow the impurity to exchange spin with the bath. More generally, the dominant charge state will have an $N$-fold degeneracy including spin and orbital degrees of freedom and virtual transitions will lead to a variety of exchange processes which may be encoded in a Hamiltonian of the Coqblin-Schrieffer form \cite{Coqblin69}
\begin{align}
H_\text{CS} = H_\text{bath}+H_\text{spin}+H_J
\end{align}
with impurity states labeled by a spin- or orbital quantum number $\alpha$ and a bath described by an energy quantum number $k$ and a spin/orbital quantum number $b$:
\begin{align}
H_\text{bath}&=\sum_{kb}\varepsilon_{k} c_{kb}^\dagger c_{kb},\\
H_\text{spin}&=\sum_\alpha E_\alpha X_{\alpha\alpha},\\
H_\text{J}&=-\sum_{\alpha\alpha',bb'\atop kk'}J_{\alpha\alpha'}^{kk',bb'}X_{\alpha\alpha'}c_{kb}c_{k'b'}^\dagger.
\end{align}
Here $X_{\alpha\alpha'} = |\alpha\rangle\langle\alpha'|$ and without loss of generality we have chosen a basis in which the impurity (spin) Hamiltonian is diagonal. The exchange parameters $J$ are typically of order $V^2/U$ and in most treatments the $k$-dependence is neglected. Furthermore, in the applications presented to date the spin-orbit quantum numbers of the bath electrons are those of the impurity states and are conserved so that $H_J \rightarrow H_J^{CS}$ where
\begin{align}
H_J^\text{CS} = -\sum_{\alpha\alpha'}J_{\alpha\alpha'}X_{\alpha\alpha'} c_{\alpha}c_{\alpha'}^\dagger, \label{CS}
\end{align}
and $c_\alpha=1/\sqrt{N}\sum_k c_{k\alpha}$.
The consequences of removing this approximation are an important open question.

The formalism of Otsuki {\it et al.} follows Eq.~(\ref{Zseries1}) with $H_J$ playing the role of the expansion term $H_b$. While formally this is a perturbative expansion in an interaction parameter, it is in a practical sense closely related to the hybridization expansion: each vertex changes the state of the impurity, just as does the $V$-term in CT-HYB; the difference is that at each event, one electron and one hole is created. In what follows we summarize the main features of the CT-J algorithm, following the presentation by \textcite{Otsuki07}.

\subsection{Partition function expansion}

The partition function $Z$ divided by the conduction electron contribution $Z_\text{bath} = \text{Tr}_{c} {\rm e}^{-\beta H_\text{bath}}$ is
\begin{align}
	\frac{Z}{Z_\text{bath}} =& \sum_{k=0}^{\infty} (-1)^k
	 \int_0^{\beta} {\rm d}\tau_1 \cdots \int_{\tau_{k-1}}^{\beta} {\rm d}\tau_k
	 \sum_{\alpha_1 \alpha'_1} \cdots \sum_{\alpha_k \alpha'_k} \nonumber\\
	 &\times J_{\alpha_1 \alpha_1'} \cdots J_{\alpha_k \alpha_k'} \nonumber\\
	 &\times s \prod_{\alpha} \Big\langle T_{\tau} c_{\alpha}^{\dag}(\tau_1') c_{\alpha}(\tau_1'') \cdots c_{\alpha}^{\dag}(\tau_{k_{\alpha}}') c_{\alpha}(\tau_{k_{\alpha}}'') \Big\rangle_{\rm c} \nonumber\\
	 &\times \text{Tr}_\text{spin} \Big[ T_{\tau} {\rm e}^{-\beta H_\text{spin}}
	 X_{\alpha_1 \alpha_1'}(\tau_1) \cdots X_{\alpha_k \alpha_k'}(\tau_k) \Big].
\end{align}
Here, the conduction electron operators are grouped by component index $\alpha$ (a resultant sign in the permutation is represented by $s$), $k_{\alpha}$ is the number of operators $c_{\alpha}^{\dag} c_{\alpha}$ for each component $\alpha$, and $\sum_{\alpha} k_{\alpha}=k$. We also used the notation $\langle \cdots \rangle_{c} = Z_\text{bath}^{-1} \text{Tr}_{c} [ {\rm e}^{-\beta H_\text{bath}} \ldots ]$. Wick's theorem for the conduction electrons implies
\begin{align}
	\frac{Z}{Z_\text{b}}  &= 
	\sum_{k=0}^{\infty} 
	 \int_0^{\beta} {\rm d}\tau_1 \cdots \int_{\tau_{k-1}}^{\beta} {\rm d}\tau_k
	 \sum_{\alpha_1 \alpha'_1} \cdots \sum_{\alpha_k \alpha'_k} w_k, \\
	w_k &= (-1)^k J_{\alpha_1 \alpha_1'} \cdots J_{\alpha_k \alpha_k'}
	\cdot s \prod_{\alpha} \det \D_{\alpha}^{(k_{\alpha})} \nonumber\\
	&\times
	\text{Tr}_\text{spin} \Big[ T_{\tau} {\rm e}^{-\beta H_\text{spin}}
	 X_{\alpha_1 \alpha_1'}(\tau_1) \cdots X_{\alpha_k \alpha_k'}(\tau_k) \Big].
\label{eq:part_func}
\end{align}
The $k_{\alpha} \times k_{\alpha}$ matrix $\D_{\alpha}^{(k_{\alpha})}$ is defined by $(\D_{\alpha}^{(k_{\alpha})})_{ij} = \mathcal{G}^0_{\alpha} (\tau_i'' - \tau_j')$ with $\mathcal{G}^0_{\alpha} (\tau) = -\langle T_{\tau} c_{\alpha}(\tau) c_{\alpha}^{\dag}(0) \rangle_{\rm c}$ and 
$w_k$ is the weight of a Monte Carlo configuration of order $k$. This configuration can be represented in terms of 
the numbers $\{ \tau_i \}=(\tau_1, \cdots, \tau_{k})$ and 
$\{ \alpha_i \}=(\alpha_1, \cdots, \alpha_{k})$, or, as illustrated in Fig.~\ref{fig:diagram}, by a decomposition of the imaginary time interval into $k$ segments $[\tau_i,\tau_{i+1})$ (modulo periodic boundary condition) with flavor $\alpha_i$. 
These variables define the sequence of $X$-operators 
\begin{align}
	X_{\alpha_{k} \alpha_{k-1}}(\tau_{k}) \cdots X_{\alpha_i \alpha_{i-1}} (\tau_i) \cdots
	X_{\alpha_1 \alpha_{k}}(\tau_1),
\end{align}
and a corresponding sequence of $c$-operators:
\begin{align}
	&(-1)^{k+1}
	c_{\alpha_{k}}^{\dag} (\tau_1) c_{\alpha_{k}}(\tau_{k}) \ldots
	c_{\alpha_{i}}^{\dag} (\tau_{i+1}) c_{\alpha_{i}}(\tau_{i}) \ldots \nonumber\\
	&\hspace{5mm}\ldots c_{\alpha_{1}}^{\dag} (\tau_2) c_{\alpha_{1}}(\tau_{1}).
\end{align}

An expression equivalent to  Eq.~\ref{eq:part_func} was presented in the landmark 1970 Anderson-Yuval study of the Kondo model \cite{Yuval70}, but at that time could not be used as a starting point for numerical calculations.

\subsection{Updates}

Updates which change the order $k$ are required for ergodicity, and updates which shift one of the operators increase sampling efficiency. 
In this section, we discuss updates which change the perturbation order by $\pm 1$. Note that if some coupling constants are 0, the straightforward sampling may not be ergodic. For example, when the interaction lacks diagonal elements in the $N=2$ model, the perturbation order must be changed by $\pm 2$. We refer the reader to \cite{Otsuki07} for a discussion of updates which insert or remove several operators.  

\begin{figure}[t]
	\begin{center}
	\includegraphics[width=0.95\columnwidth]{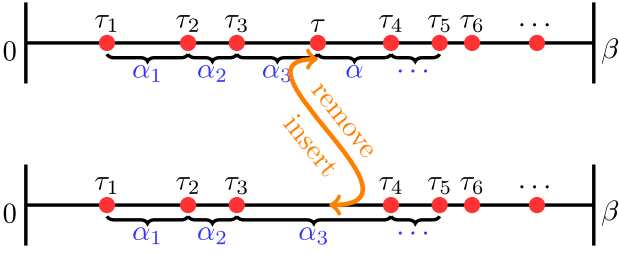}
	\end{center}
	\caption{Illustration of an insertion of a segment. The diagrams represent the configurations of $\{ \tau_i \}$ and $\{ \alpha_i \}$.}
	\label{fig:diagram}
\end{figure}

Let us consider the process of adding $\tau$ and $\alpha$, which are randomly chosen in the range $[0, \beta)$ and from the $N$ components, respectively. 
If $\tau$ satisfies $\tau_{n+1} > \tau > \tau_{n}$, $\{ \tau_i \}$ and $\{ \alpha_i \}$ change into
$(\tau_1, \cdots, \tau_n, \tau, \tau_{n+1}, \cdots, \tau_{k})$ and 
$(\alpha_1, \cdots, \alpha_n, \alpha, \alpha_{n+1}, \cdots, \alpha_{k})$, respectively. 
In other words, one of the $X$-operators is replaced by
\begin{align}
	X_{\alpha_{n+1} \alpha_n} (\tau_{n+1})
	\rightarrow
	X_{\alpha_{n+1} \alpha} (\tau_{n+1}) X_{\alpha \alpha_n} (\tau),
\end{align}
which corresponds to the change illustrated in Fig.~\ref{fig:diagram}: a segment $\alpha$ is inserted between $\alpha_n$ and $\alpha_{n+1}$ with shortening of the segment $\alpha_n$. 
In the corresponding removal process, one removes a randomly chosen segment. 

Following the  discussion in Sec. \ref{SeriesMC} and taking into account the proposal probabilities $d\tau/N\beta$ and $1/(k+1)$ for insertion and removal ($N$ is the number of local states) the ratio $R$ of Eq.~(\ref{transitionrate}) becomes
\begin{align}
	R = \frac{p_{k+1}}{p_k} \frac{N \beta}{k+1},
\end{align}
with $p_k = w_k d\tau_1 \dots d\tau_k$ as in Eq.~(\ref{pdef}), where for $k \neq 0$, the ratio $p_{k+1}/p_k$ is given by
\begin{align}
	\frac{p_{k+1}}{p_k}
	= \frac{J_{\alpha_{n+1} \alpha} J_{\alpha \alpha_n}}{J_{\alpha_{n+1} \alpha_n}}
	e^{ -l(E_{\alpha} - E_{\alpha_n}) }
	\frac{\det \D_{\alpha}^{(+)}}{\det \D_{\alpha}}
	\frac{\det \tilde{\D}_{\alpha_n}}{\det \D_{\alpha_n}}.
\label{eq:prob_seg}
\end{align}
Here $l=\tau_{n+1} - \tau$ is the length of the new segment. 
$\D_{\alpha}^{(+)}$ is the matrix with $c_{\alpha}^{\dag}(\tau_{n+1}) c_{\alpha}(\tau)$ added to $\D_{\alpha}$, and $\tilde{\D}_{\alpha_n}$ is the matrix with one of the operators shifted in time according to $c_{\alpha_n}^{\dag}(\tau_{n+1}) \rightarrow c_{\alpha_n}^{\dag}(\tau)$. 
The ratio of determinants can be evaluated in $O(k)$ using fast-update formulas (see Sec.~\ref{fastupdate}).
If $\alpha = \alpha_n$ in Fig.~\ref{fig:diagram}, the change is just an addition of a diagonal element $X_{\alpha \alpha}(\tau)$, so that eq.~(\ref{eq:prob_seg}) is reduced to
\begin{align}
	\frac{p_{k+1}}{p_k}
	= -J_{\alpha \alpha}
	\frac{\det \D_{\alpha}^{(+)}}{\det \D_{\alpha}}.
\end{align}
Here $\D_{\alpha}^{(+)}$ is a matrix in which $c_{\alpha}^{\dag}(\tau) c_{\alpha}(\tau+0)$ is added to the original one. 
The equal-time Green function in $\D_\alpha^{(+)}$ should be $\mathcal{G}^0_{\alpha} (+0)$ to keep the probability positive (for $J>0$).
For $J<0$ see the appendix in \cite{Hoshino09}.
In the case $k=0$ all states contribute to the trace, and therefore $p_1/p_0$ is given by
\begin{align}
	\frac{p_1}{p_0}
	= -J_{\alpha \alpha} \frac{e^{-\beta E_{\alpha}}}{\sum_{\alpha'} e^{ -\beta E_{\alpha'}}} g_{\alpha}(+0).
\label{eq:prob_seg0}
\end{align}

The ratios of the weights in Eq.~(\ref{eq:prob_seg})--(\ref{eq:prob_seg0}) change their signs depending on the signs of the coupling constants. 
It was found in \cite{Otsuki07} that the probability remains positive in the case of antiferromagnetic couplings, i.e., $J_{\alpha \alpha'}>0$. 
This is consistent with the fact that the CS model with antiferromagnetic couplings is derived from the Anderson model, where the minus sign problem does not appear. 

Staggered susceptibilities and other two-particle correlation functions are discussed in \cite{Otsuki09b}.

\subsection{The Kondo model}

Perhaps the most important projected model is the spin $S=1/2$ Kondo model which is typically written as
\begin{align}
	H = \sum_{{k} \sigma} \varepsilon_{{k}} c_{{k} \sigma}^{\dag} c_{{k} \sigma}
	+ J {\bf S} \cdot \vec{\bf \sigma}_c,
\label{eq:Kondo}
\end{align}
where $\vec{\bf \sigma}_c=\psi_c^\dagger \vec{\bf \sigma} \psi_c$ is the Pauli matrix for conduction electrons ($\psi^\dagger_c=(c^\dagger_\uparrow, c^\dagger_\downarrow)$).  While it is possible to simulate this model directly using CT-HYB \cite{Werner06Kondo}, it may be more convenient for some applications to re-express it in Coqblin-Schrieffer form by introducing pseudo-Fermion operators $f^\dagger$, $f$ to represent the different states of the local moment: $S=\frac{1}{2}\psi^\dagger_f \vec{\bf \sigma}\psi_f$. Rearranging gives
\begin{align}
	H = \sum_{{k} \sigma} \varepsilon_{{k}} c_{{k} \sigma}^{\dag} c_{{k} \sigma}
	 + v \sum_{\sigma} c_{\sigma}^{\dag} c_{\sigma}
	 + J \sum_{\sigma, \sigma'} f_{\sigma}^{\dag} f_{\sigma'} c_{\sigma'}^{\dag} c_{\sigma},
\label{eq:Kondo_CS}
\end{align}
which is of the Coqblin-Schrieffer form Eq.~(\ref{CS}) with $J_{\sigma\sigma'}=J$ and with an additional potential scattering given by  $v=-J/2$. 

Carrying out the CT-J expansion requires knowledge of the $c$-electron Green function ${\tilde G}(z)$ in the presence of the potential scattering $v$. 
${\tilde G}(z)$ may be expressed in terms of the bare ($v=0$) Green function $G^0(z)$ as
\begin{align}
	\tilde{G} = \frac{\mathcal{G}^0}{1 - v \mathcal{G}^0}.
\end{align}
In the simulation of the CS model, $\mathcal{G}^0(z)$ is replaced by $\tilde{G}(z)$ and the calculation yields the impurity $t$-matrix $t_J(z)$, computed with respect to $\tilde{G}(z)$.
To obtain the $t$-matrix $t(z)$ of the Kondo model, Eq.~(\ref{eq:Kondo}), the contribution of the potential scattering should be subtracted from $\tilde{G}(z)$. 
The full Green's function $G(z)$ can be expressed as
\begin{align}
	G = \tilde{G} + \tilde{G} t_J \tilde{G}
	 = \mathcal{G}^0 + \mathcal{G}^0 t \mathcal{G}^0. 
\label{eq:Kondo_G}
\end{align}
Solving Eq.~(\ref{eq:Kondo_G}) for $t(z)$ gives
\begin{align}
	t = \frac{v}{1-v\mathcal{G}^0} + \frac{t_J}{(1-v\mathcal{G}^0)^2}.
\label{eq:tmat_kondo}
\end{align}

%% file: phonons.tex
\section{Phonons and retarded interactions}\label{phononsec}
\subsection{Models}

In this section we present the application of CT-QMC techniques to models of electrons coupled to harmonic oscillators or, equivalently, to models of electrons subject to time-dependent (retarded) interactions. Such models arise in the study of electron-phonon coupling and if dynamical screening is important \cite{Werner10_frequency}. 

In Hamiltonian form the quantum impurity model $H_\text{QI}$ is supplemented by a boson Hamiltonian $H_\text{B}$ and an electron-boson coupling $H_\text{el-B}$ so that 
$H_\text{QI} \rightarrow H_\text{QI} + H_\text{B}+H_\text{el-B}$ with 
\begin{align}
H_\text{B}+H_\text{el-B} = \sum_{\nu a} g_\nu^a {\cal O}^a(b_\nu^\dagger + b_\lambda) + \sum_\nu \omega_\nu b_\nu^\dagger b_\nu.
\label{H-el-B}
\end{align}
Here $b^\dagger_\nu$ is the creation operator for a boson mode labeled by $\nu$,   $O^a$ denotes a  bilinear fermion operator, and $\omega_\nu$, $g_\nu^a$ are the boson frequency and electron-boson coupling constant respectively. In the widely studied ``Holstein-Hubbard'' model, for example, there is just one boson mode and the operator ${\cal O}$ is the on-site electron density.

An alternative representation in terms of an action may be obtained by integrating out the bosons, leading to the contribution
\begin{align}
S_\text{ret} = \sum_{ab}\int_0^\beta d\tau d\tau' {\cal O}^a(\tau) W^{ab}(\tau-\tau'){\cal O}^b(\tau'),
\label{S-el-B}
\end{align}
with
\begin{align}
W^{ab}(\tau) &= \int_0^\infty \frac{d\omega'}{\pi} (W^{ab})''(\omega')\frac{\cosh[(\tau-\beta/2)\omega']}{\sinh(\beta\omega'/2)},
\label{Wdef}\\
(W^{ab})''(\omega) &= -\pi\sum_\nu g^a_\nu g^b_\nu\big[\delta(\omega - \omega_\nu) - \delta(\omega + \omega_\nu)\big].
\end{align}
Conversely, models of electrons subject to time-dependent (retarded) interactions are defined by an action involving a term such as $S_\text{ret}$ and reversing the above arguments shows that these interactions may be represented in Hamiltonian form by adding a coupling to bosons, as defined in Eq.~(\ref{H-el-B}).

Solving $H_\text{QI} + H_\text{B}+H_\text{el-B}$ 
requires keeping track of the bosonic sector of the Hilbert space, which  in principle involves an infinite number of additional states. Previous approaches to the  problem have involved either treating the bosons semiclassically \cite{Blawid03,Deppeler02} or  truncating the boson Hilbert space, retaining only a  finite number of boson states \cite{Koller04,Koller04b,Capone04,Sangiovanni05,Sangiovanni06}. The semiclassical approach cannot account for quantal phonon effects such as electronic mass renormalization or superconductivity, while treating the boson Hilbert space directly adds very considerably to the computational overhead and therefore limits what can be done.

Two approaches have been used in the CT-QMC context. One \cite{Werner07holstein} is based on a canonical transformation applied to the CT-HYB method and is (at least in its present form) restricted to  models in which the operators ${\cal O}$ to which the phonons couple commute with the local Hamiltonian. For models (such as the single-site dynamical mean field theory of the Holstein-Hubbard model or of the dynamically screened Hubbard $U$) which fulfill these conditions an electron-boson coupling can be added at essentially no additional computation cost.  The other  method \cite{Assaad07} is a generalization of CT-INT to time-dependent interactions and can treat more general models, although at a substantially higher computational cost.

\subsection{CT-HYB}\label{phonon_hyb}

In models where  the  oscillator degree of freedom couples to a conserved quantity of the local Hamiltonian, the phonons can be decoupled from the electrons by a canonical transformation of the sort originally introduced by \textcite{Lang62}. By using the transformed variables to evaluate the trace over the phonon states,  the hybridization expansion can be performed with negligible extra computational overhead \cite{Werner07holstein}.

We present the idea in the context of the single-site dynamical mean field description of the Holstein-Hubbard model, for which the local Hamiltonian may be written as
\begin{eqnarray}
H_\text{loc} &=& -\mu(n_\uparrow+n_\downarrow)+U n_\uparrow n_\downarrow\nonumber\\
&& +\sqrt{2}\lambda (n_\uparrow+n_\downarrow-1)X+\frac{\omega_0}{2}\left(X^2+P^2\right).\hspace{5mm}
\label{H_loc_holstein}
\end{eqnarray}
Here the boson coordinate $X$ and momentum $P$ satisfying $[P, X]=i$  are related to the familiar boson creation and annihilation operators by $X=(b^\dagger +b)/\sqrt{2}$ and $P=(b^\dagger -b)/i\sqrt{2}$,  and the $\sqrt{2}$ in the coupling term and the notation of the coupling constant as $\lambda$ are  conventional.

Following \cite{Lang62}, the boson and fermion operators may be decoupled by  shifting $X$ by $X_0=(\sqrt{2}\lambda/\omega_0)(n_\uparrow+n_\downarrow-1)$. The shift is accomplished by the unitary transformation $e^{iPX_0}$  so that  the  Hamiltonian $\tilde H_\text{loc} = e^{iPX_0}H_\text{loc} e^{-iPX_0}$  becomes
\begin{eqnarray}
\tilde H_\text{loc}&=& -\tilde \mu(\tilde n_\uparrow+\tilde n_\downarrow)+\tilde U \tilde n_\uparrow \tilde n_\downarrow+\frac{\omega_0}{2}(X^2+P^2). \hspace{5mm}
\label{H_loc_transf}
\end{eqnarray}
${\tilde H}_\text{loc}$ is of the Hubbard form but
with modified chemical potential $\tilde\mu$ and interaction strength $\tilde U$, where
\begin{eqnarray}
\tilde \mu&=&\mu-\lambda^2/\omega_0,\label{mu_tilde}\\
\tilde U&=&U-2\lambda^2/\omega_0.\label{U_tilde}
\end{eqnarray}
The impurity electron creation and annihilation operators are transformed according to
\begin{eqnarray}
\tilde d^\dagger_\sigma &=& e^{iPX_0}d^\dagger_\sigma e^{-iPX_0} = e^{\frac{\lambda}{\omega_0}(b^\dagger-b)}d^\dagger_\sigma,\\
\tilde d_\sigma &=& e^{iPX_0}d_\sigma e^{-iPX_0} = e^{-\frac{\lambda}{\omega_0}(b^\dagger-b)}d_\sigma,
\end{eqnarray}
and this factor propagates into the hybridization. 

After the transformation, the electron and boson sectors are decoupled and the 
expectation value $\langle \cdots \rangle_b$ becomes the product of a term
involving electron operators which is analogous to that computed for the Hubbard
model without phonons, and a phonon term which is the expectation value of a product
of exponentials of boson operators. The total weight of a configuration 
\begin{equation}
w(\{O_i(\tau_i)\})=w_b(\{O_i(\tau_i)\})\tilde w_\text{Hubbard}(\{O_i(\tau_i)\}).
\end{equation}
Here, $\tilde w_\text{Hubbard}$ is the weight of a corresponding configuration in the 
pure Hubbard impurity model (with parameters modified according to Eqs.~(\ref{mu_tilde}) and (\ref{U_tilde})  while the phonon contribution is  
\begin{eqnarray}
w_b(\{O_i(\tau_i)\})&=&\left<e^{s_{2n}A(\tau_{2n})}e^{s_{2n-1}A(\tau_{2n-1})}\cdots e^{s_1A(\tau_1)}\right>_b\nonumber\\
\end{eqnarray}
with $0\le \tau_1<\tau_2<\ldots <\tau_{2n}<\beta$, $s_i=1$ $(-1)$ if the $n^{th}$ operator is a creation (annihilation) operator and 
$A(\tau)=\frac{\lambda}{\omega_0}(e^{\omega_0\tau}b^\dagger-e^{-\omega_0\tau}b)$.
The expectation value is to be taken in the thermal state of free bosons.  The standard  disentangling of operators leads to 

\begin{eqnarray}
&&w_b(\{O_i(\tau_i)\})=\exp\bigg[-\frac{\lambda^2/\omega_0^2}{e^{\beta \omega_0}-1}\bigg(n \big(e^{\beta \omega_0}+1\big)\nonumber\\
&&\hspace{11mm}+\sum_{2n\geq i>j\geq 1} s_is_j\big\{e^{\omega_0(\beta-(\tau_i-\tau_j))}+e^{\omega_0(\tau_i-\tau_j)}\big\}\bigg)\bigg],\nonumber\\
\end{eqnarray}

\begin{figure}[t]
\begin{center}
\includegraphics[angle=0, width=0.8\columnwidth]{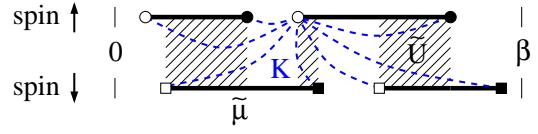}
\caption{
Illustration of an order $n=4$ diagram for the Holstein-Hubbard model. Empty (full) circles and squares represent $V^\dagger$ ($V$) hybridization events. Dashed lines indicate interactions $K(\tau)$ connecting all pairs of hybridization events.
Adapted from \cite{Werner10_frequency}.
}   
\label{diagram_K}
\end{center}
\end{figure}

The phonon contribution can be interpreted as an interaction $K(\tau-\tau')$ between all pairs of operators (see Fig.~(\ref{diagram_K}) and \cite{Werner10_frequency}) of the form
\begin{align}
K(\tau)&=-\frac{\lambda^2}{\omega_0^2}\frac{\cosh[\omega_0(\tau-\beta/2)]-\cosh[\omega_0\beta/2]}{\sinh[\omega_0\beta/2]},
\end{align}
which is the twice integrated retarded interaction (Eq.~\ref{Wdef}) produced by the phonon coupling. 
The inclusion of phonons (or more generally any operator which commutes with $H_\text{loc}$) is thus   possible without any truncation and 
with negligible extra computational cost.   

\begin{figure}[t]
\begin{center}
\includegraphics[angle=-90, width=0.9\columnwidth]{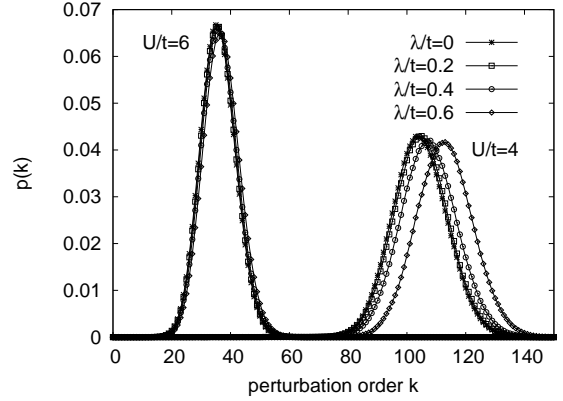}
\caption{Distribution of perturbation orders in converged single-site DMFT solutions of the half-filled
Holstein-Hubbard model with a semi-circular density of states with bandwidth $4t$, phonon
frequency $\omega_0=0.2t$, inverse temperature $\beta t=400$ and values of
electron-electron ($U$) and electron-phonon interaction strength ($\lambda$) indicated.
Both in the insulating ($U/t=6$) and metallic ($U/t=4$) phases, the distribution shifts 
little as $\lambda$ is increased except near the phase boundary to the bipolaronic phase ($\lambda/t=0.6$, $U/t=4$). 
}
\label{order}
\end{center}
\end{figure}

Figure~\ref{order} presents statistics on the perturbation order for a DMFT simulation of the Holstein-Hubbard model with semicircular density of states with bandwidth $4t$ and phonon frequency $\omega_0=0.2t$. The average perturbation order is seen to be   
little affected by the strength of the phonon coupling. Additional results on the Holstein-Hubbard model may be found in \cite{Werner07holstein}.

A closely related method has also been applied to study the consequences of dynamical screening of the Hubbard interaction by other degrees of freedom in the solid. Screening leads to a retarded interaction of the form of Eq.~(\ref{S-el-B}) with ${\cal O}^a$ the on-site density and $W{''}$ determined by the dynamical charge susceptibility of the other degrees of freedom in the solid. The passage back to Eq.~(\ref{H-el-B}) provides an oscillator representation and the formalism described above may be applied. Details are given in \cite{Werner10_frequency}.

\begin{figure}[t]
\begin{center}
\includegraphics[width=0.48\columnwidth]{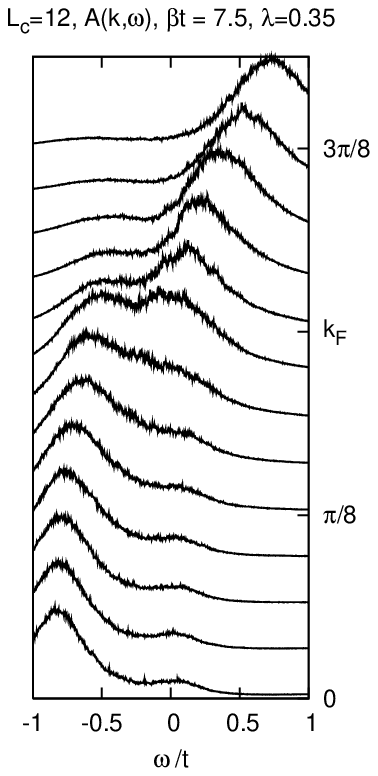}
\includegraphics[width=0.48\columnwidth]{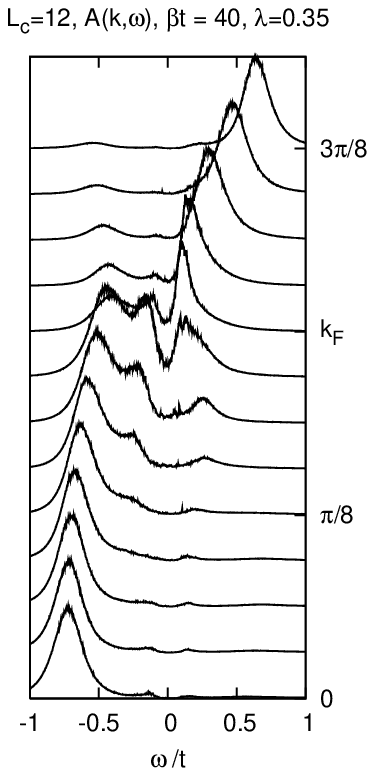}
\caption{Single particle spectral function of the one dimensional Holstein model, computed as a function of frequency using ``CDMFT'' cluster dynamical mean field methods on an $L_c=12$ cluster at filling $n=1/4$ at high temperature (left panel) and low temperature (right panel).
The spectra reveal a temperature dependent line broadening and the appearance at low $T$ of a near-Fermi-level structure associated with the development of intersite correlations.
From Ref.~\cite{Assaad08}.}\label{Assaadfig}
\end{center}
\end{figure}
\subsection{CT-INT}\label{weakphonon}

\textcite{Assaad07} showed  that the CT-INT approach, too, allows a transparent treatment of phonon degrees of freedom.  Their algorithm enables the simulation of wider classes of models than the canonical transformation approach but at much greater computational expense.  To date it has been formulated for the Holstein-Hubbard model and  applied \cite{Assaad08} to cluster dynamical mean field studies of  the one-dimensional Holstein model, and we follow this formulation in our description below.  The formalism however appears to apply also to non-Holstein couplings and further investigation along these lines would be of great interest. 

The treatment begins from an action formulation, with an interaction term which Assaad and Lang write as $S=S_{\tilde U}+S_{W}$ with $S_{\tilde U}$ the usual Hubbard interaction and 
\begin{eqnarray}
S_{W} =\sum_i \int_{0}^{\beta} d\tau d\tau' [ n_i(\tau) - 1] W(\tau - \tau') [n_i(\tau') - 1].\nonumber\\
\end{eqnarray}
with $W$ given by Eq.~\ref{Wdef}. As in other  CT-INT calculations, it is advantageous to introduce auxiliary fields in the interaction terms to eliminate a trivial sign problem.  Assad and Lang choose
\begin{equation}
   S_{\tilde U} = \int_0^\beta d\tau \frac{\tilde U}{2} \sum_{i,s } 
               \prod_{\sigma}( n_{i,\sigma}(\tau)  - \alpha_{\sigma}(s) )
\end{equation} 
with $\sigma$ the physical spin, $s=\pm1$ an auxiliary spin and  ${\alpha_{\sigma}(s) = 1/2 + \sigma s \delta}$, with $
\delta$ some constant (see also Sec.~\ref{weakcoupling}). The phonon term is shifted as
\begin{eqnarray}
   \label{Phonon}
   S_W &=& \int_{0}^{\beta} d\tau d\tau' \sum_{\sigma,\sigma'} 
         \sum_{s = \pm 1}W(\tau-\tau')  \nonumber\\
 & &    \times \left[ n_{i,\sigma} (\tau ) - \alpha_{+}(s ) \right]   
 \left[n_{i,\sigma'} (\tau') - \alpha_{+}(s ) \right] .\hspace{3mm}
\end{eqnarray} 
Assaad and Lang then perform an expansion of the CT-INT type, but employing a general vertex 
\begin{equation}
	V(\tau) = \left\{ i,\tau, \sigma, \tau', \sigma', s , \nu \right\},
\end{equation}
where $\nu$ enumerates the vertex types Hubbard (${\nu=0}$) or phonon (${\nu=1}$).
The sum over the available phase space becomes 
\begin{equation}
	 \sum_{V(\tau)} =  \sum_{i,\sigma,\sigma',s,\nu} \int_{0}^{\beta} {\rm d} \tau',
\end{equation}
and the weight of the vertex is 
\begin{equation}
	w\left[V(\tau)\right]  =  \delta_{\nu,0} \frac{\tilde U}{2} \delta(\tau-\tau') +
                         \delta_{\nu,1} W(\tau-\tau').
\end{equation}
Using furthermore the notation
\begin{eqnarray}
  H [V(\tau)]  &=&   
	\delta_{\nu,0} \delta_{\sigma,\uparrow} \delta_{\sigma',\downarrow} \delta(\tau-\tau') 
          \nonumber \\
       & &   \left[ n_{\uparrow}(\tau)-\alpha_{+} (s) \right] 
         \left[ n_{\downarrow}(\tau)-\alpha_{-} (s) \right] +  \nonumber\\
& &  \delta_{\nu,1} \left[ n_{\sigma}(\tau)  - \alpha_{+}(s) 
           \right]   
     \left[n_{i,\sigma'} (\tau') - \alpha_{+}(s) \right],
\nonumber 
\end{eqnarray}
 the partition function can be written as
\begin{align}
\frac{Z}{Z_0} &= \sum_{n=0}^{\infty} (-1)^{n} 
\int_{0}^{\beta} {\rm d} \tau_1 \sum_{V_1(\tau_1)} w[V_1(\tau_1)] \ldots 
\int_{0}^{\tau_{n-1}} {\rm d} \tau_{n}   \nonumber \\
&\times\sum_{V_{n}(\tau_{n})} w\left[V_n(\tau_{n})\right] \Big\langle T_\tau {H}\left[ V_1(\tau_1)\right] \ldots {H} \left[ V_n(\tau_n)\right] \Big\rangle_{0}.
\end{align}
The Monte Carlo procedure follows the scheme described in Chapter~\ref{weak_chapter}, with addition and removal of general vertices. 

In a cluster dynamical mean field calculation of  the one-dimensional Holstein model (Eq.~(\ref{H_loc_holstein}) with $U=0$) the method reveals interesting near-Fermi-level structures in the electron spectral function related to intermediate range correlations \cite{Assaad08}; see Fig.~\ref{Assaadfig} for representative results. 

The flexibility of the method, which seems applicable also to CT-HYB and CT-J,  and the importance of  electron-phonon couplings in materials science suggests that the implementation and investigation of more general types of electron phonon couplings would be a worthwhile effort.

%% file: nonequilibrium.tex
\section{Expansion on the Keldysh contour: Real-time and non-equilibrium physics}

\subsection{Introduction}\label{noneqintro}

In this section we describe diagrammatic Monte Carlo techniques capable of computing the real-time and nonequilibrium properties of quantum impurity models. These methods have been used to calculate the transport properties and relaxation dynamics of current-biased quantum dots and as impurity solvers for dynamical mean field studies of the nonequilibrium properties of solids. Real-time CT-QMC methods were  pioneered by M\"{u}hlbacher and Rabani who used a hybridization expansion method to study a problem of electrons coupled to phonons \cite{Muehlbacher08}. The non-equilibrium hybridization expansion was  generalized to the case of electron-electron interactions in \cite{Schmidt08}, \cite{Werner09}, \cite{Schiro09}, \cite{Schiro10}, while the real-time version of the CT-AUX  method was given  in \cite{Werner09}, \cite{Werner10} and used in \cite{Eckstein09,Eckstein10,Tsuji10,Eurich10}. It is important to bear in mind that unlike in the equilibrium case, where the algorithms have been tried, tested and optimized, the nonequilibrium extensions of CT-QMC are still in an experimental stage. The methods which have been  implemented so far are more-or-less straightforward adaptations of the equilibrium CT-QMC algorithms. Significant improvements may be possible.

While both CT-AUX and CT-HYB based methods have been studied, we will restrict our explicit discussion in this section to the CT-AUX algorithm, which has allowed an accurate study of the steady-state current-voltage characteristics of  half-filled quantum dots in the weak- and intermediate-correlation regime. 
For a discussion of the real-time CT-HYB algorithm we refer the reader to the original papers by \textcite{Muehlbacher08}, \textcite{Schiro09}, and also to Ref.~\cite{Werner10}, where both CT-AUX and CT-HYB real-time  algorithms are presented in detail.

\subsection{Keldysh formalism}

The basic theoretical task  is to evaluate the expectation value of some operator ${\cal O}$ at some time $t$, given that the system was prepared at time $t=0$ in a state described by the density matrix $\rho_0$. Using the Heisenberg representation the expectation value may be expressed mathematically as 
\begin{equation}
\langle {\cal O} (t) \rangle =Tr\Big[\rho_0 e^{i\int_0^t dt' H(t')}{\cal O} e^{-i\int_0^t dt'' H(t'')}\Big]
\label{Odef}
\end{equation}
(the generalization to operators with multiple time dependencies is straightforward and will not be written explicitly). A nonequilibrium situation may arise through a time dependence of  $H$ (as occurs for example in a system `pumped' by a laser), through nonequilibrium correlations expressed by $\rho_0$ (as occurs for a quantum dot with current flowing across it) or through  an initial density matrix $\rho_0$ which is different from the long-time (thermal equilibrium) limit, as occurs if a system is `quenched' into a different state.

\begin{figure}[t]
\begin{center}
\includegraphics[angle=0, width=0.8\columnwidth]{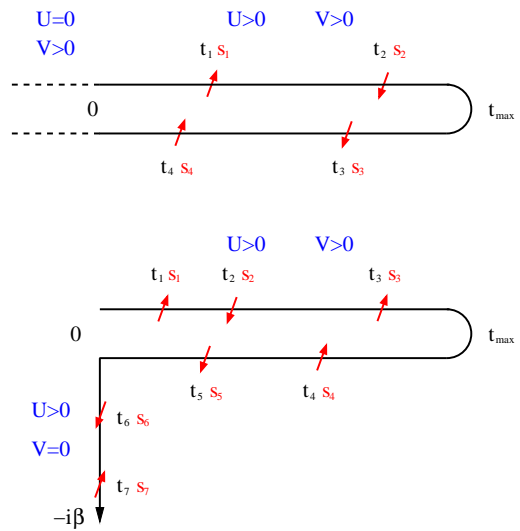}
\caption{Illustration of the Keldysh contour for a CT-AUX study of the Anderson model with interaction quench (top panel) and voltage quench (bottom panel). In an interaction quench starting from $U=0$, the imaginary time branch of the contour is shifted to $t=-\infty$ and need not be explicitly considered in a weak-coupling Monte Carlo simulation. The red arrows represent auxiliary Ising spin variables. The top panel shows a Monte Carlo configuration corresponding to perturbation order $n_+=2$, $n_-=2$, and the bottom panel a configuration corresponding to $n_+=3$, $n_-=2$, $n_\beta=2$. From Ref.~\cite{Werner10}.}
\label{Lcontour}
\end{center}
\end{figure}

One may \cite{Kadanoff} view the expectation value in Eq.~(\ref{Odef}) as an evolution on Schwinger-Keldysh contours, examples are given in Fig.~\ref{Lcontour}. In each panel the  upper contour represents the evolution from initial time $t=0$ to measurement time $t$ via $e^{-iHt}$, the operator  ${\cal O}$ is positioned at the bend where the lower and upper contours meet, and the lower contour represents the evolution from $t$ back to $t=0$ via $e^{iHt}$.  

The two panels of Fig.~\ref{Lcontour} also indicate two ways to prepare the initial state of the system.  The upper panel indicates the standard approach, which we call the {\em interaction quench}. In this approach one imagines that at times $t<0$ the interactions are turned off, so that $\rho_0$ is the density matrix of the non-interacting, but potentially non-equilibrium system. At $t=0$ the interactions are turned on, and one studies the subsequent evolution of the system. The lower panel indicates an alternative approach, the {\em voltage quench}.  In this approach one prepares the system by performing an equilibrium simulation of the interacting model (accomplished formally by propagating along the imaginary branch of the contour shown in the figure) and then turns on the nonequilibrium effects at time $t=0$.

The general strategy for evaluating Eq.~(\ref{Odef}) is the same as in the equilibrium case, namely to write $H$ as a sum of two terms: one, $H_a$, for which the time evolution can be treated exactly and another, $H_b$, which is treated by a formal perturbative expansion. The expansion in $H_b$ generates diagrams which are sampled stochastically, using an importance sampling which accepts or rejects proposed diagrams on the basis of their contributions to $\langle \tilde{ \cal O}\rangle$, with for example $\tilde {\cal O}=1$. Time plays the role of $\beta=1/T$. 

There are crucial differences.   In equilibrium calculations, the expansion can be formulated on the imaginary time axis $0\le \tau < 1/T$ as an expansion of $\Tr {T_\tau} e^{-\beta H_a}\exp[-\int_0^\beta d\tau H_b(\tau)]$. Thus one can work with real (or Hermitian) quantities and only one exponential must be expanded.  In the nonequilibrium situation one must expand two exponentials, doubling the perturbation order required to reach a given time. Also, the result of a measurement at a finite time depends on the initial preparation of the system. It is thus essential that the computation proceed for long enough to build up the correct entanglement between the impurity and the bath before steady-state quantities are measured. The main difficulty of nonequilibrium calculations, however, is that  the expansion must be done for real times, so diagrams come with factors of $i$ raised to  powers determined by the perturbation order. The terms in the expansion are complex so a 'phase' problem exists, but in all cases known to date the expansion can be arranged so that all terms are real. A sign problem however remains. Convergence of the perturbation theory is thus oscillatory rather than exponential and the result is a sign problem which severely  limits the maximum perturbation order that can be attained and hence the maximum time which can be reached.

\subsection{Real-time CT-AUX}

Here we present the formalism needed for a nonequilibrium application of the CT-AUX method. For simplicity we focus on a nonequilibrium version of the single-impurity Anderson model, Eq.~(\ref{AIM}), where the local Hamiltonian is coupled to two leads (``left'' and ``right'') which may be at different chemical potentials $\mu_\alpha$. Thus the bath and hybridization terms in the Hamiltonian become
\begin{eqnarray}
H_\text{bath} &=& \sum_{\alpha=L,R} \sum_{p} \big(\varepsilon^\alpha_{p,\sigma}-\mu_\alpha \big)c^{\alpha \dagger}_{p,\sigma} c^\alpha_{p,\sigma},\label{H_bath}\\
H_\text{hyb} &=& \sum_{\alpha=L,R} \sum_{p,\sigma} \big(V_p^\alpha c^{\alpha \dagger}_{p,\sigma}d_\sigma+h.c. \big),\label{H_hyb}.
\end{eqnarray}

A crucial parameter is  the level broadening 
\begin{equation}
\Gamma^\alpha(\omega)=\pi\sum_p|V_p^\alpha|^2\delta(\omega-\varepsilon_p^\alpha)
\label{Gamdef}
\end{equation}
associated with lead $\alpha$. The total level broadening is 
\begin{equation}
\Gamma=\Gamma^L+\Gamma^R.
\label{Gammatotal}
\end{equation}
$\Gamma$ is the imaginary part of the real axis hybridization function. It plays a crucial role in what follows so we identify it by a separate symbol.

In nanoscience applications one is interested in the current flowing through the impurity.  The flow of charge into, say, the left lead may be determined from the time derivative of the number of left lead electrons ${\hat N}^L=\sum_{p\sigma}a^{L\dagger}_{p\sigma}a^L_{p\sigma}$. Taking the commutator of ${\hat N}^L$ with the Hamiltonian shows that the current flowing through the impurity into the left lead is determined by the $t\rightarrow t{'}$ limit of the quantity
\begin{equation}
A(t,t{'})=\sum_{p\sigma}V^L_p\left<\text{T}_\mathcal{C}c^{L\dagger}_{p\sigma}(t)d_\sigma(t{'})\right>.
\label{adef}
\end{equation}
${\text T}_\mathcal{C}$ is the contour ordering operator, 
which exchanges the 
product $A(t) B(t')$  of two operators if $t$ is earlier on the contour than $t'$ (a 
minus sign is added if the exchange involves an odd number of Fermi operators).
Finding an efficient means of measuring $A$ is an important part of the algorithm.

In the nonequilibrium Anderson model  an {\it interaction quench} corresponds to taking $U=0$ for times $t<0$ with an instantaneous step to a non-zero $U$ at $t=0$ while the chemical potential difference is time independent and the initial density matrix that appropriate to noninteracting electrons in the given bias voltage.  A {\it voltage quench} corresponds to taking  $\mu_L=\mu_R$ for time $t<0$ with an instantaneous step to a nonzero $\mu_L-\mu_R$ at $t=0$.  One assumes  that the lead electrons equilibrate instantly to the new chemical potential so that the  equal time correlators of lead operators are $\langle c^{\alpha\dagger}_{p,\sigma}c^{\beta}_{p',\sigma'}\rangle=\delta_{\alpha,\beta}\delta_{p,p'}\delta_{\sigma,\sigma^{'}}f_{T_\alpha}(\varepsilon^\alpha_{p,\sigma}-\mu_\alpha)$, with $f_T(x)=(e^{x/T}+1)^{-1}$ the Fermi distribution function for temperature $T$ and $\mu_\alpha$ the value of the chemical potential for lead $\alpha$ at the appropriate time.

A compact derivation of all measurement formulae for both voltage and interaction quenches may be obtained from manipulations of the ``partition function'' (more precisely, an expression for the expectation value of the operator ${\cal O}=1$) on the  contour shown in the lower panel of Fig.~(\ref{Lcontour}):

\begin{align}
Z &= e^{-K_\beta}\text{Tr}  \big[ e^{-\beta(H_\text{bath}^\text{eq}+H^0_\text{dot}+H_\text{hyb}+H_{\tilde U}-K_\beta/\beta)} 
\nonumber \\
&\hspace{12mm}\times e^{it(H_\text{bath}^\text{neq}+H^0_\text{dot}+H_U+H_\text{hyb}-K_t/t)} \nonumber\\
&\hspace{12mm}\times e^{-it(H_\text{bath}^\text{neq}+H^0_\text{dot}+H_U+H_\text{hyb}-K_t/t)}\big].
\label{Z}
\end{align}
The notation $H_\text{bath}^\text{neq}$ indicates that on the real-time portion of the contour the two leads may have different chemical potentials, whereas $H_\text{bath}^\text{eq}$ means that on the imaginary time portion of the contour the two leads have the same chemical potential.  At this stage $K_\beta$ and $K_t$ are arbitrary  constants. Convenient choices for $K_{\beta,t}$ will be discussed below.  

In Eq.~(\ref{Z}) the interaction ${\tilde U}$   on the imaginary time branch need not be the same as the interaction $U$   on the real-time branches. 
The generalization to time-dependent $U(t)$ or $\mu_{L,R}(t)$ is straightforward \cite{Tsuji10, Eurich10}.
In the voltage quench ${\tilde U}=U$ while in the interaction quench ${\tilde U}=0$ and the imaginary time portion of the contour drops out of the problem. 

The time evolution along the real-time and imaginary-time contours is expanded in powers of $H_U-K_t/t$ and $H_U-K_\beta/\beta$, respectively. Each interaction vertex is then decoupled using Ising spin variables ($x=t$ or $\beta$)
\begin{eqnarray}
H_U-K_x/x &=& -\frac{K_x}{2x}\sum_{s=-1,1}e^{\gamma_x s (n_{d,\uparrow}-n_{d,\downarrow})},\\
\cosh(\gamma_x)&=&1+(xU)/(2K_x),
\label{decouple}
\end{eqnarray}
as in Eq.~(\ref{auxdec}).
The resulting collection of Ising spin variables on the contour represents the Monte Carlo configuration $\{ (t_{1}, s_1),(t_{2}, s_2), \ldots (t_{n}, s_{n}) \}$, with $t_i$ denoting the position of spin $i$ on the L-shaped contour (see illustration in Fig.~\ref{Lcontour}). There are $n_+$ spins on the forward branch, $n_-$ spins on the backward branch and $n_\beta$ spins on the imaginary-time branch of the contour ($n=n_++n_-+n_\beta$).
The weight of such a configuration is obtained by tracing over the dot and lead degrees of freedom and can be expressed in terms of two determinants of $n\times n$ matrices $N_\sigma^{-1}$: 
\begin{align}
&p(\{ (t_{1}, s_1),(t_{2}, s_2), \ldots (t_{n}, s_{n}) \})=\nonumber\\
&\hspace{3mm}(-i^{n_-})(i^{n_+})(K_t dt/2t)^{n_-+n_+}(K_\beta d\tau/2\beta)^{n_\beta}\prod_\sigma \det N_\sigma^{-1},\label{weight}\\
&N_\sigma^{-1} = e^{S_\sigma}-(iG_{0,\sigma})(e^{S_\sigma}-I).
\end{align}
Here $(G_{0,\sigma})_{ij}=G_{0,\sigma}(t_i,t_j)$ is the $ij$ element of the $n\times n$ matrix of non-interacting Green functions 
\begin{equation}
G_{0,\sigma}(t,t')=-i\langle \text{T}_\mathcal{C} d_\sigma(t)d^\dagger_\sigma(t')\rangle_0
\label{eqn:G0input}
\end{equation}
computed using the possibly time-dependent chemical potentials and evaluated at the time arguments defined by the Ising spins. 
The quantity 
$e^{S_\sigma}=\text{diag}(e^{\gamma_1 s_1\sigma}, \ldots, e^{\gamma_n s_n \sigma})$ is a diagonal matrix depending on the spin variables (with $\gamma_i=\gamma_t$ for spins located on the real-time branches and $\gamma_i=\gamma_\beta$ for spins on the imaginary time branch). 

A Monte Carlo sampling of all possible spin configurations is then implemented based on the absolute value of the weights (\ref{weight}). 
The contribution of a specific configuration $c=\{ (t_{1}, s_1),(t_{2}, s_2), \ldots (t_{n}, s_{n}) \}$ to the Green function ($G^c_\sigma$) and current ($A^c_\sigma$) is given by \cite{Werner09} 
\begin{align}
&G_\sigma^c(t,t')=G_{0,\sigma}(t,t')\nonumber\\
&\hspace{2mm}+i\sum_{i,j=1}^n G_{0,\sigma}(t,t_{i})[(e^{S_\sigma}-I)N_\sigma]_{i,j}G_{0,\sigma}(t_{j}, t'),\label{tildeG}\\
&A_\sigma^c(t,t')=A_{0,\sigma}(t,t')\nonumber\\
&\hspace{2mm}+i\sum_{i,j=1}^n G_{0,\sigma}(t,t_{i})[(e^{S_\sigma}-I)N_\sigma]_{i,j}A_{0,\sigma}(t_{j}, t'),\label{tildeA}
\end{align}
with the first term on the right hand side giving the contribution to the non-interacting Green function or current and the second term a correction due to the interactions. 
In Eq.~(\ref{tildeA})
\begin{equation}
A_{0,\sigma}(t,t')=\sum_{p\sigma}V^L_p\langle \text{T}_\mathcal{C} c^{L\dagger}_{p\sigma}(t')d_\sigma(t)\rangle_0
\label{defA0}
\end{equation}
denotes a dot-lead correlation function of the noninteracting model. 
The Green function and current expectation value is 
\begin{align}
G_\sigma(t,t')&=\langle G^c_\sigma(t,t') \phi_c\rangle/\langle \phi_c\rangle,\\
I(t)&=-2\text{Im} \sum_\sigma[\langle A^c_\sigma(t,t) \phi_c\rangle/\langle \phi_c\rangle],
\end{align}
where $\langle . \rangle$ denotes the Monte Carlo average and $\phi_c$ the phase of the weight of the configuration $c$.
As in Eq.~(\ref{g_reduc}), it is advantageous to accumulate the quantity
\begin{equation}
X_\sigma(s_1, s_2)= i \sum_{i,j=1}^n \delta_\mathcal{C}(s_1,t_{i})[(e^{S_\sigma}-1)N_\sigma]_{i,j}\delta_\mathcal{C}(s_2,t_{j}), \label{X}
\end{equation}
which is related to the self-energy $\Sigma$ by $X\star G_0=\Sigma\star G$ (with $\star$ denoting a contour convolution).
Furthermore, it follows from Eq.~(\ref{weight}) that the weight of a Monte Carlo configuration changes sign if the last spin (corresponding to the largest time argument) is shifted from the forward contour to the backward contour or vice versa. Since the absolute value of the weight does not change, these two configurations will be generated with equal probability. As a result, all the terms in Eq.~(\ref{X}) which do not involve the last operator on the contour will cancel. It is therefore more efficient to accumulate 
\begin{align}
&X_\sigma(s_1, s_2)=
i(1-\delta(\{t_i\}))\sum_{i,j=1}^n x(s_1, i; s_2, j)\nonumber\\
&\hspace{5mm}+i\delta(\{t_i\})\sum_{l \neq \text{last}}^n [ x(s_1,
\text{last}; s_2, l)+x(s_1, l; s_2, \text{last})],
\end{align}
with $x(s_1, i; s_2, j) \equiv \delta_\mathcal{C}(s_1,t_i)[(e^{\Gamma_\sigma}-1)N_\sigma]_{i,j}\delta_\mathcal{C}(s_2,t_j)$ and $\delta(\{t_i\})=1$ if $\max_i \text{Re}(t_i)>0$ and 0 otherwise.

In an interaction quench starting from $U=0$, the imaginary-time evolution is not explicitly considered in the Monte Carlo simulation and  temperature appears only as a parameter in the noninteracting Green functions (see Fig.~\ref{Lcontour}). Moreover, the latter depend only on time differences, and thus can be easily expressed in terms of their Fourier transform. 
Assuming a large band cutoff and neglecting the real part of the lead self-energy we find \cite{Jauho94, Werner09} 
\begin{widetext}
\begin{eqnarray}
G_0(t,t')&=&2i\sum_{\alpha=L,R}\int \frac{d\omega}{2\pi}e^{-i\omega(t-t')}\frac{\Gamma^\alpha(\omega)(f(\omega-\mu_\alpha)-\Theta_\mathcal{C}(t,t'))}{(\omega-\varepsilon_d-U/2)^2+\Gamma^2},\label{G0}\\
A_0(t,t')&=&-2i\int \frac{d\omega}{2\pi}e^{-i\omega(t-t')}\frac{\Gamma_L(\omega) \Gamma_R(\omega) (f(\omega-\mu_L)-f(\omega-\mu_R))}{(\omega - \varepsilon_d-U/2)^2+\Gamma(\omega)^2}\nonumber\\
&&+2\int \frac{d\omega}{2\pi}e^{-i\omega(t-t')}\frac{\Gamma_L(\omega)(\omega - \varepsilon_d-U/2)(f(\omega-\mu_L)-\Theta_\mathcal{C}(t,t'))}{(\omega - \varepsilon_d-U/2)^2+\Gamma^2}. 
\label{A0}
\end{eqnarray}
\end{widetext}

In the voltage quench, on the other hand, the interaction is non-vanishing on the imaginary time portion of the contour (Fig.~\ref{Lcontour}), while the chemical potential difference jumps instantaneously from zero (on the imaginary branch) to $V$ (on the real branches). Because of the time dependence of the chemical potentials, the noninteracting Green functions are not time translation invariant and we cannot express $G_{0,\sigma}$ and the dot-lead  correlator $A_{0,\sigma}$ in the form of a Fourier transform. 
Instead, those functions must be computed numerically from their equations of motion, as explained in \cite{Werner10}.

\subsection{Sign problem}

The sign (phase) problem in the real-time CT-QMC methods grows exponentially with the average perturbation order on the real-time branches, which in turn is proportional to the simulation time. Operators on the imaginary time branch do not add significantly to the sign problem.  While accurate results can be obtained for average signs down to $10^{-3}$, this threshold is reached if the expected number of operators on the real-time contour is approximately ten. To reach long times or strong interactions, it is therefore important to reduce the average perturbation order on the real-time branches as much as possible. In this context it is worth noting that in the particle-hole symmetric case, the parameters $K_x$ of the CT-AUX algorithm can be chosen such that only even perturbation orders appear in the expansion. In fact, for 
\begin{equation}
K_x=-xU/4
\end{equation}
the spin degree of freedom effectively disappears ($e^{\gamma s \sigma}=-1$) and the algorithm becomes the real-time version of the weak-coupling solver (Section~\ref{weak_chapter}) for the particle-hole symmetric interaction term $H_U-K_x/x=U(n_{d,\uparrow}-\frac{1}{2})(n_{d,\downarrow}-\frac{1}{2})$. 
The odd perturbation orders are continuously suppressed as $K_x$ approaches $-xU/4$.
This suppression of odd perturbation orders was essential in the nonequilibrium dynamical mean field calculations of \cite{Eckstein09,Eckstein10} and the current calculations of \cite{Werner10}.

%% file: technical.tex
\section{Comparison of the efficiency of the different methods}
\begin{figure}[ht]
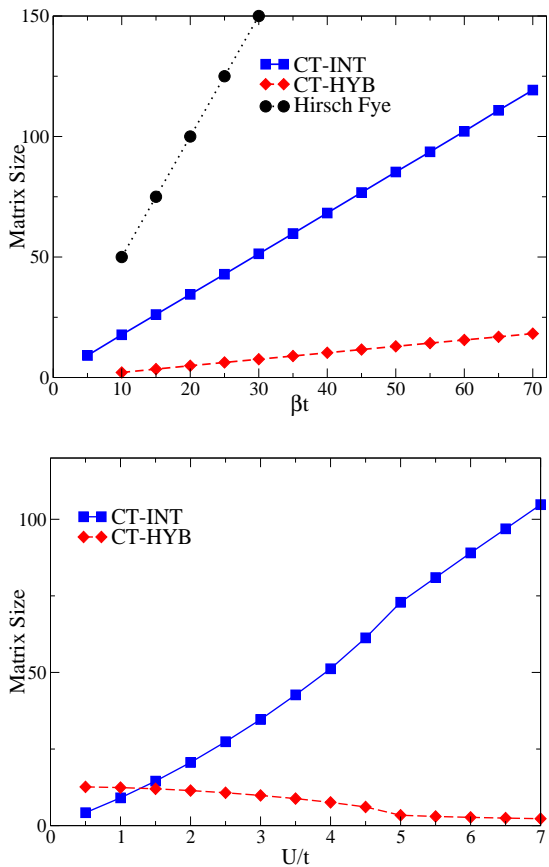

\begin{center}
\includegraphics[width=.4\textwidth,height=0.23\textheight]{figures/MatrixSizes.eps} \\
\vspace{5mm}
\includegraphics[width=.4\textwidth,height=0.23\textheight]{figures/MatrixU.eps} \\
\end{center}
\caption[MatrixU]{Upper panel: Bethe lattice, single site DMFT, scaling of matrix size with temperature at $U/t=4$ for the Hirsch-Fye, CT-INT and CT-HYB algorithms. For Hirsch-Fye, the resolution $N=\beta U$ has been chosen as a compromise between reasonable accuracy and acceptable speed, while the average matrix size is plotted for the continuous-time solvers. Lower panel: scaling of matrix size with  $U/t$ for fixed $\beta t=30$. The solutions for $U \leq 4.5$ are metallic, while those for $U \geq 5.0$ are insulating. The much smaller matrix size in the relevant region of strong interactions is the reason for the higher efficiency of the hybridization expansion method. From Ref.~\cite{Gull07}.}
\label{Matrix.fig}
\end{figure}

\subsubsection{Average expansion orders and matrix sizes}

For all diagrammatic quantum Monte Carlo algorithms discussed here, the computational effort scales as the cube of the expansion order or matrix size, as discussed in detail in \cite{Gull07}.  For a \textcite{Hirsch86} solver the matrix size is determined by the time discretization $\Delta\tau=\beta/N$. In the case of the continuous-time solvers it is determined by the perturbation order $k$, which is peaked roughly at the mean value $\langle k\rangle$ determined by the probability distribution $p(k)$. In Fig.~\ref{Matrix.fig}, we plot these matrix sizes as a function of inverse temperature $\beta$ for fixed $U/t=4$ and as a function of $U/t$ for fixed $\beta t=30$, for a semi-circular density of states with bandwidth $4t$.

It is obvious from the upper panel of Fig.~\ref{Matrix.fig} that the matrix size in all three algorithms scales linearly with $\beta$. The Hirsch-Fye data are for a number of time slices $N=\beta U$, which is apparently a common choice, although Fig.~\ref{sigma_lowest_freq.fig} shows that it may lead to considerable systematic errors. Thus, the grid size should in fact be chosen much larger ($N\gtrsim5\beta U$).

While the matrix size in the CT-INT approach is approximately proportional to $U/t$, as in Hirsch-Fye, the $U$-dependence of the hybridization expansion algorithm is very different: a decrease in average matrix size with increasing $U/t$ leads to much smaller matrices in the physically interesting region $4\lesssim U/t \lesssim 6$, where the Mott transition occurs in this model. The results in Fig.~\ref{Matrix.fig} and the cubic dependence of the computational effort on matrix size show why the continuous-time solvers are much more powerful than Hirsch-Fye and why the hybridization expansion is best suited to study strongly correlated impurity models with density-density interactions.

There is of course a prefactor to the cubic scaling, which depends on the computational overhead of the different algorithms and on the details of the implementation. However, the results presented here indicate large enough difference between the methods that the effects of optimization are of secondary importance.

\subsubsection{Accuracy for constant CPU time}
The CT-INT, CT-HYB, and Hirsch-Fye algorithms considered here work in very different ways. Not only are the configuration spaces and hence the update procedures entirely different, but so also are the measurement procedures  of the Green's functions and other observables.

Ref.~\cite{Gull07} proposed that the performance of solvers should be compared by measuring the accuracy to which physical quantities can be determined for fixed CPU time. This is the question which is relevant for implementations and avoids the tricky (if not impossible) task of separating the different factors which contribute to the uncertainty in the measured results. Because the variance of the observables measured in successive iterations of the self-consistency loop turned out to be considerably larger than the statistical error bars in each step, the mean values and error bars were determined by averaging over 20 DMFT iterations starting from a converged solution.

The Hirsch-Fye solver suffers from additional systematic errors due to time discretization. These systematic errors are typically much larger than the statistical errors. In order to extract meaningful results from Hirsch-Fye simulations it is essential to do a careful (and time-consuming) $\Delta\tau\rightarrow 0$ analysis \cite{Blumer}. The continuous-time methods are free from such systematic errors.

The high precision of the hybridization expansion results for the kinetic energy indicate that this algorithm can accurately determine the shape of the Green's function near $\tau=0$ and $\beta$.

\begin{figure}[t]
\begin{center}
\includegraphics[width=.4\textwidth,height=0.23\textheight]{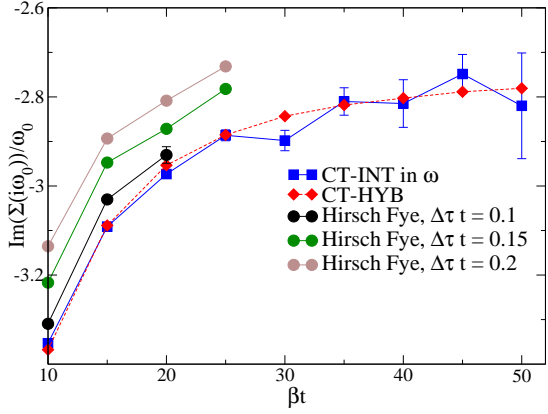} \\
\end{center}
\caption[]{Self-energy $\text{Im} \Sigma(i\omega_0)/\omega_0$ at the lowest Matsubara frequency $\omega_0=\pi T$ as a function of $\beta$ for $U/t=4.0$. The Hirsch-Fye results exhibit large discretization errors, while the continuous-time methods CT-INT and CT-HYB agree within error bars. CT-HYB is particularly suitable for measuring quantities which depend on low-frequency components, such as the quasi-particle weight. From Ref.~\cite{Gull07}.}
\label{sigma_lowest_freq.fig}
\end{figure}

For the self-energy,
\begin{equation} \Sigma(i\omega_n)\ =\ \mathcal{G}_0(i\omega_n)^{-1}-G(i\omega_n)^{-1},\end{equation}
the Matsubara Green's functions have to be inverted and subtracted. This procedure amplifies the errors of the self-energy 
especially in the tail region where $\mathcal{G}_0(i\omega_n)$ and $G(i\omega_n)$ have similar values. Fig.~\ref{sigma_lowest_freq.fig}, in contrast,  shows low frequency results $\text{Im} \Sigma(i\omega_0)/\omega_0$ for $U/t=4$ and several values of $\beta$. This quantity is related to the quasi-particle weight $Z\approx 1/(1-\text{Im} \Sigma(i\omega_0)/\omega_0)$. Again, the Hirsch-Fye results show  systematic errors due to the time discretization which must be extrapolated. The results from the continuous-time solvers agree within error-bars, but the size of the error bars is very different. The hybridization expansion approach yields very accurate results for low Matsubara frequencies in general. 

The advantage of measuring in Matsubara frequencies as opposed to imaginary time in the CT-INT and CT-AUX algorithms becomes apparent for large $\omega_n$. Only the difference of $G$ to the bare Green's function $\mathcal{G}_0$ has to be measured in this algorithm (see the detailed discussion in Sec.~\ref{weakmeas}, in particular Eq.~\ref{matsubarameas}). These differences 
decrease with $1/\omega_n$ for large $\omega_n$ and Eq.~(\ref{matsubarameas}) yields an accurate high frequency estimate, so that the tail of the self energy can be computed without amplification of errors.

Further discussion of the relative advantages of different methods can be found in \cite{Gull07}.

\section{Technical aspects}\label{technicalsec}
The following sections, independent from each other, are referenced from the algorithm sections and explain aspects of updates, measurements, 
and numerical methods needed for the efficient implementation of these continuous-time algorithms. Efficient  and accurate measurement, especially of the high frequency behavior, remains a bottleneck in the computations; further progress in this area would be desirable. 
\subsection{Inverse matrix formulas}\label{fastupdate}
The dominating computational task in most continuous-time quantum Monte Carlo impurity solver algorithms 
is the computation
of ratios $r$ of determinants of matrices $\D^k$ of size $k$ and $\D^{k+1}$ of size $k+1$,
\begin{align}
r = \frac{\det \D^{k+1}}{\det \D^k},
\end{align}
with matrices that have one row and one column (sometimes two rows and two columns) changed, added, or removed.
The only exceptions are given by the hybridization expansion in its general formulation (Sec.~\ref{hyb_matrix_section}), which is dominated by a trace computation, and the bold CT-QMC method \cite{Gull10_bold}, where the determinant structure is replaced by an analytic resummation of diagrams.

To compute the determinant of large matrices directly, it is best to first perform a factorization like the  $LU$ or $QR$ factorization, where
the matrix $A$ is written as the product
of a matrix of which the determinant is known, and another matrix where the determinant is easy to compute, e.g. the
diagonal of an upper / lower triangular matrix.
The cost of such a straightforward factorization is $O(k^3)$.

Determinant ratios of two matrices that differ only by one or two rows and columns can be computed much more efficiently if the
inverse of one of the matrices is known. This is the reason for computing the inverse Green's function matrix $\M = \D^{-1}$ in
the CT-INT algorithm, the inverse hybridization function matrix $\M = \mathbf{\Delta}^{-1}$ in the hybridization algorithm, and the matrix $\N$ in the
CT-AUX algorithm.
We illustrate the linear algebra at the example of the  CT-AUX matrix $\N$ of Eq.~(\ref{Dyson1}) introduced in
Sec.~\ref{ctaux_chapter}. The formulas for Eq.~(\ref{wick})
(CT-INT) and Eq.~(\ref{fcdet}) (CT-HYB) are computed analogously.

We start by considering a configuration at expansion order $k$, characterized by an $\N$-matrix of size $k\times k$, and consider the insertion of a vertex, thereby enlarging the configuration to $k+1$ vertices.
For ease of writing we choose a basis such that the rows and columns changed are the last ones, though of course in the code any row/column can be changed.
Inserting a vertex into the configuration of order $k$ leaves most of the \emph{inverse} of $\N$ unchanged (Eq.~(\ref{Dyson1})): it adds one row (here called $R$) and one column $Q$ to it, enlarging it to a $(k+1)\times(k+1)$ - matrix. However, changes to the new $\N^{k+1}-$ matrix, denoted by quantities with a tilde, are dense:
\begin{align}
(\N^{k+1})^{-1} &= \begin{pmatrix} (\N^k)^{-1}&Q \\R&S \end{pmatrix}, \\
\N^{k+1} &= \begin{pmatrix}\tilde{\fatP} & \tilde{Q} \\ \tilde R & \tilde S \end{pmatrix}.
\end{align}
$\tilde{\fatP}$ is of size $k\times k$, the vectors $Q, \tilde{Q}$ and $R, \tilde{R}$ have size $(k\times 1)$ and $(1\times k)$, and $S, \tilde{S}$ are scalar. 
A block calculation shows that the elements of the matrix $\N^{k+1}$ may be computed from $\N^k, R, S,$ and $Q:$
\begin{subequations}
\begin{align}
 \tilde S &= (S-[R] [\N^{(k)}Q])^{-1}, \label{newS}\\
 \tilde Q &= -[\N^{(k)}Q] \tilde S, \label{newQ}\\
 \tilde R &= -\tilde S [R \N^{(k)}], \label{newR}\\
 \tilde \fatP &= \N^{(k)} + [\N^{(k)}Q]\tilde S [R \N^{(k)}].\label{newP}
\end{align}
\end{subequations}

The determinant ratios needed to accept or reject an update in Eq.~(\ref{RSIAM}), Eq.~(\ref{metro_ctaux}), or Eq.~(\ref{segacrat}) are given by
\begin{align}\label{detratinv}
\frac{\det (\N^{k+1})^{-1}}{\det (\N^{k})^{-1}} = \frac{1}{\det \tilde S} = \det (S - R \N^{k}Q),
\end{align}
as can e.g. be seen from an $LU$ decomposition of the block-matrix $(\N^{k+1})^{-1}$.

The computational effort for computing the insertion probability $W^\text{acc}_{\x\y}$ of a spin (or vertex, or segment) is $O(k^2),$ or a matrix-vector multiplication followed by an inner product, as in Eq.~(\ref{newS}). The removal probability
is computed in $O(1),$ as $\tilde S$ is an element of $\N^{k+1}$ and therefore already known.
If an update is accepted, an $O(k^2)$ rank one update has to be performed for Eq.~(\ref{newP}). As
approximately $k$ updates are needed to decorrelate a configuration, the overall algorithm scales as $O(\langle k\rangle^3),$ with $\langle k\rangle$ the average expansion order.
Note that the acceptance probabilities for vertex insertions or removals are more expensive  than spin-flips in the case of the Hirsch-Fye algorithm ($O(k^2)$ vs. $O(1)$), while accepted updates require rank one updates $(O(k^2))$ in both cases.

\subsection{Spin-flip updates}\label{spinflips}
In CT-AUX, if only the value of an auxiliary spin is changed and not the imaginary time or site index of a vertex, a Dyson equation similar to the Hirsch-Fye Dyson equation may be employed.
For a spin-flip from interaction $V_{p_q}$ to $V'_{p_q}$ of spin $p_q$ at Monte Carlo step $q$ we obtain:
\begin{align}
(\N\G^0)_{ij}^{q+1} &= (\N\G^0)_{ij}^q + ((\N\G^0)_{ip_q}^q - \delta_{ip_q}) \lambda^q (\N\G^0)_{p_qj}^q,\\
N_{ij}^{q+1} &= N_{ij}^q + ((\N\G^0)_{ip_q}^q - \delta_{ip_q}) \lambda^q (N)_{p_qj}^q,\label{rankone}\\
\lambda^q &= \frac{\gamma^q}{1+(1-(\N\G_0)^q_{p_qp_q})\gamma^q} = \frac{\gamma^q}{R^q},\\
\gamma^q &= e^{V_{p_q}' -V_{p_q}} - 1\\
R^q &= 1+(1-(\N\G_0)^q_{p_qp_q})\gamma^q. \label{ratio}
\end{align}
$R^q$ is the spin-flip acceptance ratio. The expression
$(\N\G_0)^q_{lm}=G_{lm}^q$ can easily be computed from the identity
\begin{align}
N_{iz}G^0_{zj}(e^{V_j}-1) &= N_{ij}e^{V_j}-\delta_{ij}, \\
(\N\G_0)_{lm} &=(N_{lm}e^{V_m}-\delta_{lm})/(e^{V_m}-1) \label{NG0}
\end{align}
Spin-flip proposals are $O(1)$ (as in Hirsch-Fye), and the same linear algebra applies. Spin-flip updates are not ergodic in continuous-time algorithms; updates which change the expansion order and vertex times are needed.

\subsubsection{Delayed spin-flip updates}\label{delayed}
Spin-flip updates can be separated into two parts: the computation of the acceptance ratio $R$ (Eq.~(\ref{ratio})), and the update of the Green's function
after an accepted spin-flip move. ``Delayed'' updates, a concept developed by \textcite{Alvarez08} for the Hirsch-Fye algorithm and applied to continuous-time methods in \textcite{GullPhD}, delay the (expensive and slow) update of the Green's function to a later time, while computing a sequence $R^1, \cdots, R^{q_\text{max}}$
of Monte Carlo spin-flip acceptance ratios.
In analogy to \textcite{Alvarez08}, we define two vectors $a^q_i$ and $b^q_j$ (compare to Eq.~\ref{rankone}) as
\begin{align}
a^q_i &= \lambda^q ((\N\G^0)^q_{ip_q} -\delta_{ip_q}), \label{aj}\\
b^q_j &= N^q_{p_qj}. \label{bj}
\end{align}
For $R^q$ we need to know $(\N\G_0)^q_{p_qp_q} = G^q_{p_qp_q}$, which is computed by Eq.~(\ref{NG0}) from $N^q_{p_qp_q}$.

At the first step ($q=1$) $\N=\N^0$ is known. We start by selecting a spin $p^1$. We then compute $R^1$ according to
Eq.~(\ref{ratio}), and accept or reject the update.

In a next step ($q=2$), we choose the spin $p_2$. In order to compute $R^2$, we need to know $N^1_{p_2p_2},$ which we compute as
\begin{align}
N_{p_2p_2}^1 = N_{p_2 p_2}^0 + a_{p_2}^1 b_{p_2}^1.
\end{align}

More generally, the $j$th diagonal element $d_j^q = N_{jj}^q$ after $q$ (accepted) spin-flips is given by
\begin{align}
d^q_j = N^0_{jj} + \sum_{l=1}^q a^l_j b^l_j.
\end{align}
We define two vectors, $col^q$ and $row^q$, that iteratively recompute the elements of the matrix $\N$ for the row and column $p_q$:
\begin{align}
col^q_j = N^0_{jp_q}+ \sum_{l=1}^{q}a^l_j b^l_{p_q} &= N^{q}_{jp_q} \label{roweq}\\
row^q_j = N^0_{p_qj}+ \sum_{l=1}^{q}a^l_{p_q} b^l_j &= N^{q}_{p_qj}.\label{coleq}
\end{align}
These are sufficient to compute the new $q$-th column (row) of the matrices $a_j^q$ ($b_j^q$) (Eq.~(\ref{aj}) and (\ref{bj})) and the new diagonal vector $d_j$:
\begin{align}\label{aeq}
a^q_i &= \lambda_q ((\N\G^0)^q_{ip_q} -\delta_{ip_q}) \\ \nonumber &= \lambda_q ( (N_{ip_q}e^{V_{p_q}} - \delta_{ip_q})/(e^{V_{p_q}} -1) - \delta_{ip_q}) \\ \nonumber
      &= \lambda_q ( (col^q_ie^{V_{p_q}} - \delta_{ip_q})/(e^{V_{p_q}} -1) - \delta_{ip_q}),\\
b^q_j &= row^q_j, \label{beq}\\
d^{q+1}_j &= d^q_{jj} + a^q_j b^q_j = N^{q+1}_{jj}. \label{deq}
\end{align}
As seem in Eq.~(\ref{ratio}), $d^{q+1}_{p_{q+1}}$ is needed to accept or reject the next spin-flip at the next proposed position $p_{q+1}$.
After some steps $q_\text{max}$ we retrieve the full $\N$-matrix by computing
\begin{align}
N_{ij}^{q_\text{max}} = N_{ij}^0 + \sum_{l=1}^{q_\text{max}} a_{il} b_{lj}.\label{delayedmatrix}
\end{align}

A complexity analysis shows the cost of the delayed spin-flip updates: Eq.~(\ref{roweq}) and (\ref{coleq}) are $O(qk)$, Eq.~(\ref{deq}) is $O(k),$ and Eq.~(\ref{delayedmatrix}) is an $O(k^2q_\text{max})$ matrix-matrix multiplication.
The reason for performing delayed spin-flip operations instead of straightforward spin-flips or insertion and removal updates is that, on current hardware architectures, the final matrix multiplication
in Eq.~(\ref{delayedmatrix}) is about a factor 10 faster for large (i.e. out-of cache) matrices than successive rank one updates, as fast matrix operations that reuse data can be employed. The additional overhead of computing $a$, $b$, $d$, $row,$ and $col$ will dominate the algorithm for large $q^\text{max}$. We therefore recompute $\N$ often enough that the overhead does not dominate, but that we can still take advantage of the matrix operations.
In practice  $q^\text{max}=32$ or $64$ are reasonable values \cite{Alvarez08}, also in the continuous-time algorithms. For more information see also \cite{MikelsonsThesis} and \cite{submatrix}.

\subsection{Efficient measurements in the CT-AUX and CT-INT formalism}\label{weakmeas}
In the CT-AUX and CT-INT algorithms the Green's function measurement formula Eq.~(\ref{GSreduc}) and Eq.~(\ref{Gweakconf}) in imaginary time, for sites $i$ and $j$ and at times $\tau_i$ and $\tau_j$, is
\begin{align}\label{taumeas}
&G_{ij,\sigma}(\tau_i-\tau_j) = \mathcal{G}^0_{ij,\sigma}(\tau_i-\tau_j) \\ \nonumber&-\Big\langle\sum_{pq}\mathcal{G}^0_{ix_p,\sigma}(\tau_i-\tau_p)\mathcal{G}^0_{x_qj,\sigma}(\tau_p-\tau_j) M_{pq}\Big\rangle_\text{MC},
\end{align}
where $x_p (x_q)$ and $\tau_p$ ($\tau_q$) denote the site and time of the vertex at row (column) $p$ ($q$) of $M$.
Fourier transformed to Matsubara frequencies, the Green's function is estimated as
\begin{align}\label{matsubarameas}
&G_{ij,\sigma}(i\omega_n) = \mathcal{G}^0_{ij,\sigma}(i\omega_n) \\ \nonumber&-\frac{1}{\beta} \Big\langle\sum_{pq}{\mathcal{G}^0_{ix_p,\sigma}(i\omega_n)\mathcal{G}^0_{x_qj,\sigma}(i\omega_n)} e^{i\omega_n \tau_p} M_{pq} e^{-i\omega_n\tau_q}\Big\rangle_\text{MC}.
\end{align}

Measurement using Eq.~(\ref{taumeas}) in the imaginary time domain has a crucial drawback:
To sample the smooth function $G_{ij}(\tau)$ the formulas need to be evaluated for definite $\tau$ on some grid (which may be chosen non-equidistant). Further processing, e.g. Fourier transforms, may introduce discretization errors caused by this grid.
As the cost computing $\G$ straightforwardly is proportional to the number of
imaginary time points at which it needs to be evaluated, a fine grid of time points becomes prohibitively expensive.
In addition, $\G$ estimated by Eq.~(\ref{taumeas}) has a further drawback: the observable
average $G(\tau_i-\tau_j)$ is translationally invariant, while the estimator explicitly depends on two times $\tau_i, \tau_j$,
so that translation symmetry needs to be restored by the random walk.

In the Matsubara frequency domain, Eq.~(\ref{matsubarameas}), there already is a discrete grid of frequencies $\omega_n = (2n+1)\pi/\beta, n=0,1,2, \ldots$. The summands inside $\langle \cdot \rangle_\text{MC}$ decay as $1/\omega_n^2$.
A measurement method implemented directly in frequency space measures all frequencies up to a maximum cutoff $\omega_\text{max}$. 
To obtain the number of frequency points needed
we use information from a high frequency expansion of the self energy or the Green's function, and automatically adjust the cutoff
frequency such that systematic errors from the cutoff are much smaller than statistical (Monte Carlo) errors.
This controllability makes this method the preferred one for high accuracy measurements of the Green's function.

In translationally invariant clusters, only diagonal entries of the Green's function in k-space are non-zero. For a cluster with $N_c$ sites this implies
that only $N_c$ independent $k$-space Green's functions need to be measured (instead of $N_c^2$ real space Green's functions), at the small cost of performing a (real space) Fourier transform.

Computing the exponential factors $\exp(\pm i\omega_n\tau)$ needed for the frequency measurement is expensive.
Even with fast vectorized functions available as part of
numerical libraries, these operations are so time consuming that they may dominate computer time in large simulations.
An obvious simplification consists of creating a fine imaginary time grid.
At the start of the simulation,  $\exp(i\omega_n\tau)$ is computed for all $\omega_n$ needed and all $\tau$ on that grid, and the exponentials in Eq.~(\ref{matsubarameas}) are taken from it.
This eliminates the expensive calculation of $e^{i\omega_n \tau}$ at runtime at the cost of some (but relatively little) additional memory.
We did not observe any inaccuracies introduced by this discretization.

\subsubsection{Self energy binning measurement}\label{efficient_measurements}
An efficient measurement method, presented in \textcite{Gull08_ctaux}, is based on measuring $\Sigma G \equiv S$. $M$ plays the role of a $T$-matrix: $M \mathcal{G}^0 = \Sigma G$. This measurement method works in imaginary time but does not have the drawbacks described in the previous section.
The measurement formula (omitting the spin index) is rewritten as
\begin{align}
&G_{ij}(\tau) = \mathcal{G}^0_{ij}(\tau) - \Big\langle \sum_{pq} \mathcal{G}^0_{ix_p}(\tau - \tau_p) M_{pq} \mathcal{G}^0_{x_qj}(\tau_q)\Big\rangle_\text{MC}\\
&=\mathcal{G}^0_{ij}(\tau) -\int d\tau_z\sum_l \mathcal{G}^0_{il}(\tau - \tau_z) \nonumber \\ &\ \ \times\Big\langle \sum_{pq}\delta(\tau_z - \tau_p)\delta_{x_pl}M_{pq}\mathcal{G}^0_{x_pj}(\tau_q)\Big\rangle_\text{MC}\nonumber \\ \nonumber
&= \mathcal{G}^0_{ij}(\tau) -\int_0^\beta d\tau_z \sum_l\mathcal{G}^0_{il}(\tau-\tau_z) \langle S_{lj}(\tau_z)\rangle_\text{MC}.\label{g_reduc}
\end{align}
The Matsubara Green's function can similarly be extracted directly from the expectation value of $S$:
\begin{align}
G_{ij}(i\omega_n) = \mathcal{G}^0_{ij}(i\omega_n) - \mathcal{G}^0_{il}(i\omega_n) \int_0^\beta \sum_ld\tau_z e^{i\omega_n \tau_z}\langle S_{lj}(\tau_z)\rangle_\text{MC}.
\end{align}
In the Monte Carlo process only the quantity $\langle S\rangle_\text{MC}$ is measured and binned into fine (typically $10000$) bins.
The cost of this binning process is independent of the number of time discretization points of $S$, and only requires the evaluation of $M\mathcal{G}^0$ at runtime.
In practice we employ the translational invariance in the time - domain to obtain multiple estimates of the Green's function at the same step, and perform a matrix-matrix multiplication
of the matrix $M_{pq}$ and a matrix $\mathcal{G}^0_{qj}= \mathcal{G}^0_{s_qs_j}(\tau_q - \tau_j)$ with randomly chosen $\tau_j$ to obtain estimates for $S$.
The method is accurate and significantly faster than the other methods presented here and is therefore the measurement
method of choice for the CT-AUX and CT-INT algorithms, unless access to large clusters or high precision in the high frequency part of the self energy is needed (e. g. for analytic continuation), in which case we use the frequency measurement.

\subsection{Green's function (``worm'') --sampling}\label{wormchapter}
In all algorithms discussed so far, diagrams or configurations were generated with the weight that they contribute to the partition function (Sec.~\ref{SeriesMC}). For measurements, Green's function diagrams were then obtained by modifying such partition function configurations. A priori it is not clear that the Green's function configurations with large weight are the ones created by modifying configurations important for the partition function. 
If the Green's function estimates generated by importance sampling of partition function configurations are not the ones with large contributions to the Green's function, the Green's function estimator obtains a large variance and therefore the measurement of the Green's function becomes inefficient.
This problem may be overcome by employing a ``worm'' algorithm. The concept was originally developed in the bosonic context by Prokof'ev and collaborators \cite{Prokofev98A} and, among other problems, applied to the attractive - $U$ Hubbard model \cite{Burovski06}. The name ``worm'' refers to two dangling Green's function lines in the diagrams that build the head and tail of the ``worm''. 

Instead of generating configurations of $Z$ and measuring $G$, a worm method stochastically samples both the series for $Z$ and the series for $G_{ij,\sigma}(\tau_i-\tau_j)$ simultaneously.  To this end the configuration space $\mathcal{C}$ is enlarged to include both the set of  diagrams for $Z$, ${\mathcal C}_Z$  and the set of  diagrams ${\mathcal C}_G$ for $G$: 
\begin{align}
\mathcal{C} = \mathcal{C}_Z \cup \mathcal{C}_G.
\end{align}
A new partition function of the combined system is defined by extending $Z$ by the sum over all Green's function diagrams, with an arbitrary factor $\eta$ that controls the relative importance the Green's function sector ${\mathcal C}_G$ and the partition function sector ${\mathcal C}_Z$,
\begin{align}
Z_\text{tot} = Z + \eta \sum_{ij,\sigma}\iint d\tau_1 d\tau_2G_{ij,\sigma}(\tau_1,\tau_2).
\end{align}
In practice, $\eta$ is chosen such that both summands for $Z_\text{tot}$ have non-vanishing weight.

The random walk and updates are modified such that the entire space $\mathcal{C}$ is sampled: In addition to the partition function updates, ``worm insertion'' and ``removal'' updates, i.e updates that transition between ${\mathcal C}_Z$ and ${\mathcal C}_G$ by inserting or removing Green's function operators, as well as updates in the Green's function space like the shift of Green's function lines or vertex insertions (and removals) in the Green's function space need to be considered. 
Updates that change the vertex part of a Green's function configuration are important, as they allow importance sampling for all elements of a Green's function diagram.

Measurements in real space and imaginary time  are straightforward: A histogram of worm positions with the appropriate sign needs to be recorded. Such an imaginary time measurement yields estimates for the Green's function with continuous times. These estimates are best measured on a fixed, but preferably non-uniform, grid by proposing ``worm shift'' updates onto measurement locations.

Worm methods have been implemented both for CT-HYB and CT-INT algorithms \cite{GullPhD}. For equilibrium DMFT simulations, without reweighing, the worm algorithm did not result in much better statistics than the partition function algorithm, as sampling problems appear to be minimal. 
Combined e.g. with ``Wang Landau'' techniques the worm method offers the possibility to perform reweighing of the Green's function to obtain better statistics. Worm updates are however crucial in a ``bold'' sampling method \cite{Gull10_bold} where, due to a partial resummation of some diagrams, there is no direct relation between Green's function and partition function diagrams.

\subsection{Wang Landau sampling}\label{wanglandau}
In the usual sampling process of the partition function CT-INT algorithm, as well as in the other algorithms described in Sec.~\ref{ctaux_chapter} and \ref{hyb_seg_chapter},
 diagrams of the expansion (elements of the configuration space) are sampled with the weight
that they contribute to the partition function. Observables are
then measured in this ensemble. However, we are free to sample any
arbitrary ensemble -- as long as the proper reweighing according to 
Eq.~(\ref{reweighing}) is performed. While the samples generated
are likely to have a larger variance [Eq.~(\ref{variance})], there may be other
advantages, in particular smaller autocorrelation times. 
Here we describe so-called flat-histogram sampling
methods. These methods are particularly useful for problems, such as first order phase transitions, with barriers in the
configuration space of the Markov walker.

\begin{figure}[hbt]
\begin{center}
\includegraphics[width=0.3\columnwidth]{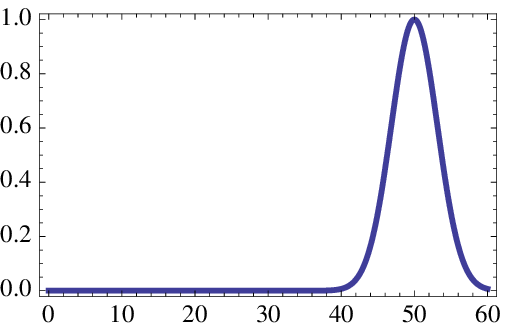}
\includegraphics[width=0.3\columnwidth]{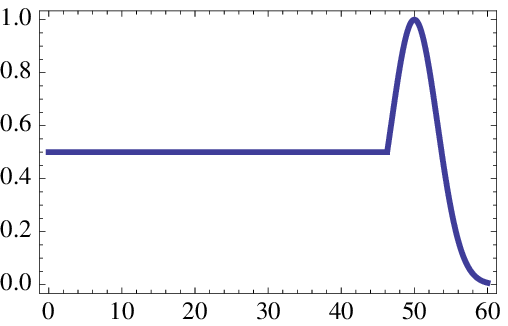}
\includegraphics[width=0.3\columnwidth]{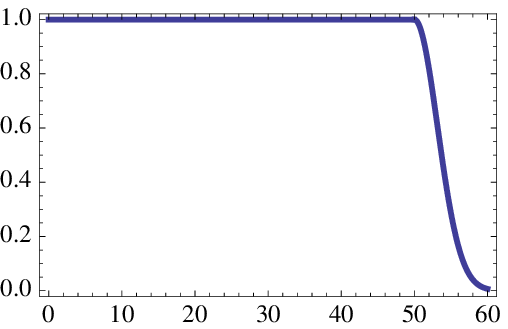}
\end{center}
\caption{Sketch of the histogram of the expansion order for flat histogram sampling. $x$-axis: expansion order $k$. $y$-axis: histogram $h(k)$.
Left panel: no flat histogram sampling. Middle panel: flat histogram sampling up to half of the maximum order.
Right panel: flat histogram sampling up to the maximum contributing order. From Ref.~\cite{GullPhD}}\label{WLPix2}
\end{figure}

\textcite{Wang01,Wang01b} presented a general reweighing scheme that is designed to find and overcome barriers and phase
transitions without prior knowledge of where in phase space they are. The method was extended to quantum problems by \textcite{Troyer03}.  In the quantum Wang-Landau method the key is to reweigh the perturbation series so that all orders up to some $k_\text{max}$ are sampled with approximately equal probability  (see Fig \ref{WLPix2}). $k_\text{max}$ has to be chosen in such a way that all local minima of phase space have some overlap with order $k_\text{max}$, so they can be reached by flat histogram sampling. 
For the reweighed system, the acceptance ratio (\ref{RSIAM}) is replaced with $R \frac{\lambda_{k+1}}{\lambda_{k}}$ with
\begin{align}
\lambda(k) = \left\{ {\frac{1}{p(k)}, k < k_{\max} \atop 1, k > k_{\max}}\right.,
\end{align}
where $p(k)$ is the probability of having expansion terms at order $k$ in a non-reweighed sampling.
Reweighing factors need also be taken into account while calculating averages (Eq. (\ref{MCreweight})).
 
The probability $p(k)$ is unknown at the start of the simulation. Therefore the reweighing coefficients are adjusted as the  simulation proceeds: the value of $\lambda(k)$ is slightly decreased for the frequently-visited values of $k$
and increased for rarely-visited ones, until the histogram is flat. As the ensemble $\lambda(k)$ does not enter
the expectation value of the observables, it is not important to have a very accurate estimate of it, as long as it is sufficient to
ensure ergodicity.

The reweighed algorithm  generates diagrams both at the physically interesting orders and at orders that are very close to zero, {\it i.e.} the bare Green's function or non-interacting partition function in CT-INT. Deliberately generating configurations that contribute little weight to the partition function may seem inefficient,
as the idea behind importance sampling is to generate the diagrams with the importance they contribute to the partition function. However, when revisiting the noninteracting case at $0^{th}$ order of the series, all vertices and therefore all correlations are removed, and when the series is rebuilt it will likely end up in a different part of phase space -- for example in a different global symmetry sector, thereby avoiding trapping in local minima. In other words, the method aims to provide a reduction of the {\it autocorrelation time} for the Markov walker. Closer analysis shows that the algorithm can be improved by minimizing the round-trip time between low and high order states \cite{Dayal04,Trebst04}.

In practice, flat-histogram sampling turned out to be very efficient at obtaining symmetrized, paramagnetic Green's functions \cite{GullPhD}.
The fact that most configurations sampled have
low order and contribute next to nothing to the observables is compensated by the fact that
they are quick to sample due to the $O(k^2)$ scaling of the matrix operations.

\begin{figure}[t]
\begin{center}
\includegraphics[width=\columnwidth]{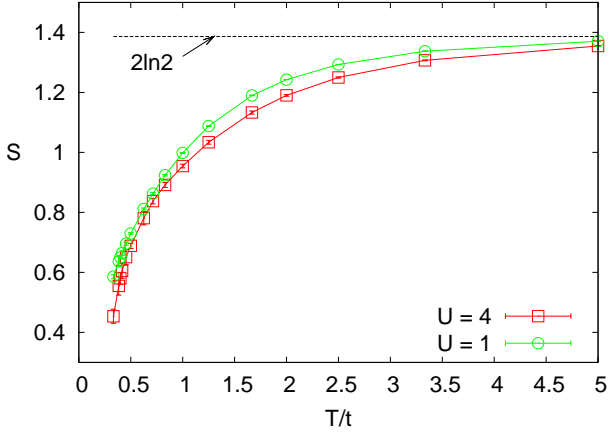}
\end{center}
\caption{Entropy per site of the half-filled Hubbard model with $t=1$,
computed for a periodic $4\times4$ cluster at different temperatures. From \cite{Li09}.}
\label{entropy}
\end{figure}

A further important application of the flat-histogram methods is the calculation of thermodynamic potentials. The grand potential of the  Hubbard model on a finite lattice at temperature $T$, for example, can be found by the integration of the equation  $d\Omega =  - N d\mu + D dU$, because the quantities on the right hand side can be measured in a standard simulation.  Such a procedure, however, requires several simulations for a finite $U$ range. For the Hubbard model (with no DMFT self-consistency) a more elegant and efficient way \cite{Li09} is to employ  flat-histogram methods to obtain the partition function $Z$ directly, as the zero order term for $Z$ is just 1. Given the reweighing factors $\lambda(k)$ and frequency $P_k$ with which different perturbation orders have been visited during the random walk, the partition function is computed as $Z=(P_0 \lambda(0))^{-1} \sum P_k \lambda(k)$. Knowing $Z$, all thermodynamic potentials can be calculated. As an example, we present in Figure \ref{entropy} the graph for the entropy of a $4\times 4$ Hubbard cluster. If a converged solution of an impurity model is available the same technique may be used to compute the impurity model partition function, but relating this to the  thermodynamic properties of the full lattice model requires taking into account the variation  of the bath density of states. This has not yet been explored.

\subsection{Computation of the trace for general interactions in the hybridization expansion}\label{tracesec}
\begin{figure}[htb]
\begin{center}
\includegraphics[width=0.95\columnwidth]{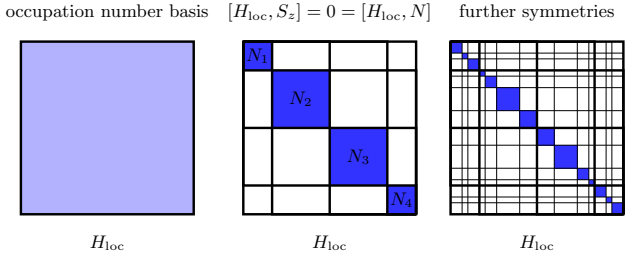}
\end{center}
\caption{Sketch of results of applying rotation / block diagonalization operations to the local Hamiltonian. The Hamiltonian in the occupation number basis (depicted in the left panel) is sparse but not blocked.
A first permutation operation builds blocks according to the
occupation number and spin of the local Hamiltonian, leading to a Hamiltonian (depicted in the middle panel) which is nearly block diagonal with dense blocks. A second (rotation) matrix further reduces block size by considering rotational
 and translational invariance of the impurity Hamiltonian, leading to the sparse block structure shown in the right panel.}
\label{LocalHamRotation.fig}
\end{figure}
In the general formulation of the hybridization expansion, as derived in Eq.~(\ref{FullExpHyb}), the principal computational difficulty is the evaluation of the trace of a product of operators and exponentials of the local Hamiltonian.
In a given basis this corresponds to taking the trace of a product of O($k$) (large) matrices that have the linear size $n_\text{loc}$ of the local Hilbert space.
Matrix-matrix multiplications of matrices with size $n_\text{loc}$ scale as $O(n_\text{loc}^3)$.
It is therefore important to find a way both to reduce the size of the matrices that need to be multiplied 
as well as the number of matrix-matrix multiplications that have to be performed.

Computing the exponential of a matrix \cite{MolerVanLoan03} is an expensive operation. In the following we transform to the eigenbasis of the local Hamiltonian
by diagonalizing it. In the eigenbasis, $\exp (-H\tau)$ is diagonal. The (formerly sparse) local creation and annihilation operators become dense matrices.

\subsubsection{Block diagonalization}
The local Hamiltonian $H_\text{loc}$ has symmetries. While these symmetries are dependent on the exact form of the local Hamiltonian, usually
the total particle number $N_\text{tot}$, the total spin z-component $S_z$ and rotational or translational symmetries of the impurity Hamiltonian
are conserved: $[H_\text{loc}, N_\text{tot}] = 0 = [H_\text{loc},S^z_\text{tot}].$
This implies that the local Hamiltonian may be decomposed into a block-diagonal form, containing several blocks with size $n_\text{block} \ll n_\text{loc}$.
This procedure is illustrated in Fig.~(\ref{LocalHamRotation.fig}).

\begin{figure}[tbh]
\begin{center}
\includegraphics[width=0.95\columnwidth]{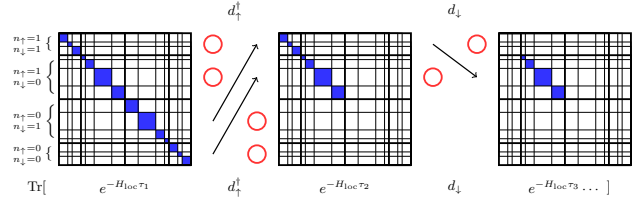}
\end{center}
\caption{
Sketch of one of the optimizations in the general representation: Four symmetry sectors are drawn, for which $S_z$ and $N$ are different. After the trace of $d_\uparrow^\dagger$ and $d_\downarrow$ is taken only one of the symmetry sectors still contributes.
In the implementation, we first identify which symmetry sectors contribute, and then compute the matrix product and trace only for these sectors. Additional symmetries vastly simplify the computation.}
\label{BlockDiagonalOperatorTrace.fig}
\end{figure}

The advantage of changing to a block-diagonal form \cite{Haule07} is that operators $d_i,d_j^\dagger$ are also in block-matrix form (see Fig.~\ref{LocalHamRotation.fig}, \ref{BlockDiagonalOperatorTrace.fig}).
The operator $d_{i\uparrow}^\dagger,$ for example, raises both the total particle number and the total $S_z$-component by one and therefore consists of off-diagonal blocks connecting the
$(S_z,n)$ - symmetry sector with the $(S_z+1, n+1)$ - sector.
As the most expensive part of the code is the
computation of matrix products, which scales as $O(\sum_\text{block}n_\text{block}^3)$ or $O(n_\text{max block}^3)$ instead of $O(n_\text{loc. Ham}^3),$
the advantage of using symmetries is obvious \cite{Haule07,Gull08_plaquette}.

A typical example is the four-site Hubbard plaquette with next-nearest neighbor ($t'-$) hopping. The local Hamiltonian has a size
of $256 \times 256$ elements ($4^4$ local states). However, $H$ commutes with $n_\uparrow,n_\downarrow$ and has a four-fold rotational
symmetry (or a couple of inversion and mirror symmetries). This allows us to split up
the $256\times256$ matrix into $84$ small
blocks that have at most $16 \times 16$ elements. 

An appropriate basis choice also allows insight into the physics. The CT-HYB formalism allows one to determine which impurity model states make the dominant contribution to the computation of an observable, and the matrix of eigenstate occupation probabilities is the projection of the density matrix onto the localized orbital basis.
This information is much more easily interpreted if a physically motivated, symmetry-related basis choice is made. For examples see Figs.~\ref{pdos_sectors} and \ref{Probab}.

\subsubsection{Basis truncation}\label{truncation_sec}
As noted in \textcite{Haule07}, in situations where the local Hilbert space is prohibitively large, e.g. in the case of large multi-orbital problems or clusters, the computation of the trace is only feasible if the size of the local Hilbert space is reduced by an appropriate truncation of the basis. In systems with very highly excited states that are unlikely to contribute (e.g. the $5$, $6$ or $7$ -electron states in Cerium), it is common practice to simply truncate the local Hilbert space and eliminate these states entirely. The same can be done for high energy / high momentum states in clusters. In addition to that, the highest few excited states of the local Hamiltonian in a particular symmetry sector may be truncated.

Simple truncation based on some a priori criterion is an uncontrolled approximation justified only by a need to solve a particular problem with available resources. Truncation based on the eigenvalues of the local Hamiltonian only is especially dangerous, as the hybridization may broaden and shift levels.  Truncation is likely to introduce systematic errors. Short excursions into infrequently visited states are often needed to produce transitions between frequently visited states. For example, in the large $U$ Anderson impurity model it is the rare transitions into the states with $n=0, 2$ that produces spin flips. If truncation is to be used, it is advantageous to do so in two steps: First one does a short simulation, keeping as many states as feasible, while keeping a histogram of visited states. Even if this simulation is not fully thermalized or long enough to allow accurate measurements, it will enable an identification of frequently and infrequently visited states, which may be used to construct a truncated Hilbert space for extensive simulation.

A ``dynamic'' truncation method that speeds up the calculation of the trace without introducing errors involves checking if $\exp(-\Delta\tau H_\text{loc})$ falls below machine precision or some other threshold, and if so not computing the remainder of the product of that particular part of the trace. Unlike in the ``static'' truncation case described above, short excitations into highly excited states are still possible, but the computational gain is significantly smaller.

\subsubsection{Binning and tree algorithms for the hybridization expansion}
\begin{figure}[tb]
\begin{center}
\includegraphics[width=0.95\columnwidth]{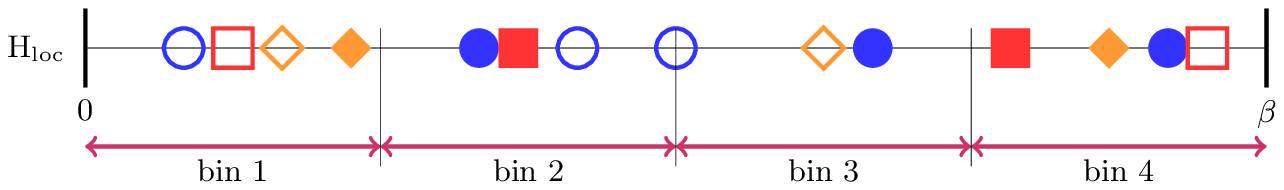}\\
\includegraphics[width=0.95\columnwidth]{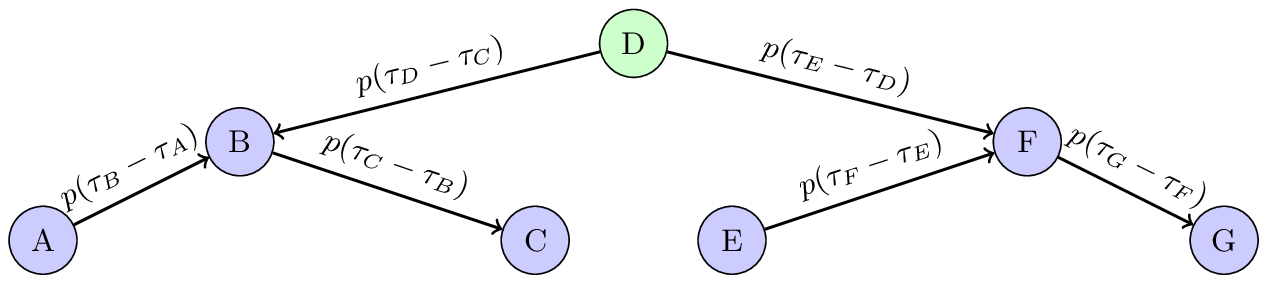}
\caption{Top panel: Binning algorithm: binning of the $k$ operators into $\sqrt{\langle k\rangle}$ bins, each having approximately $\sqrt{k}$ elements, reduces the effort of computing the trace after inserting or removing an operator to $O(\sqrt{\langle k \rangle})$. Bottom panel: binary tree for the tree algorithm, $O(\log \langle k\rangle)$. (Ref.~\cite{GullPhD})}\label{binpix}
\label{TreeStructure}
\end{center}
\end{figure}

The most expensive part of the algorithm is the computation of the trace, which is linear in the numbers of
hybridization operators present in the configuration.
Computing the complete trace in the general case will be $O(k)$, as each operator matrix must be accessed at least once. However, {\it recomputing } the trace after
an operator insertion or removal update allows simplifications: A first step is trivial to implement and reduces the effort to $O(\sqrt{k})$: the
operator trace is chopped into around $\sqrt{\langle k\rangle}$ intervals between zero and $\beta$. We then store the matrix product of
all the operators within this interval, such that  each  sub-interval contains approximately $\sqrt{\langle k\rangle}$ operators. If we insert two operators, we will
change the matrix product of one or two intervals - which need to be recomputed at the cost of $\sqrt{k}$ operations. The whole recompute operation
is therefore of $O(\sqrt{k})$, and a sweep of $O(k^{3/2})$. This algorithm is illustrated in the upper panel of Figure \ref{binpix}.

A better, but more complicated algorithm uses the properties of self-balancing binary trees. 
AVL \cite{AVL62,AVL62b,Knuth97} trees are one possibility.
Denoting dense matrices from the hybridization operators with capital letters and the exponential vectors $p(\tau_{i+1} - \tau_i) = e^{\Delta \tau H_0} = p_{i,i+1}$ with
lower case letters, we can write the trace in Eq.~(\ref{FullExpHyb}) as
\begin{align}\label{operatorchain}
\Tr \left[p_{0A}^i A_{ij} p_{AB}^j B_{jk} p_{BC}^k C_{kl} p_{CD}^l \cdots Z_{pi} p_{Z\beta}^i\right],
\end{align}
and arrange all the operators in (\ref{operatorchain}) in a binary tree. It is easy to see that for every exponential $p(\tau \rightarrow \tau_{i+1}) = e^{H_0(\tau_i - \tau_{i+1})}$ between the first and last operator we can assign one of the branches  of the tree. These ``propagators'' from time $\tau_i$ to time $\tau_{i+1}$,
where a right branch contains the propagator from the node to the smallest time of the right subtree, and a left branch contains the propagation
from the largest time of the left subtree to the node (Fig. \ref{TreeStructure}).
The main idea of the algorithm is that each node stores products of the matrix product of the left subtree times the propagator to the left, the operator, and the propagator to the right times the matrix product of the right
subtree. This requires an extra storage cost of $O(k)$ in memory and additional functions for the re-balancing of binary trees, but reduces the computational effort of a sweep to $O(k\log k)$.

\subsection{Use of symmetries, global updates}\label{globalupdates}
Global updates are updates that affect many or all of the vertices of a
configuration. Two simple examples are a spin-flip of all auxiliary spins in a CT-AUX simulation, and the exchange of all the segments of two orbitals in CT-HYB.

If the update corresponds to an exact symmetry (global spin-flips in a paramagnetic system, segment interchange for degenerate orbitals, \textellipsis), the weight of a configuration remains unchanged and a proposal of the global update will
always be accepted. In cases with exact symmetries the same effect may be achieved by enforcing the symmetry at the end of the simulation.

Global updates are useful for the systems with weakly
broken symmetries, in particular near a phase transition, where they may help to radically reduce autocorrelation times.
An example are weakly spin-polarized states.
In all instances considered in the literature so far, global updates required
the recalculation of determinants to estimate acceptance
probabilities, at the cost of $O(k^3)$
operations. Hence  they should be performed at most once per
$\langle k\rangle$ update steps. The concept has proved to be useful to describe an insulating state with a small polarization in Refs.~\cite{Poteryaev07,Poteryaev08}, and similarly in \cite{Kunes09,Chan09}.

\subsection{Vertex functions}
For some applications, in particular the determination of phase boundaries, response functions, or ``dual Fermion'' \cite{Rubtsov08}, ``D$\Gamma$A'' \cite{Toschi07} and other \cite{Slezak09,Kusunose06} extensions beyond dynamical mean field theory, the expectation value
of observables with four (or even six)
creation and  annihilation operators are needed. An example is $\Gamma^4_{abcd}(\tau_1, \tau_2, \tau_3, \tau_4) = \langle T_\tau d_a^\dagger(\tau_1) d_b(\tau_2) d_c^\dagger(\tau_3) d_d(\tau_4)\rangle$. These correspond to reducible vertices.

Time translation invariance implies that the four-point vertex is dependent on three time differences or three frequencies. Orbital or cluster symmetries of the impurity model may further reduce the number of independent indices $abcd$. Nevertheless,
for most these problems the number of observables that need to be measured -- especially for clusters or multi-orbital problems -- is overwhelming:
In a single orbital model, retaining $100$ Matsubara frequencies in each of the three momentum indices requires obtaining and storing results of $10^6$ measurements. In a four-site cluster calculation, the same number would lead to $64$ million observables.

In the CT-AUX and CT-INT algorithms, the four point correlation functions are computed using the fact that for a fixed auxiliary spin configuration
the problem is Gaussian and Wick's theorem can therefore be used together with Eq.~(\ref{GSreduc}).
Thus the problem reduces to the accumulation of
the determinant of a $2\times2$ matrix \cite{Gull08_ctaux,Rubtsov05}
\begin{equation}
\left\langle
\left | 
\begin{matrix}
(\mathcal{G}_0^{12}+\mathcal{G}_0^{1k}M_{kl}^{\stau}\mathcal{G}_0^{l2}) & (\mathcal{G}_0^{14}+\mathcal{G}_0^{1k}M_{kl}^{\stau}\mathcal{G}_0^{l4}) \\
(\mathcal{G}_0^{32}+\mathcal{G}_0^{3k}M_{kl}^{\stau}\mathcal{G}_0^{l2}) & (\mathcal{G}_0^{34}+\mathcal{G}_0^{3k}M_{kl}^{\stau}\mathcal{G}_0^{l4})
\end{matrix}
\right |
\label{fpdet}\right\rangle
\end{equation}
with $M_{kl}^{\stau}$ defined in Eq.~(\ref{M_ij}).
If only a few correlation functions are measured, Eq.~(\ref{fpdet}) is best evaluated at run-time. If many or all correlation functions have to be
measured at $n_\tau$ time points and the size $n_M$ of $M$ is comparatively small, it is advantageous to accumulate only
$\langle M_{ij}^{\stau}\rangle$ and $\langle M_{ij}^{\stau}M_{kl}^{\stau}\rangle$
and reconstruct the correlation function at the end of the computation.
While binning the latter expression is $O(n_\tau^3)$ in memory, it is only $O(n_{M}^3)$ computationally
(using the time translation symmetry).

For larger problems, in particular cluster problems, $G_{ab}(\omega_1, \omega_2)$, the instantaneous single-particle Green's functions, are computed directly in frequency (and in DCA: momentum) space for a given spin configuration.
The four-point functions are then obtained by computing the Monte Carlo average of the instantaneous Green's function products $\Gamma_{abcd}=\langle G_{ab}G_{cd}-G_{ad}G_{cb}\rangle$ and using symmetries to reduce the number of observables.
For many problems, only some of the four-point functions are needed (e.g. only the ones with energy transfer $0$, or diagonal cluster momenta in DCA).
Direct frequency measurement allows selective measurement of only these observables.

In the CT-HYB, configurations with four local operators are generated by removing two hybridization lines from configurations of the partition function, similar to how Green's function configurations are generated by removing one hybridization line.

Some applications (see, e.g.,  \cite{Toschi07,Slezak09}) require the computation of \emph{irreducible} two-particle quantities. In order to obtain these vertices from the reducible ones, Bethe-Salpeter equations have to be inverted. This process can be numerically unstable, and how it is best done is currently still an open question.

Our experience is that vertices measured directly in frequency space by the CT-AUX and CT-INT methods are most accurate, so that this is the method that should be used.

\subsection{High frequency expansions of the self energy}
In many of the CT-QMC algorithms  the self energy is obtained as the difference between the inverses of the full (${\mathbf G}$) and bare  ($\boldsymbol{ \mathcal{G}^0 }$) Green functions. Because both of these become small at high frequencies while their errors stay constant, the errors of the difference of the inverses become large and ${\mathbf \Sigma}(\omega)$ is difficult to measure accurately for large $\omega$.  It is therefore useful to have an analytical representation of the self energy at high frequencies. In a general $N$ orbital model the self energy is an $N\times N$ matrix ${\mathbf \Sigma}$ and its high frequency expansion is
\begin{equation}
{\mathbf \Sigma}(i\omega_n)={\mathbf \Sigma}_\infty+\frac{1}{i\omega_n}{\mathbf \Sigma}_1+{\mathcal O}\left(\frac{1}{\omega_n^2}\right).
\end{equation}
The coefficients ${\mathbf \Sigma}_{0,1}$ may be obtained by from the coefficients in the high frequency expansions of the full and bare  Green functions \cite{Potthoff97} using
\begin{equation}
{\mathbf G}(i\omega_n)=\frac{1}{i\omega_n}{\mathbf G}_0+\frac{1}{(i\omega_n)^2}{\mathbf G}_1+\frac{1}{(i\omega_n)^3}{\mathbf G}_2,
\end{equation}
the analogous equation for $\boldsymbol{ \mathcal{G}^0 }$, and for ${\mathbf \Sigma}=\boldsymbol{ \mathcal{G}^0 }^{-1}-{\mathbf G}^{-1}$.

The coefficients in the high frequency expansions of ${\mathbf G}$ and $\boldsymbol{ \mathcal{G}^0 }$  are in turn obtained from the discontinuities in the derivatives of ${\mathbf G}$ and $\boldsymbol{ \mathcal{G}^0 }$   across $\tau=0$ as
\begin{equation}
{\mathbf G}_n=\partial^{(n)}_\tau{\mathbf G}(\tau=0^+)-\partial^{(n)}_\tau{\mathbf G}(\tau=0^-).
\end{equation}
The time derivatives themselves may be computed from the definition
\begin{equation}
G^{ab}(\tau-\tau{'})=-\left<T_\tau d_a(\tau)d^\dagger_b(\tau{'})\right>
\end{equation}
by performing a small-time expansion of 
\begin{equation}
d_a(\tau)=e^{H\tau}d_ae^{-H\tau}
\end{equation} 
and its conjugate. The structure of the second derivative term is simplified by exploiting time translation invariance to place the derivative on the first or second operator as appropriate. The result is
\begin{eqnarray}
G_0^{ab}&=&\left\langle\{d_a,d^\dagger_b\}\right\rangle=\delta_{ab},
\\
G_1^{ab}&=&\left\langle\{\left[H,d_a\right],d^\dagger_b\}\right\rangle,
\\
G_2^{ab}&=&\left\langle\{\left[H,d_a\right],\left[H,d^\dagger_b\right]\}\right\rangle.
\end{eqnarray}
The coefficients for $\boldsymbol{ \mathcal{G}^0 }$ are obtained using the Hamiltonian without the interaction term. 

We illustrate the procedure for the generic Hamiltonian 
\begin{align}
H &= \sum_{ab}E^{ab}d^\dagger_ad_b+\sum_{a_1a_2b_1b_2}I^{a_1a_2b_1b_2}d^\dagger_{a_1}d^\dagger_{a_2}d_{b_1}d_{b_2}\\ \nonumber &+\sum_{k\alpha b}\left(V^{\alpha b}_kc^\dagger_{k\alpha} d_b +H.c.\right) +\sum_{k\alpha} \varepsilon_{k\alpha}c^\dagger_{k\alpha}c_{k\alpha}
\end{align}
(fermion antisymmetry implies that $I^{a_1a_2b_1b_2}=-I^{a_2a_1b_1b_2}$).
Important for the self energy are the commutators with the interaction term, which are (bearing in mind the antisymmetry)
\begin{eqnarray}
{\hat J}_a&\equiv\left[ {\hat I},d_a\right]=2\sum_{a_1b_1b_2}I^{aa_1b_1b_2}d^\dagger_{a_1}d_{b_1}d_{b_2},
\\
{\hat J}^\dagger_b&\equiv\left[ {\hat I},d^\dagger_b\right]=2\sum_{a_1a_2b_1}I^{a_1a_2b_1b}d^\dagger_{a_1}d^\dagger_{a_2}d_{b_1}.
\end{eqnarray}

Expanding and comparing terms we find that the constant term in the self energy is the familiar Hartree term
\begin{equation}
\Sigma_\infty^{ab}=4\sum_{a_1b_1}I^{aa_1b_1b}\left<d^\dagger_{a_1}d_{b_1}\right>,
\label{sigmainfty}
\end{equation}
while 
\begin{equation}
 \Sigma_1^{ab}=\left\langle\left\{{\hat J}_a,{\hat J}^\dagger_b\right\}\right\rangle.
 \label{sigma1}
\end{equation}

The expectation values in Eq.~(\ref{sigmainfty}) and (\ref{sigma1}) must in general be measured. 

For the single-orbital Anderson impurity model we find (with a chemical potential shift of $U/2$ usually employed) \cite{Knecht03,Blumer,ArminPhD}
\begin{align}
\Sigma(\omega) = U \left(\langle n_{-\sigma}\rangle-\frac{1}{2}\right) + \frac{U^2}{i\omega_n}\langle n_{-\sigma}\rangle(1-\langle n_{-\sigma}\rangle) + \mathcal{O}\left(\frac{1}{i\omega_n^2}\right).
\end{align}

Expressions for multi-orbital models with density density interactions are derived in \cite{GullPhD}, for plaquette CDMFT in \cite{Haule07plaquette,Haule07}, and for multi-orbital models with the Slater-Kanamori form of interactions in \cite{WangPhD}.

%% file: applicationsdmft.tex
\section{Applications I: DMFT}

\subsection{Overview}
Dynamical mean field theory (DMFT) provided an important initial motivation for the development of CT-QMC impurity solvers and is perhaps the domain to which the new solvers have made the most important contributions. We therefore consider  dynamical mean field applications in some detail. We do not review the dynamical mean formalism in detail here,  instead referring the reader to reviews of the original ``single-site''  \cite{Georges96} and subsequent ``cluster'' \cite{Maier05} formulations (see also \cite{Potthoff05}) and to reviews of  the combination of the formalism with modern electronic structure theory which provides an important step towards an ab-initio description of strongly correlated compounds \cite{Kotliar06,Held07}. However, for clarity we provide a brief explanation of the essential ideas.

A common strategy in theoretical physics is to obtain an approximation to the solution of a problem in terms of a solution of a more tractable auxiliary problem, which is specified by a self-consistency condition. Weiss mean-field theory and density functional band theory are examples.  Dynamical mean field theory provides an approximate solution of a lattice fermion problem in terms of an auxiliary quantum impurity model with interaction terms specified by the interactions in the original lattice model and single particle energies and hybridization functions determined by a self-consistency condition. One may think of it as based on an approximation of the full self energy $\Sigma^{ab}(k,\omega)$, which depends on a discrete set of orbital labels $a,b$ and continuous momentum and frequency variables $k,\omega$, in terms of $N$ functions of frequency $\Sigma_{j=1...N}(\omega)$ which are the self energies of an $N$-orbital impurity model. Different 'flavors' of dynamical mean field theory correspond to different prescriptions for reconstructing the lattice self energy from the $\Sigma_j(\omega)$ and to different forms of the self-consistency condition. All formulations require a solution of the  quantum impurity model which is of high and reasonably uniform accuracy over a wide frequency range.  It is not unfair to say that it is the development of CT-QMC techniques that  has given DMFT   the computational power needed to address the full range of problems arising in the physics of  correlated electron physics.

The first applications of dynamical mean field theory were ``single-site DMFT'' approximations to the physics of   model systems such as the one orbital Hubbard model and the one orbital Anderson models. \cite{Georges96} For these two cases  the auxiliary impurity model is the single-impurity Anderson model (Eq.~(\ref{AIM})) which can be solved to sufficient accuracy for most purposes by  pre-CT-QMC techniques, in particular the Hirsch-Fye approach. While the greatly improved efficiency of CT-QMC methods has enabled a more refined study of some aspects of the physics and has shed light on some special cases, the single-site, single-orbital case has mainly served  as a test-bed for investigating and evaluating CT-QMC methods.

A second class of DMFT applications is the ``cluster'' extensions \cite{Maier05}, which can treat the short ranged correlations characteristic of high temperature superconductors and other low dimensional systems. In the single-site DMFT method the self energy is replaced by  its average over the Brillouin zone. Cluster DMFT methods allow for some coarse-grained momentum dependence and include some aspects of intersite correlations. They  are thus of interest in the context of understanding the strong momentum space differentiation observed in high-$T_c$ cuprates and other low dimensional systems.  As of this writing, most of the ``cluster-DMFT'' literature has focused on models with a ``Hubbard'' interaction. For models with Hubbard interactions considerable progress has been made by the use of Hirsch-Fye \cite{Jarrell01,Vidhyadhiraja09,Maier05,Maier05_dwave,Macridin06} and exact-diagonalization\cite{Liebsch08,Liebsch09,Kyung06,Kancharla08} methods (but see \cite{Koch08}). However, the more efficient CT-QMC methods have permitted the examination of much wider regions of phase space, which has led to new results and insights. 

A third class of DMFT applications is to  the study of materials such as transitional metal oxides and actinides with partially filled $d$ or $f$ shells.\cite{Held06,Kotliar06,Held07,Kotliar04} In these materials multiplet interactions such as Eq.~(\ref{HSK}) are crucial to many aspects of the physics.   CT-QMC methods have provided the first reliable solvers for this class of  models and have yielded new insight into their physics.

A fourth class of DMFT applications are extensions such as the ``dual fermion'' and ``dynamical vertex'' approximations \cite{Kusunose06,Toschi07,Rubtsov08,Slezak09}. These methods require the accurate calculation of the full four-point vertices of impurity models, and this computationally challenging task seems feasible only with CT-QMC methods.

In the rest of this section we summarize the applications in the order presented above, and close with remarks about future challenges.

\subsection{Single-site DMFT approximation to the single orbital Hubbard model }

An important early success of single-site dynamical mean field theory was an improved understanding of  the ``Mott'' or correlation-driven metal insulator transition. This  is one of the fundamental questions in electronic condensed matter physics \cite{Mott49,Imada98}. The essential physics is captured by the one-band Hubbard model, specified by a hopping $t_{ij}$ between sites $i$ and $j$ and an on-site interaction $U$:
\begin{equation}
H=\sum_{ij}t_{ij}c^\dagger_{i\sigma}c_{j\sigma}+U\sum_in_{i\uparrow}n_{i\downarrow}.
\label{HHubbard}
\end{equation}

It has been known for many years \cite{Imada98} that at a carrier concentration $n=1$ per site the model exhibits a paramagnetic metal to paramagnetic ('Mott') insulator transition as the interaction strength $U$ is increased above a critical value  of the order of the bandwidth. The state obtained by doping  the large $U$ Mott insulating state has many unusual properties.

\begin{figure}[t]
\includegraphics[width=0.8\columnwidth,clip=]{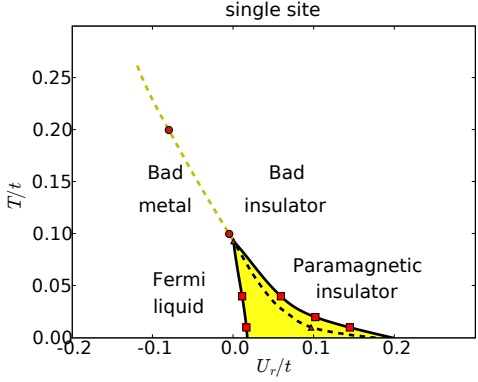}
\caption{ (Color online) Metal-insulator phase diagram of paramagnetic two dimensional Hubbard model in the single site DMFT approximation, plotted against normalized interaction strength $U_{r}=\frac{U-U_{MIT}}{U_{MIT}}$ with $U_{MIT}=9.35t$ for this model. The transition is first order, with coexistence region indicated by shading (yellow online). The dashed line indicates the bad metal/bad insulator crossover determined from  the condition that the imaginary part of the self-energy at few lowest Matsubara frequencies is flat at the crossover value of $U$. From Ref.~\cite{Park08plaquette}.
 }
\label{phasedss}
\end{figure}

A single-site dynamical mean field theory of the Hubbard model was formulated in Ref.~\cite{Georges92b}. As shown by \textcite{MH89} and by \textcite{Metzner89}, it becomes exact in a limit of spatial dimensionality $d\rightarrow \infty$ and is believed to be reasonably reliable in $d=3$. \cite{Kotliar04} corrections are significant in $d=2$ and $d=1$. The corresponding quantum impurity model is Eq.~(\ref{AIM}). Studies prior to the advent of CT-QMC established that in the single-site dynamical mean field approximation the phase diagram at half filling involves a first-order transition  with a critical end-point in the $T-U$ plane and a higher temperature crossover regime, as shown in Fig.~\ref{phasedss}. Physics beyond the single-site approximation will correct the phase diagram. In two spatial dimensions the change is qualitative, but in higher dimension the changes are less severe and the single-site phase diagram remains relevant.  The scales are low, presenting a challenge to computational methods. 

The metal-insulator  transition may be characterized by the `kinetic energy', essentially $\langle\sum_{ij}t_{i-j}c^\dagger_{i\sigma}c_{j\sigma}\rangle$, which gives a measure of the degree to which electron motion is blocked by the interaction $U$ \cite{Millis04}.  At low $T$ the transition from insulator to metal is marked by the appearance of a very narrow band of quasiparticle states inside the gap, which itself  remains well formed for a range of $U$ below the transition. These states form a Fermi liquid, but with very low Fermi temperature.  Theoretical arguments \cite{Fisher95,Kotliar02} established that the doping driven transition is also first order at low $T$, marked by the sudden appearance of states inside the Mott gap. However, the transition in this case is only weakly first order and for many years proved difficult to observe.  These and other somewhat unusual features of the phase diagram occur because in the single-site approximation the paramagnetic insulating state has an extensive entropy of $\ln 2$ per site \cite{Georges96}.  In the physical situation the entropy will be quenched below some scale, but in real three dimensional materials the scales may be low enough that the single-site phase diagram remains experimentally relevant.\cite{Kotliar04}

\begin{figure}[t]
\includegraphics[angle=-90,width=0.85\columnwidth]{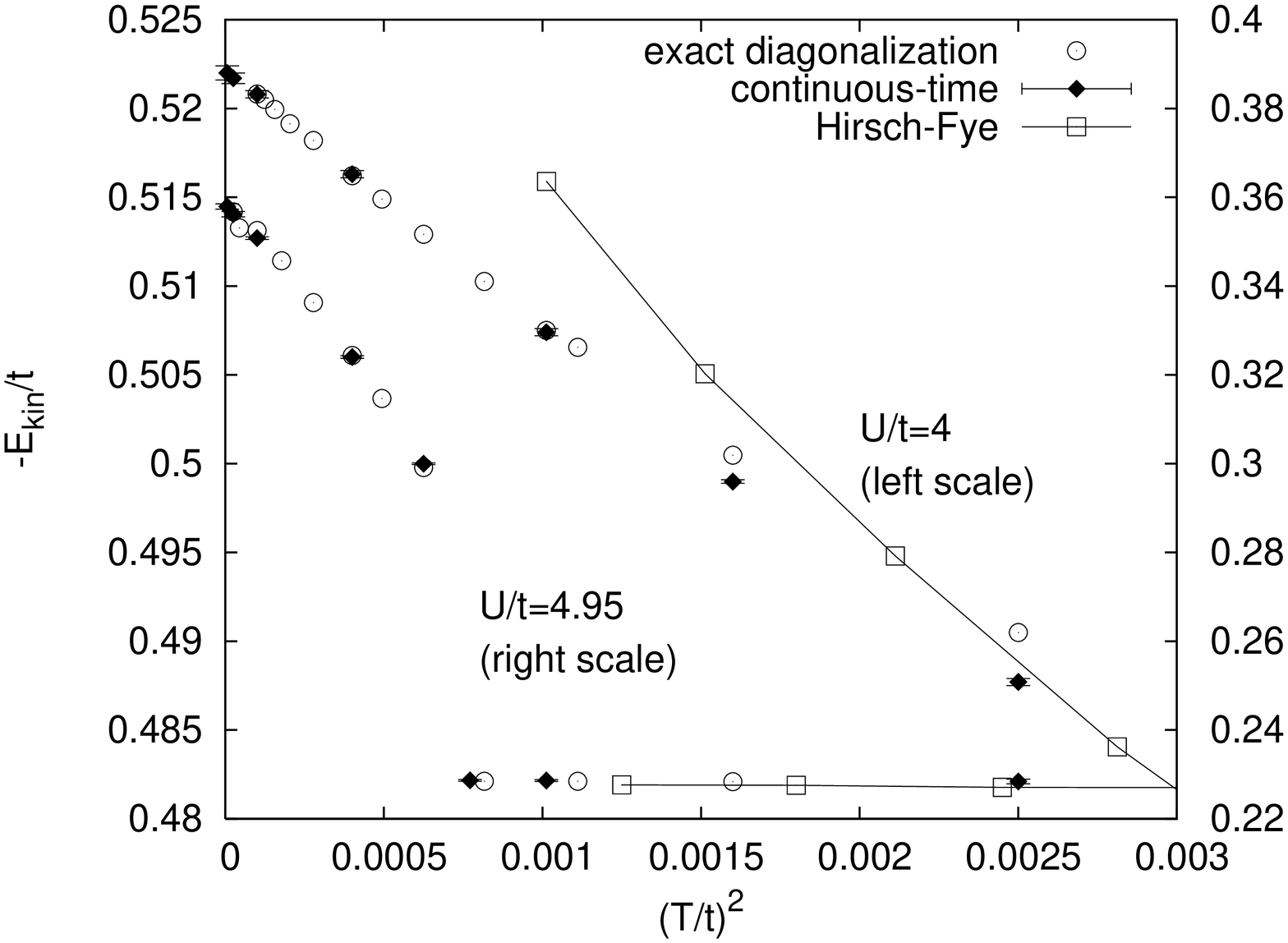}\\
\includegraphics[angle=-90,width=0.85\columnwidth]{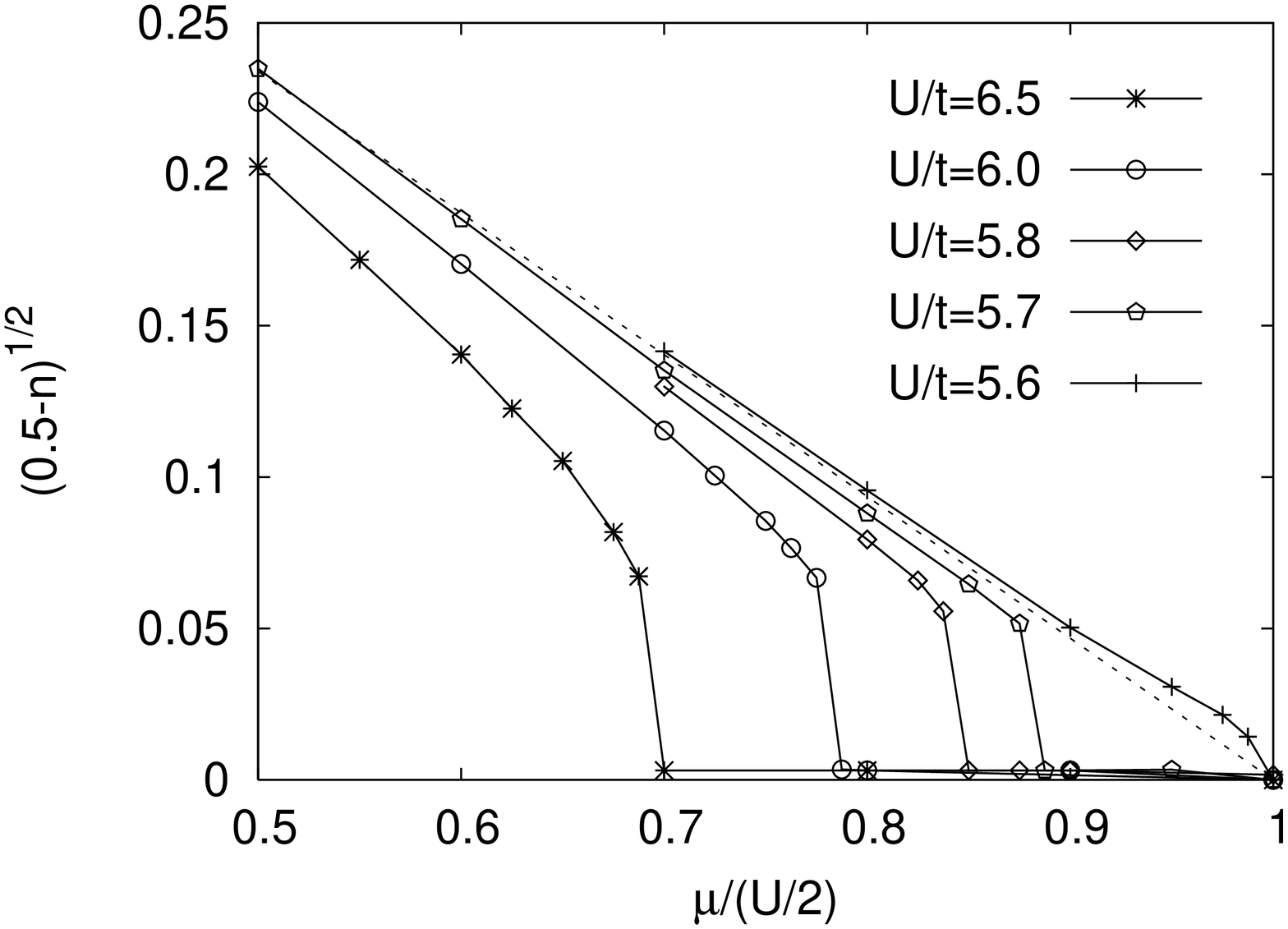} %
\caption{Top panel: Kinetic energy obtained using the indicated impurity solvers plotted as a function of temperature for the Hubbard model with a semicircular density of states and bandwidth $4t$ and interactions indicated.  For $U=4t$ the model is in a strongly renormalized metallic phase, for $U=4.95t$ a low-T metal to higher T insulator transition occurs, visible as a jump in kinetic energy at $(T/t)^2\approx 0.0007$.  From Ref.~\cite{Werner06}. Bottom panel:
Doping per spin, $0.5-n$,  as a function of chemical potential for $\beta t=400$ and indicated values of $U/t$.
At this temperature, the transition at half-filling ($\mu=\mu_1=U/2$) occurs at $U_{c}(T)\approx 5.65$.
For $U>U_c$ the $n=1$ state is insulating and shifting the chemical potential induces a first order metal-insulator transition visible as a discontinuity in $n(\mu)$. From Ref.~\cite{Werner07doping}.}
\label{CTQMCHubbard}
\end{figure}
The top panel of Fig.~\ref{CTQMCHubbard} shows results from the first CT-HYB study of the interaction driven metal-insulator phase transition \cite{Werner06}.  It compares the kinetic energy calculated in the single-site dynamical mean field theory for the one band Hubbard model  via a Hirsch-Fye simulation, an exact-diagonalization method, and the CT-HYB method. One see that the CT-HYB method agrees with the other methods (where there is overlap), allows access to very low temperatures, clearly reveals the $T^2$ behavior associated with a strongly renormalized Fermi liquid and captures the first-order Mott transition. The bottom panel, taken from \cite{Werner07doping} shows the dependence of carrier concentration on chemical potential for interaction strengths above and below the Mott transition providing  the first clear verification that the doping-driven Mott transition is first order. 
Results such as these established that the CT-HYB method provides a successful description even of subtle, low temperature properties of impurity models.

Another long-standing question in correlated electronic theory was Nagaoka's prediction \cite{Nagaoka66} of ferromagnetism in the Hubbard model at carrier concentrations very near to half filling and very strong interactions. The status of this result was unclear for many years because Nagaoka's original arguments applied rigorously only to one hole in a Mott insulator, not to a thermodynamic density of holes. Park, Haule, Marianetti and Kotliar used the CT-HYB method to establish the existence of a thermodynamic Nagaoka phase \cite{Park08}, at least in the $d = \infty$ limit.

While quantum Monte Carlo methods are most effective for imaginary time (thermodynamic) simulations, it is of course very important to attempt to obtain spectra which can be compared to experimental response functions. The standard method is maximum-entropy analytical continuation of the imaginary time data \cite{Jarrell96}. One question of particular importance has been the value of the insulating gap in the strong correlation limit at half filling.  Here a weakness of the CT-QMC methods reveals itself: because the Green function is numerically very small in the middle of the imaginary time window, the simulation does not visit this region much and the statistics are relatively poor. But it is precisely this region which is important for the value of the insulating gap. Straightforward analytical continuation of the Green function leads to broadened gap edges.  Ref.~\cite{Wang09gap} discusses the issue in detail, arguing that one should instead continue the self energy and construct the green function from the continued self energy.

\subsection{Cluster dynamical mean field theory of the single orbital Hubbard model.}

The single-site dynamical mean field theory neglects spatial correlations and while it becomes exact in an appropriately defined infinite dimensional limit  \cite{Metzner89} it  is known to  provide an insufficient description of the metal insulator transition in finite dimensional models. Deviations from the single-site dynamical mean field picture are particularly large in the case of two spatial dimensions relevant for high temperature superconductivity. A striking  and still ill-understood feature of hole-doped high temperature cuprate superconducting materials is the  `pseudogap', a suppression of the electronic spectral function occurring for momentum states along the Brillouin zone face but not for states along the zone diagonal.  (In electron doped cuprates a phenomenologically somewhat different effect, confusingly also sometimes termed ``pseudogap'' is now understood as arising  from proximity to a state with long-ranged two sublattice antiferromagnetic order \cite{Kyung04,Zimmers05,Motoyama07,Armitage10}).
The pseudogap is a dramatic example of the more general phenomenon of `momentum space differentiation': an increase in the variation of physical quantities around the Fermi surface as the insulating phase is approached. Its origin and consequences remain hotly debated topics.

\begin{figure}[tb]
\includegraphics[width=0.8\columnwidth]{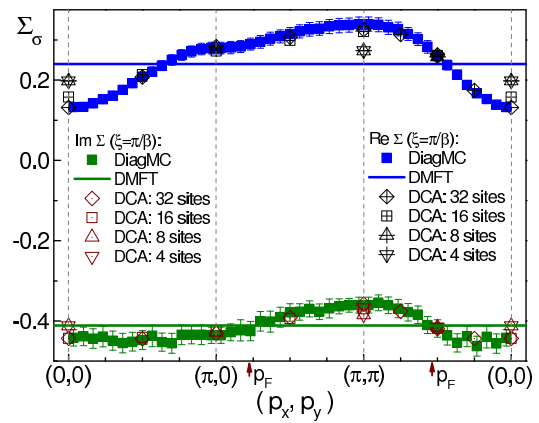}
\caption{Comparison of momentum-dependence of the self-energy of the  two-dimensional Hubbard model with parameters $U/t=4$, $\mu/t=3.1$ and $T/t=0.4$  calculated at the Matsubara frequency $\omega_0 = \xi = \pi/\beta$ calculated for  along the cut $(0,0)-(\pi,0)-(\pi,\pi)-(0,0)$ in the first Brillouin zone using a numerically exact ``diagrammatic Monte Carlo'' procedure and using CT-AUX simulations of single site and $4$, $8$, $16$, and $32$-site DCA DMFT approximations. From Ref.~\cite{Kozik09}.}
\label{diagmc}
\end{figure}
Attention in recent years has focused on cluster dynamical mean field theories \cite{Maier05}, which capture at least some aspects of spatial correlations. These methods have produced a range of very exciting results with strong qualitative similarities to the cuprates \cite{Tremblay06} but are computationally very demanding.To date, cluster dynamical mean field approximations have mainly been used to study the single-band, two dimensional  Hubbard and $t-J$ \cite{Zhang88} models, although some work on Hubbard-like  models related to heavy fermions has appeared \cite{Sun05}. Significant results were obtained with approximate analytical and semi-analytical methods \cite{Parcollet04,Kyung06,Chakraborty08}, exact diagonalization \cite{Civelli05,Kancharla08,Liebsch08,Koch08,Liebsch09} and Hirsch-Fye QMC \cite{Lichtenstein00,Maier05,Maier05_dwave,Maier06,Maier07,Maier07B,Jarrell01,Huscroft01,Macridin06,Vidhyadhiraja09} approaches.  While ED results have been reported only for clusters up to $4$ sites, Hirsch-Fye approaches have been extended up to clusters of size $64$ (at  weak interaction strength) \cite{Moukouri01} and $16$ \cite{Vidhyadhiraja09} (at moderate to strong interaction strength) although the magnitude of the computations required meant that studies were restricted to select dopings.

The advent of CT-QMC methods  greatly increased the ranges of parameters that could be studied with reasonable computational resources. Scans of parameter space became feasible and phase diagrams have been established.  Hybridization expansion methods have been used to study $2$-site \cite{Ferrero09,Ferrero09b} and $4$-site \cite{Gull08_plaquette,Park08plaquette} clusters. In the case of these small clusters, the analysis of cluster eigenstate occupation probabilities has provided new insights. For larger clusters the dimension of the local Hilbert space is so large that the hybridization expansion method has not been successfully applied.   CT-AUX methods have been used to study $8$ site clusters \cite{Werner098site,Gull09_8site} and, at $U=4t$, a range of cluster sizes up to $32$ \cite{Kozik09}. 

We present here a few representative CT-QMC cluster DMFT results which illustrate the power of the methods and the nature of the new results which have been obtained. The convergence of cluster schemes with cluster size is shown in Fig.~\ref{diagmc}, which compares CT-AUX  cluster DMFT results are to a direct Monte Carlo evaluation (``diag-MC'') of diagrams of the lattice problem \cite{Kozik09}. While ``diagMC'' as of now only works for relatively weak interactions, the results do not contain a $k$-space discretization of the self energy, so that the results are exact within error bars. As seen in Fig.~\ref{diagmc} for $U=4$, convergence of the cluster DMFT results to the exact ones is achieved with $32$ sites.

We now turn to results relating to stronger coupling physics, beginning with results obtained for four site clusters, which have been studied using CDMFT \cite{Kotliar01} and DCA \cite{Hettler98} versions of cluster dynamical mean field theory. The four-site cluster calculations may be thought of as approximating the full momentum dependence of the self energy by its value at the four points $S=(0,0)$, $P_y=(0,\pi)$, $P_x=(\pi,0)$ and $D=(\pi,\pi)$.
\begin{figure}[t]
\includegraphics[width=0.8\columnwidth,clip=]{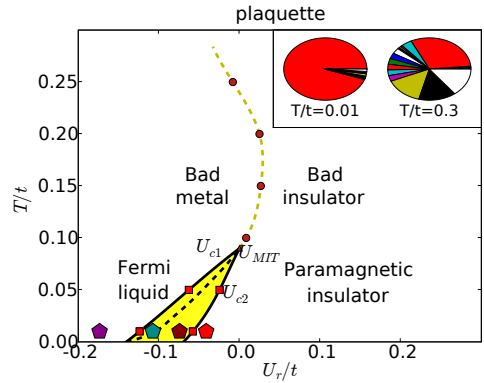}
\caption{
(Color online) Metal-insulator phase diagram of the paramagnetic phase of the two-dimensional Hubbard model in the plane of temperature $T/t$ and interaction $U/t$ measured relative to the critical end-point value $U_\text{MIT}=6.05t$ in the 4 site CDMFT cluster approximation. Band parameters are identical to those used in Fig. \ref{phasedss} Inset: pie-chart histogram of occupancy probability of the two insulating states at low and high temperatures. From Ref.~\cite{Park08plaquette}.
 }
\label{phased}
\end{figure}

The physics brought by the added momentum dependence changes the character of the Mott transition in dimension $d=2$. Fig.~\ref{phased} shows the phase diagram of the two dimensional Hubbard model obtained in a detailed CT-HYB study  of the 4-site CDMFT approximation \cite{Park08plaquette}. It should be compared to Fig.~\ref{phasedss} which presents single-site DMFT results for the same model. The interaction-driven transition was  found to be first order, as in the single-site case. However, not only is the critical interaction strength much less than in the single-site approximation, but the phase boundary bends in the opposite direction from that found in the single site calculation, indicating that in the multi-site approximation the insulating phase has lower entropy than the metallic phase. The narrow band of in-gap states whose appearance characterizes the Mott transition in high dimension\cite{Fisher95,Kotliar02}  is not found in cluster calculations for 2D systems. 

Insight into the metal-insulator transition is enhanced  by the ability of CT-HYB to provide sector occupation statistics \cite{Haule07}. These are indicated in Fig.~\ref{phased} by pie-chart insets. The low temperature insulating phase was found to be characterized by a strongly dominant occupation of one state, corresponding to a singlet configuration of the four electrons on the plaquette. This correlation was argued by Gull {\it et al.} \cite{Gull08_plaquette} to indicate that in the cluster dynamical mean field methods  the metal-insulator transition was driven by the appearance of strong short ranged order (most likely related to a columnar dimer phase).  By contrast, the high temperature ``bad insulator'' state, which has entropy of the order of $\ln(2)$, populates many states of the plaquette with significant probability.

\begin{figure}[tb]
\includegraphics[width=0.8\columnwidth]{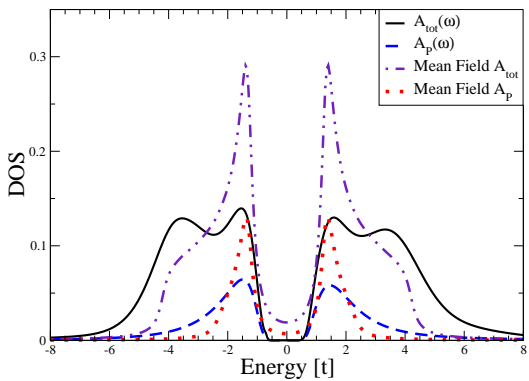}
\caption{Solid line: on-site spectral function computed for different momentum sectors by maximum entropy analytical continuation of QMC data for $U=6t$ and doping $x=0$. Dashed line: spectral function in the $P=(0,\protect\pi) , (\protect\pi,0)$%
-momentum sector. Dotted and dash-dotted lines: $P=(0,\protect\pi) , (%
\protect\pi,0)$ and local spectral functions obtained by performing the DCA
momentum averages of the standard SDW mean field expressions for the Green
function, with gap $\Delta=1.3t$. From Ref.~\cite{Gull08_plaquette}.}
\label{spectral}
\end{figure}

Further evidence of the importance of short ranged order was obtained from the electron spectral functions \cite{Gull08_plaquette,Park08plaquette} computed by maximum entropy analytical continuation and shown in Fig.~\ref{spectral}. The insulating state has a gap. The dotted line gives the spectral function calculated in a mean field approximation based on a two sublattice order; the strong similarity indicates that short ranged order is responsible for the insulating behavior.\begin{figure}[bt]
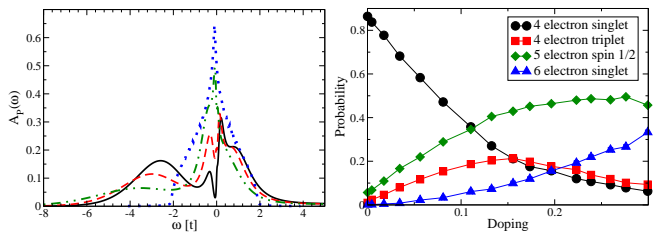

\includegraphics[width=0.49\columnwidth]{figures/appl_pdosofmu.eps}
\includegraphics[width=0.49\columnwidth]{figures/appl_sectorsvsn.eps}
\caption{Left panel: Doping dependence of $P=(0,\protect\pi) , (\protect\pi,0)$-sector density of states obtained by analytical continuation of quantum Monte Carlo data obtained from DCA approximation at $U=5.2t$, temperature $T=t/60$ and dopings  $x=0.04$ (solid), $x=0.08$ (dashed), and $x=0.15$ (dash-dotted line). Dotted line denotes the noninteracting density of states. Right panel: Evolution of the occupation probabilities with doping at $U=5.2t$ and temperature $T=t/30$. From Ref.~\cite{Gull08_plaquette}.}\label{pdos_sectors}
\end{figure}

The left panel of Fig.~\ref{pdos_sectors} presents the changes in  the density of states in the $P=(0,\pi) , (\pi,0)$-sector as electrons are added. The curves are  obtained by analytical continuation of  quantum Monte Carlo data.  The `Mott' gap visible in Fig..~\ref{spectral} has filled in even at the lowest doping shown, but for the lower dopings a small `pseudogap' (suppression of density of states) appears near the Fermi level while for $x=0.15$ the value of the spectral function at the Fermi level approaches that of the noninteracting model, indicating the restoration of Fermi liquid behavior, consistent with experiment and with many previous theoretical results. 

Examination of the sector statistics shown in the right panel of Fig.~\ref{pdos_sectors} indicated that the transition  from pseudogapped to Fermi liquid behavior occurred at the doping at which the plaquette singlet state ceased to dominate the physics. An intriguing and still open question concerns the degree to which the level crossing in sector statistics is related to the `avoided criticality' discussed by Haule and Kotliar \cite{Haule07c}.

\begin{figure}[t]
\includegraphics[angle=0, width=0.42\columnwidth]{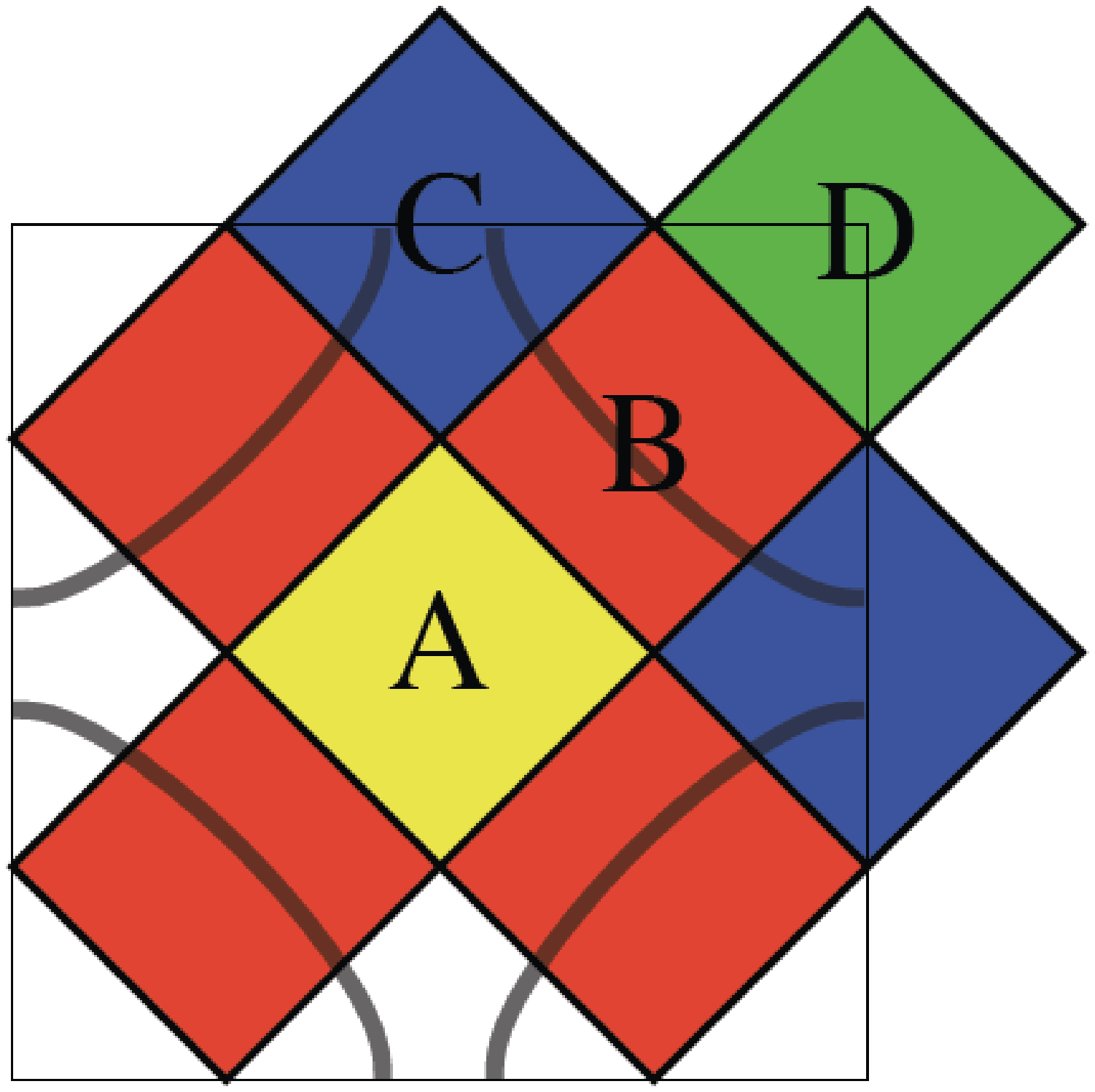}
\includegraphics[angle=0, width=0.51\columnwidth]{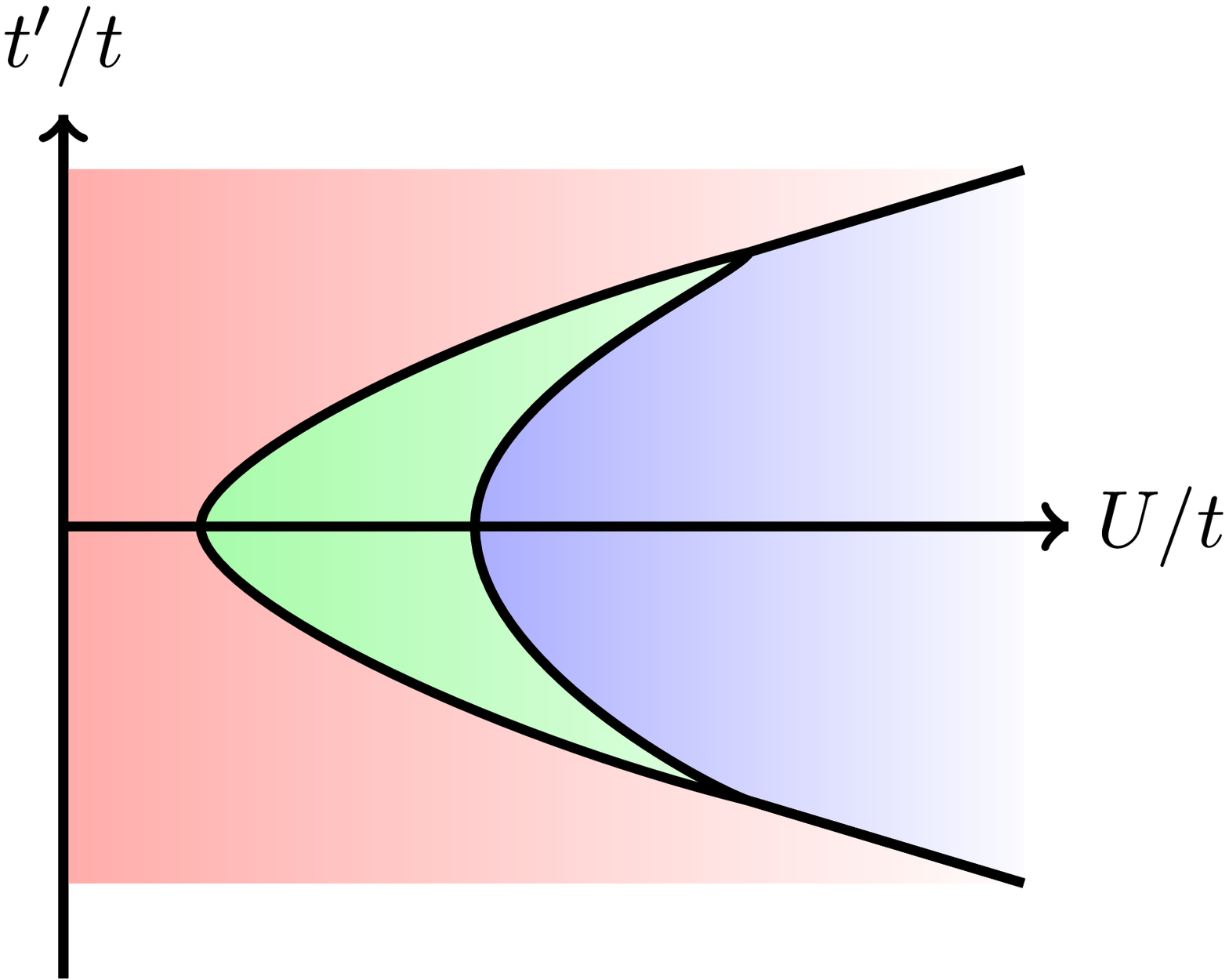}
\caption{ Left panel: Brillouin zone partitioning associated with the 8-site cluster DCA approximation with definition of the four inequivalent momentum sectors $A$, $B$, $C$ and $D$. The noninteracting Fermi surface for $t^{\prime}=-0.15t$ and density $n=1$ is indicated by the gray line. Right panel: sketch of the paramagnetic state DCA phase diagram of the Hubbard model, calculated for the cluster shown in the left panel at half filling, as a function of interaction strength $U$ and next-nearest neighbor hopping $t^{\prime}$. A Fermi liquid metal phase (left, red on-line), a sector selective
intermediate phase (middle, green on-line) in which the sectors labeled as C are gapped but those labeled as B remain gapless, and a fully gapped insulating phase (right, blue on-line) are shown. From Ref.~\cite{Gull09_8site}.}
\label{cluster}
\end{figure}

Very recently CT-AUX methods have been used to examine the larger $8$ site cluster shown in Fig.~\ref{cluster}. The greater efficiency of the CT-AUX method permitted a comprehensive examination of the behavior as a function of interaction strength, carrier concentration, second neighbor hopping and temperature \cite{Werner098site,Gull09_8site}.    A striking new result is that both the interaction-dependent and doping-dependent metal insulator transitions are multi-staged, with different regions of the Fermi surface are successively gapped as carrier concentration or interaction strength are varied. (Similar behavior was also found in a $2$ site cluster with a clever choice of momentum-space patching \cite{Ferrero09}).  The phase diagram for the interaction-driven transition is shown in the right-hand panel of Fig.~\ref{cluster}.

\begin{figure}[htb]
\begin{center}
\includegraphics[width=0.85\columnwidth]{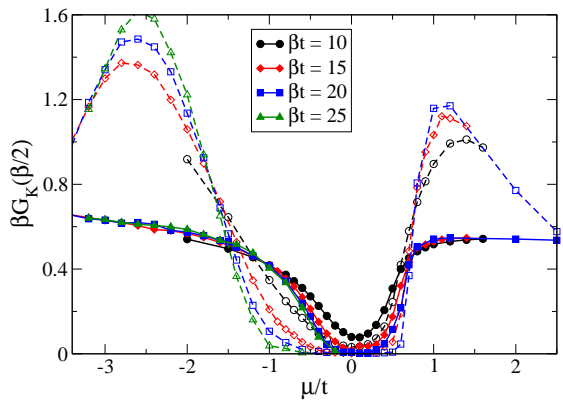}
\end{center}
\caption{
$\beta G(\frac{\beta}{2})$ calculated in DCA approximation to $8$-site cluster for sectors $B$ (full symbols) and $C$ (empty symbols), at $U/t = 7$ and $t'/t=-0.15$. The strong temperature dependence in the sector $C$ curves arises from the Van Hove divergence in the density of states. The crossing points indicate the onset of gapping in the sectors.
From Ref.~\cite{Gull09_8site}.}
\label{ghalf_mu_tprime0.15.fig}
\end{figure}

Identification of a gapped region in a spectrum can be based on analytical continuation.  However, obtaining data of the requisite quality for analytical continuation is very expensive, and analytical continuation is in any event a notoriously ill-posed problem. Methods for identifying metal-insulator phase boundaries directly from imaginary time are therefore valuable. At present it appears  that the most reliable method is to plot $\beta G_K(\beta/2)$ in momentum sector $K$, related to the density of states at the Fermi energy by $\beta G_K(\beta/2)=\int \frac{d\omega}{2\pi T}\frac{A_K(\omega)}{\cosh[\omega/(2T)]}$. This is shown as a function of interaction strength or chemical potential for several temperatures as shown in Fig.~\ref{ghalf_mu_tprime0.15.fig}. The sector gapping transitions were identified from the temperature dependence of $\beta G_K(\tau=\beta/2)$. One sees from Fig.~\ref{ghalf_mu_tprime0.15.fig} that a gap opens in sector $C$ at lower $\mu$ than in sector $B$.  Remarkably, this sector-selectivity occurs on the hole doped but not on the electron-doped side of the phase diagram. The successive gapping bears an intriguing similarity to the behavior of high-$T_c$ cuprates in the pseudogap regime. The interpretation and implications of the CT-QMC results are at present the subject of active investigation.

\subsection{Dual-fermion calculations for the single-orbital Hubbard model}
Cluster dynamical mean field methods suffer from several drawbacks. A cluster of a given size corresponds to a coarse-graining either in real or reciprocal space, which may bias the physics. At model parameters relevant for high-$T_c$ cuprates, the sign problem limits the range of cluster sizes that can be studied, even with CT-QMC methods, so that systematics of scaling with cluster size has been established only for weak interactions (Fig.~\ref{diagmc}, \cite{Maier05,Maier05_dwave,Kozik09}).

Alternative ways to handle nonlocal correlations have been proposed \cite{Kusunose06,Slezak09,Rubtsov08,Toschi07}.  These methods are systematic expansions around the single-site DMFT approximation, and have the advantage that both short- and long-range fluctuations are treated simultaneously, but require evaluation of vertex functions.  We discuss here the dual fermion approach where CT-QMC methods have been extensively applied; the computational issues for the other methods are similar. The dual fermion approach \cite{Rubtsov08} is formulated as a standard diagrammatic technique in terms of auxiliary, so-called dual variables, introduced via a continuous Hubbard-Stratonovich transformation. The corrections to single-site DMFT appear as diagrams containing the \emph{reducible} vertex parts of single-site DMFT impurity problems at nodes, whereas lines are propagators for dual Green's functions corresponding to non-local parts of the DMFT lattice Green's function.

Technically, the method requires an impurity solver that can provide not only single-electron Green's functions of the (single-site) impurity problem, but also the full four point vertices (also of the single-site impurity problem) as a function of all frequencies. The CT-QMC algorithms allow such calculations in both the interaction and hybridization expansion formalisms and have been employed for dual-fermion analyses of the Hubbard model.

\begin{figure}
\begin{center}
\includegraphics[width=0.7\columnwidth]{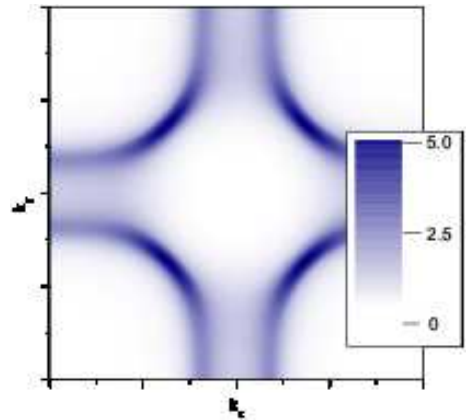}
\end{center}
\caption{Spectral function $A_{\omega=0, k}$ at the Fermi level calculated for the  Hubbard model at  $t=0.25,~ t'=-0.075,~ U=4.0,\beta=80$ and doping $x=0.14$  using the lowest order momentum-dependent diagram in the dual fermion method, with analytical continuation performed by polynomial extrapolation from Matsubara frequencies. An anisotropic destruction of the Fermi surface in the pseudogap regime is clearly visible. From
\cite{Rubtsov09}.}\label{DualArcs.fig}
\end{figure}

In Ref.~\cite{Rubtsov09} the pseudogap regime of the doped $t-t'$ Hubbard model was studied. A CT-INT solver was used to obtain both the Green's function $G$ and the 4-point vertex $\Gamma^{(4)}$ in the Matsubara-frequency domain. The spectral function $A_k=-1/\pi {\rm Im}G_{\omega=0, k}$ for the entire Brillouin zone is shown in Fig.~\ref{DualArcs.fig} for 14\% doping. The phenomenon of momentum space differentiation is clearly seen: the Fermi surface in the antinodal direction is relatively diffuse, whereas sharp quasiparticles appear near the nodal points.

\begin{figure}
\begin{center}
\includegraphics[width=0.95\columnwidth]{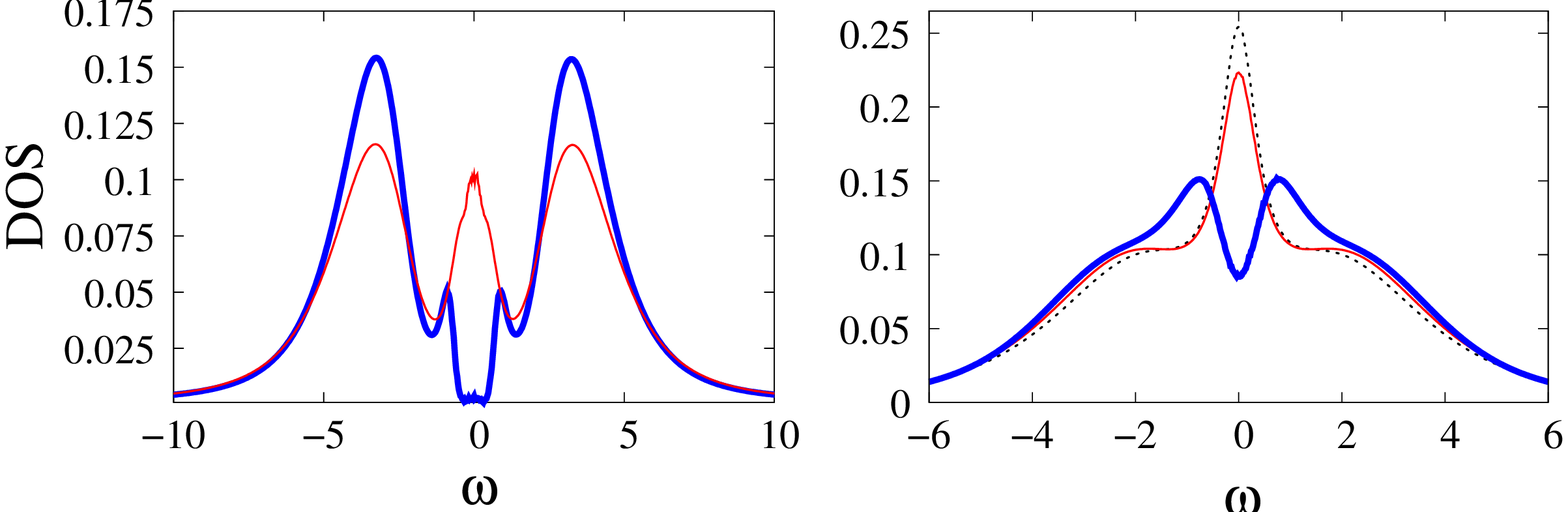}
\end{center}
\caption{Momentum-integrated spectral function (DOS)  of half-filled single-band Hubbard model calculated using the dual-fermion method. Left: Metallic (thin lines, red on-line) and insulating (heavy line, black on line) spectral function within the coexistence region of the Mott transition $U/t=6.25$, $T/t=0.08$. Right: DMFT (dashed), lowest-order dual fermion correction  (thin line) and ladder dual fermion (thick line) DOS at $U/t=4$, $T/t=0.19$. The ladder dual fermion result exhibits a pseduogap of antiferromagnetic origin. From \cite{Hafermann09}.}\label{DualDosLadder.fig}
\end{figure}

CT-INT was used in \cite{Hafermann09} to sum the particle-hole ladders in dual diagrams for the half-filled Hubbard model, revealing a pseudogap formed by antiferromagnetic correlations even in the absence of a explicit symmetry breaking. Further investigation of this and related approximations is an active area of research. Some of the results are presented in Fig.~\ref{DualDosLadder.fig}.

\section{Applications II: DMFT for multi-orbital and Kondo models}

Most ``correlated electron'' materials involve transition metal, rare-earth or actinide states with multiply degenerate levels. The electrons in these levels are subject to complicated interactions such as the ``Slater-Kanamori'' couplings shown in Eq.~(\ref{HSK}) and exhibit a richer variety of physical effects than found in the single-orbital Hubbard model, including high-spin to low-spin transitions, orbital ordering and orbitally selective Mott transitions. Until recently investigation of these models was hampered by a lack of good numerical methods: there were no good auxiliary field transformations,  so Hirsch-Fye methods could not be used unless  the rotational symmetry of the interaction was broken  so only the $J_z$ (density) component of the spin exchange was retained and the ``pair hopping''  terms in the Slater-Kanamori Hamiltonian were neglected. There were too many states for exact diagonalization or NRG methods. The situation has now changed. Studies of realistic models of materials involving electrons in two or three-fold degenerate orbitals are straightforward, $5$ orbital problems (i.e. the full $d$ multiplet, needed e.g. for the pnictides) are manageable, and problems involving one electron or hole in the $7$-fold degenerate $f$ shell are becoming possible.   However, a complete single-site DMFT computation for materials such as Pu  in which the $f$ shell is multiply occupied and rotationally invariant (exchange and pair-hopping) interactions are important cannot be done with present methods:  basis truncations (Sec.~\ref{truncation_sec}) or other approximations (for example the Krylov techniques discussed in section ~\ref{Krylov})   are required. The calculations are generally done with the hybridization solver because the interactions are strong and multiple interactions are important and so far have been restricted to the single-site dynamical mean field approximation because the proliferation of orbitals means that multisite models involve too many states to be practical at present. 

\begin{figure}[t]
\begin{center}
\includegraphics[angle=-90, width=0.9\columnwidth]{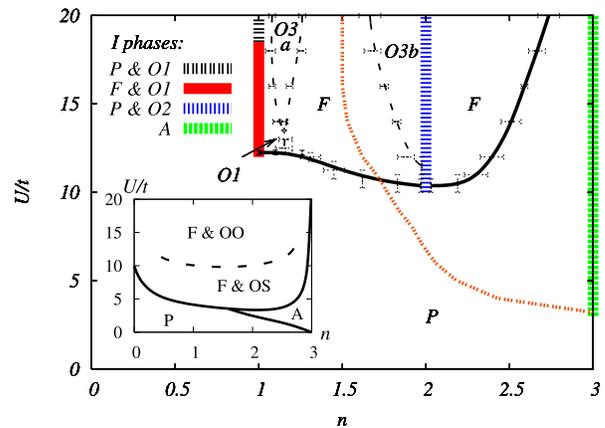}
\end{center}
\caption{(Color online) Main panel: phase diagram of the 3 band model with semicircular density of states at $\beta t=50$ and $J=U/6$ in the plane of particle density $n$ and interaction strength $U$. The vertical lines  indicate the Mott insulating
phases at integral values of $n$. The magnetic state is labeled by P (paramagnetic), F (ferromagnetic) and A (two sublattice
antiferromagnetic) while the labels O(N) denote  the 3 classes of orbital ordering discussed in \cite{Chan09}. The heavy dashed line (orange on-line) gives the boundary of the non-Fermi-liquid frozen-moment phase discovered in \cite{Werner08nfl}. Inset:  Hartree-Fock phase diagram for magnetic phases of  the same model. Magnetic phase boundaries are indicated by solid lines and orbital ordering boundaries by dashed lines. OO and OS stand for the orbitally-ordered and orbitally-symmetric phases respectively. All transitions are second order except the FM-AFM transition and
the orbital ordering transitions at $U\gtrsim 12t$ and small $n$. From Ref.~\cite{Chan09}.}
\label{phasediagram}
\end{figure}

Figure~\ref{phasediagram} shows the phase diagram \cite{Chan09} calculated  for a model of electrons moving among three degenerate orbitals with the full rotationally invariant interactions. The effect of the Hund's coupling on the multiorbital Mott transition was determined and a rich multiplicity of phases has been found.   The orbital degree of freedom is important to stabilize the metallic phase at relevant interaction strengths (the two orbital model with two electrons and $J/U=1/6$ is insulating for $U\gtrsim 3.7t$ \cite {Werner07crystal}). Suppressing  the $L=1$ orbital angular momentum states by applying a crystal field rapidly leads to an insulator.

A remarkable feature of the phase diagram is the line indicating an apparent quantum  ``spin freezing'' transition with unusual properties \cite{Werner08nfl}.  The phase exists only in the window $0<J<U/3$. For $J=0$ the frozen moment phase does not exist while for $J>U/3$ the term $ U^{\prime}-J=U-3J$ in Eq.~(\ref{HSK}) changes sign and the physics of the model becomes different. The spin freezing transition was originally identified from an unusual behavior of the self energy  and its nature was confirmed by  an examination of the local spin and orbital correlation functions.

\begin{figure}[t]
\begin{center}
\includegraphics[angle=-90, width=0.9\columnwidth]{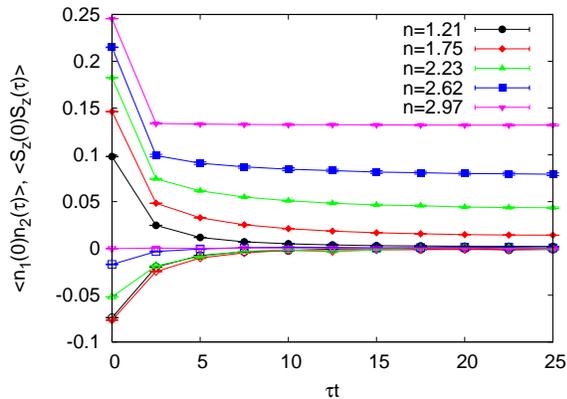} %
\end{center}
\caption{
Imaginary time dependence of the spin-spin correlation function $\langle S_z(0)S_z(\protect\tau)\rangle$ (positive correlation
function, full symbols) and orbital correlation function $\langle n_1(0)n_2(\protect\tau)\rangle$ (negative correlation function, open symbols) calculated for a three-band model with $U=8t$ and $J=U/6$ using CT-HYB at carrier concentrations $n$ indicated.  The convergence of the spin correlations to a value different from $0$ while the orbital correlation converge to zero indicates spin  but not orbital freezing in the model.
From Ref.~\cite{Werner08nfl}.}
\label{szsz_ninj_doping}
\end{figure}

Figure~\ref{szsz_ninj_doping} presents results for the imaginary-time impurity-model spin-spin and orbital-orbital correlators $C_{\mathcal{O}\mathcal{O}}(\tau)=\langle \mathcal{O}(\tau)\mathcal{O}(0) \rangle$ with $\mathcal{O}$ representing either the electron spin density $S_z=\frac{1}{3}\sum_\alpha
\frac{1}{2}(d^\dagger_{\alpha,\uparrow}
d_{\alpha,\uparrow}-d^\dagger_{\alpha,\downarrow}d_{\alpha,\downarrow})$ or
the orbital density ${\hat n}_{\alpha}=\sum_\sigma
d^\dagger_{\alpha,\sigma}d_{\alpha,\sigma}$.
In a Fermi liquid at low temperature $T$, $C_{SS}(\tau)\sim (T/\sin(\pi \tau T))^2$ for imaginary times $\tau$ sufficiently far from either $\tau=0$ or $\tau=1/T$. The DMFT results are consistent with this form in the Fermi liquid phase, but in the non-Fermi-liquid phase the spin-spin correlator $C_{SS}$ is seen to approach a constant at long times indicating the presence of frozen moments whereas the orbital correlator  is seen to decay rapidly with time on both sides of the phase transition.

The hybridization expansion solver yields information \cite{Haule07} on  which of the different eigenstates of $H_\text{loc}$ are represented in the partition function.  At $J>0$ at couplings ($U \gtrsim 4t$) only a few states are relevant. The large-$U$ density-driven transition is marked by a change in the dominant states from the one-electron states $S=1/2$, $L=1$ to a nine-fold degenerate manifold of two electron states with $S=1$ and $L=1$, with the two manifolds becoming degenerate at the transition. The interaction-driven transition is on the other hand marked by a change in the weight of the two subleading states $S=1/2$, $L=1$ and $S=3/2$, $L=0$, implying a change in the magnitudes of coupling strengths. The ability  to combine measurements of response functions with an analysis of which states contribute appreciably to the partition function is a great advantage of the CT-HYB method. This ability has been used in a recent paper to gain important new insights into the ``hidden order'' phase of the heavy fermion material URu$_2$Si$_2$ \cite{Haule09_URu2Si2}.

\subsection{Heavy Fermion compounds and the Kondo Lattice Model}

``Heavy fermion'' compounds pose one of the great conceptual challenges of correlated electron physics \cite{Stewart84}. These materials are intermetallic compounds in which one element is a rare earth (such as Ce or Yb) or actinide (such as U or Pu) with a partially filled $f$-shell, while the other elements contribute $s,p,d,$ electrons to broad, weakly correlated bands.  The $f$ electrons are weakly hybridized to the other bands and are subject to strong interactions, so that typically one $f$ valence state is strongly dominant.  At temperatures of the order of room temperature, the materials appear as two-component systems, with magnetic moments (arising from the $f$ shells) embedded in and weakly coupled to a Fermi sea of $s,p,d$ electrons. At low temperatures, however, the spins and conduction electrons combine into a new object, which may become a heavy mass Fermi liquid or a narrow gap Kondo insulator, or may become unstable to unconventional superconductivity, magnetic order, or may exhibit a variety of  quantum critical behavior \cite{Stewart01,Lohneysen07}. Our understanding of the heavy fermion state has been hampered by a lack of unbiased numerical methods.  While numerics is still far from being able to address the full richness of heavy fermion physics, the combination of dynamical mean field theory and methods including the CT-QMC  approach is beginning to have an impact on the field. 

CT-HYB methods have been applied to the study of heavy fermion materials \cite{Shim07Pu,Haule09_URu2Si2} but it appears at present that difficulties arising in the course of  dealing with realistic models of heavy fermions are sufficiently large that CT-HYB methods have been mainly used to spot-check the results of other,  approximate but much less computationally expensive, solvers.  A realistic treatment of heavy fermion materials must deal with the full complexity of the $f$-shell   and is characterized by a multiplicity of interactions, all of which are strong, a strong spin orbit coupling and (in many of the interesting materials) a low point group symmetry, leading to a complicated multiplet structure imposed on a local Hilbert space of dimension $4^7$. This HIlbert space is too large to treat directly by a straightforward application of the  CT-HYB method.  However, in many if not all  cases only a small portion of the Hilbert space is relevant to the physics, so truncation schemes in which only a portion of the Hilbert space is retained may be appropriate. In some cases, such as elemental $Ce$ or $Ce$-based heavy fermion compounds the relevant valence states are $f^0$, $f^1$ and perhaps $f^2$ and a straightforward truncation in which all higher occupancies of the $f$ state are forbidden works well. In other situations, such as $Pu$, more elaborate schemes involving truncation in energy, in valence and in size of sub-matrices is required.   CT-J methods, which in effect reduce the Hilbert space of the local problem  to the minimum possible size, are a promising alternative route. 

In a very interesting first step in this direction, Otsuki and collaborators \cite{Otsuki09,Otsuki09c} have used the CT-J method to perform a detailed study of the single-site dynamical mean field solution of the spin $1/2$ Kondo lattice model, defined by the Hamiltonian 
\begin{equation}
H_{KL}=\sum_{k\sigma}\varepsilon_kc^\dagger_{k\sigma}c_{k\sigma}+J\sum_i{\vec S}_i\cdot {\vec \sigma}_i,
\label{HKL}
\end{equation}
while Matsumoto and collaborators \cite{Matsumoto09}  have used the CT-J method along with input from ab-initio band theory to describe trends across families of heavy fermion compounds. 

\begin{figure}[tb]
\includegraphics[width=0.85\columnwidth]{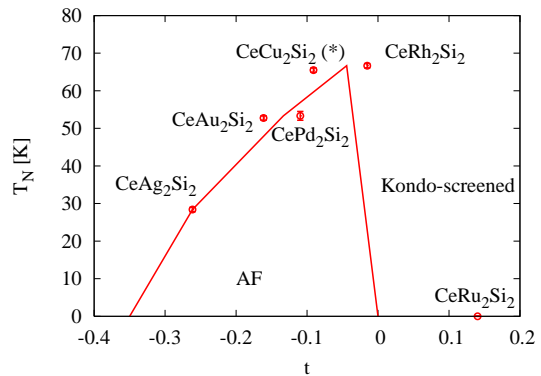}
\caption{
Phase diagram calculated using CT-J methods with band theory input for the material family CeX$_2$Si$_2$. The abscissa $t$ is defined as the Kondo coupling measured relative to the critical Kondo coupling for the $T=0$ magnetic transition. From \cite{Matsumoto09}.  }
\label{fig:phase_diagram}
\end{figure}

This model with antiferromagnetic $J$ is a minimal model for heavy fermion physics and also may be used to address other theoretical issues, for example by changing the sign of $J$.  In the physically relevant antiferromagnetic $J$ case the model is believed to have a small $J$ magnetic phase and a larger $J$ non-magnetic phase: a Fermi liquid if the density of the conduction band is different from $1$ per site and a Kondo insulator if the conduction-band density is one per site. The  qualitative form of the phase diagram has been understood since the work of Doniach \cite{Doniach77}. The band theory phase diagram calculated by the CT-J method is shown in Fig ~\ref{fig:phase_diagram}. It has the qualitative form proposed by Doniach but the CT-J method enables one to understand in detail how the high temperature local moment phase crosses over to the Fermi liquid \cite{Otsuki09}, and provides insight into the relation of the Fermi liquid coherence to the magnetic phase diagram and allows one to include material-specific information. 

\begin{figure}[tb]
\includegraphics[width=\columnwidth]{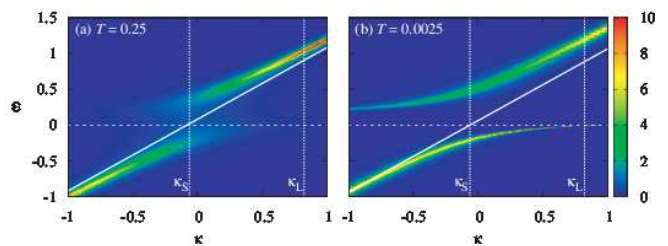}
\caption{The single-particle excitation spectrum $A(\protect\kappa, \protect%
\omega)$ for $J=0.3$ and $n_{\mathrm{c}}=0.9$ at (a) $T=0.25$ and (b) $%
T=0.0025$. The slanted line represents the non-interacting spectrum $\protect%
\omega=\protect\kappa -\protect\mu$ which is realized for $J=0$. From Ref.~\cite{Otsuki09}.}
\label{fig:spectra}
\end{figure}

Figure~\ref{fig:spectra} shows the single-particle spectral function $A(\kappa,\omega)$  computed  by \textcite{Otsuki09} using an analytical continuation based on Pad\'{e} approximants. This continuation method requires data of extremely high precision, available only with the CT-QMC methods. The vertical white lines labeled $\kappa_S$ indicate the positions of the Fermi surface  defined by the conduction band electrons in the absence of any Kondo effect; the line labeled $\kappa_L$ indicates where the Fermi surface would be if the local moment became an itinerant electron and were folded into the conduction band. The left panel shows $A(\kappa, \omega)$ for the high  $T=0.25$. The spectrum exhibits a behavior of almost non-interacting electrons at high energies. However,  a suppression of density of states is seen near the conduction electron Fermi surface $\kappa_{\mathrm{S}}$.

The right panel shows the spectral function at  $T=0.0025$, which is much lower than the impurity Kondo temperature defined by $
T_{\mathrm{K}}=\sqrt{g}e^{-1/g} \sim 0.1$ with $g=2J\rho_{\mathrm{c}}(0)$. Here the spectral function takes a form closely resembling that expected if the conduction band is weakly hybridized with a very flat band near the Fermi level and the Fermi surface has shifted to the point $\kappa_L$, indicating that the local moments in fact contribute to the Fermi volume. This behavior was expected, based on the detailed understanding which has been obtained for the single-impurity Kondo problem, but it is remarkable to see the phenomenon clearly exhibited in a lattice calculation. The fact that the bands are well defined at all $k$, and that the Kondo hybridization gap which opens up at $\kappa_S$ is well defined, are new and somewhat unexpected. 

\begin{figure}[tbp]
\par
\begin{center}
\includegraphics[width=\columnwidth]{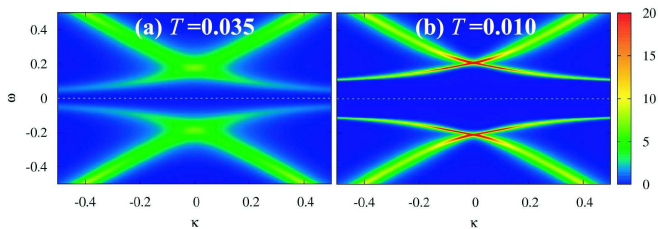}
\end{center}
\caption{False-color plot of electron spectral function in frequency $\omega$ and scaled momentum $\kappa$ plane for Kondo lattice model 
with antiferromagnetic coupling $J=0.2$ in (a) the
paramagnetic phase at $T=0.035$ and (b) the antiferromagnetic phase at $T=0.010$. From Ref.~\cite{Hoshino09}.
}
\label{fig_spect1}
\end{figure}

In a related study \cite{Hoshino09}, Hoshino and co-workers considered the Kondo lattice model at conduction band densities $n=1$ where at larger $J$ the ground state is a paramagnetic Kondo insulator.  At smaller $J$ the paramagnetic Kondo insulator is unstable to an antiferromagnetic insulator ground state. Figure~\ref{fig_spect1} shows the
spectrum for an intermediate  $J=0.2$ where a Kondo insulator phase is established at intermediate temperatures (left panel) and at lower $T$ becomes unstable to antiferromagnetism (right panel). In the region of the Brillouin zone presented in the figures the  form of the spectral function is remarkably similar in the two phases; the magnetism merely sharpens the spectral function and increases the gap size.  Again the hybridization of the local moment into the conduction band is the only reasonable interpretation of the formation of the paramagnetic insulating state. 

\begin{figure}[tbp]
\par
\begin{center}
\includegraphics[width=\columnwidth]{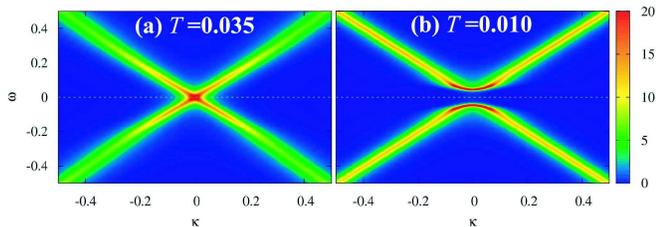}
\end{center}
\caption{False-color plot of electron spectral function in frequency $\omega$ and scaled momentum $\kappa$ plane for Kondo lattice model 
with ferromagnetic interaction $J=-0.2$ in (a) the paramagnetic phase at $T=0.035
$ and (b) the antiferromagnetic phase at $T=0.010$. From Ref.~\cite{Hoshino09}.}
\label{fig_spect2}
\end{figure}

It is interesting to contrast these results with those obtained for {\em ferromagnetic} Kondo coupling, shown in fig~\ref{fig_spect2}. Here we see that in the paramagnetic state there is a band crossing the Fermi level: the material is not an insulator because the Kondo effect does not occur (a similar effect was demonstrated in \cite{Werner06Kondo} using the CT-HYB method) and it is only when the antiferromagnetic instability occurs that a gap opens up. 

The application of CT-J methods to Kondo-like problems is still in its early stages, and it seems likely that  further extensions to more realistic models, and to cluster dynamical mean field approaches, will yield further insights.

\subsection{Dynamical mean field theory  for realistic models of correlated materials}

Dynamical mean field methods are more and more widely used in ab-initio based studies to model in a realistic way the properties of interesting materials.  These studies involve many subtle issues relating to mapping the orbitals and energies  derived e.g. from a density functional band theory calculation onto a theoretical model appropriate for solution with dynamical mean field methods.  The subject is reviewed in \cite{Kotliar06} and we will not attempt to summarize the discussion here. For the purposes of the present review it is enough to note that the correlated electron aspects of real materials typically involve multiple orbitals and several interaction parameters, so a mapping onto a simple one-band Hubbard model is typically not appropriate, while  the demanding nature of the band theory computations places a premium on having efficient impurity solvers for the dynamical mean field calculations. The development of CT-QMC methods has therefore had a significant impact on the field. The range of applications is large and growing rapidly; it will not be summarized here.  Rather, we will focus on recent results pertaining to one particularly challenging, and particularly topical system, the iron-based superconductors, where CT-HYB methods have made an important contribution to understanding the physics.  These calculations may be considered as reflective of the present ``state of the art'' of the ``realistic DMFT'' field. 

The unusually high superconducting critical temperatures  together with unusual normal state properties   are generally agreed to place the iron oxypnictides  in the broad category of strongly correlated superconductors, which also includes the $\kappa$ organics, cerium and plutonium based heavy fermions, and cuprate high temperature superconductors.   The correlated electrons reside mainly on $d$-orbitals associated with the Fe site and it appears to be necessary to retain all $5$ of the states in the $d$-multiplet and to treat carefully both the effects of the $U$ interaction which constrains charge fluctuations and the $J$-type interactions which select different states at fixed total charge. Because the couplings are neither extremely large nor extremely small, approximate methods may not be reliable: the full interacting problem must be treated by a numerically exact method. The  low point symmetry of each Fe site means that ligand field effects compete non-trivially with the interaction effects while the hybridization function is  complicated, and must be determined using band theory input. From the dynamical mean field side the complexity of the problem is such that only single-site DMFT calculations have been attempted, sometimes with a further restriction to density-density interactions.

In order to investigate the correlation effects in such complicated compounds it is important to have consistent one-electron and many-body parts of the LDA+DMFT Hamiltonian. For example, Aichhorn and collaborators studied the material LaO$_{1-x}$F$_x$FeAs  using an optimized basis of the localized $dpp$  Wannier functions which was constructed from the $22$ Bloch bands, corresponding to the $10$ Fe-$3d$,  $6$ As-$p$ and $6$ O-$p$ states (note each unit cell contains two formula units and the point symmetry of the two Fe is the same) \cite{Aichhorn09LOF}. The Green's function and hybridization function are constructed from the matrix elements of the Kohn-Sham Hamiltonian in the Wannier basis, while matrix elements of the Coulomb interactions  were calculated from the static limit of a constrained random phase approximation. The dynamical mean field theory was constructed by retaining the on-site intra-d interactions and projecting the $k$-integrated Green function onto the subspace of $d$ Wannier functions. Other groups use slightly different procedures; for example Kutepov {\it et al.} used a self consistent $GW$ procedure to compute the interaction  and an orbital-based procedure rather than a Wannier function-based procedure to define the basis of local states \cite{Kuterov10}.  

\begin{figure}[t]
\centering
\includegraphics[width=0.9\columnwidth]{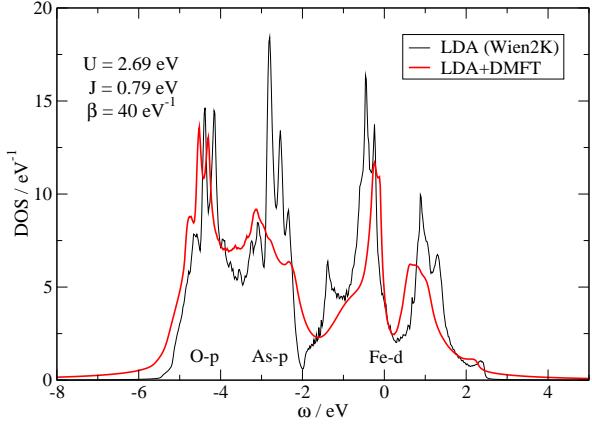}
\caption{(Color online) Full (all bands) spectral function for $dpp$ Hamiltonian description of LaOFeAs. Black
line: LDA. Red line: LDA+DMFT (computed retaining only density-density interactions). From Ref.~\cite{Aichhorn09LOF}.}
\label{fig:laofeas_dpp_tot}
\end{figure}

Aichhorn {\it et al.} then used CT-HYB simulations (but with only density-density interactions) at room temperature  to obtain the full local spectral function for the $dpp$ Hamiltonian corresponding to the experimental crystal structure of LaFeAsO and the realistic Coulomb matrix elements  \cite{Aichhorn09LOF}. Results  are shown in Fig.~\ref{fig:laofeas_dpp_tot}: The LDA+DMFT DOS near the Fermi level displays characteristic features of a metal in an intermediate range of correlations. Both occupied and empty states are shifted towards the Fermi level due to the Fermi-liquid renormalizations. No high-energy features that would correspond to lower or upper Hubbard bands can be seen in this LDA+DMFT electronic structure. 

\begin{figure}[!ht]
\includegraphics[width=0.98\columnwidth]{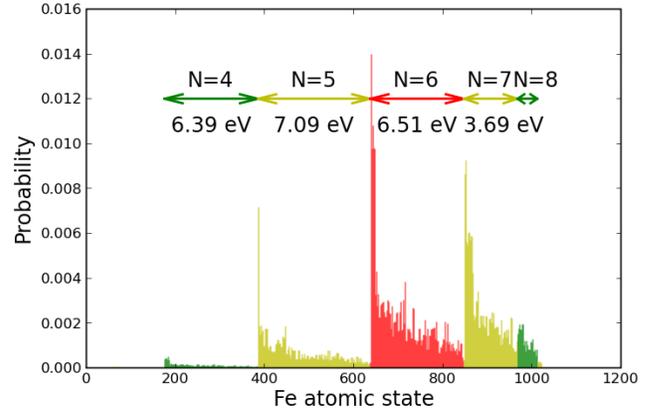}
\caption{
Histogram of occupation probabilities for each $3d$ atomic state in DMFT calculation for  BaFe$_2$As$_2$ at $T=150K$. The states are sorted by total d occupancy and within each manifold of fixed occupancy by energy. From Ref.~\cite{Kuterov10}.}
\label{Probab}
\end{figure}

In untangling the physics of the materials the ability of the CT-HYB method to provide the components of the  local density matrix, in particular the probability that any one of the atomic states of the iron $3d$ orbital is occupied, is important.  This is plotted for the material BaFe$_2$As$_2$  in Fig.~\ref{Probab}  \cite{Kuterov10}. Even the most probable atomic states have a probability of only a few percent, hence a naive strong correlation  atomic limit is qualitatively wrong for this compound. The wide spread of energies within a given submanifold is a consequence of the additional ``J-like'' interactions.

\begin{figure}[!ht]
\centering{\ \includegraphics[width=0.9\linewidth]{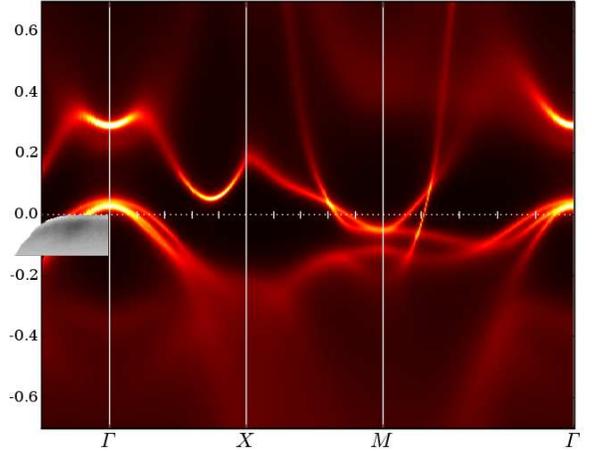} }
\caption{ Momentum resolved spectral function $A(\vk,\protect\omega)$
calculated for  BaFe$_2$As$_2$. Gray inset: ARPES intensity from Ref.~\cite{Brouet10}. From Ref.~\cite{Kuterov10}.}
\label{DMFT2}
\end{figure}

Figure~\ref{DMFT2} shows a false-color representation of  the momentum resolved spectral function $\sum_{L}A(\vk,\omega)_{LL}$ in the near-fermi-surface energy range \cite{Kuterov10}. Near the Fermi level the quasiparticle bands are well defined, while at higher energies the structures become blurred, reflecting the increased phase space for scattering. The quasiparticle velocities are renormalized relative to the band theory result (not shown) by factors of 2 for $x^2-y^2$ and $3z^2-r^2$ orbitals and 3 for the $xy$, $xz$, $yz$ orbitals.   The momentum space positions of the Fermi surface crossings are in good agreement with photoemission results, as are the renormalized velocities.
Comparison of these sorts of calculations to the rapidly growing body of experimental data are enabling a comprehensive understanding of the physics of novel materials.

%% file: applicationsnano.tex
\section{Applications III: Nanoscience}
\subsection{Transport through quantum dots: linear response and quantum phase transitions}

One important application of quantum impurity models is as representations of ``single molecule'' conductors  and other nano-devices \cite{Hanson07}. Much of the attention in the nanoscience community has been focused  on weakly interacting systems or on simple Hubbard-like dots. Standard perturbative or Hirsh-Fye QMC methods suffice for these situations, although CT-QMC methods have been used, e.g. in a study of the accuracy of the GW approximation \cite{Spataru08}. As the field moves towards consideration of quantum dots with richer physics, other approaches including  CT-QMC methods are likely to become important.

An example is provided by the two-level two lead quantum dot system uncovered by Yacoby {\it et al.} \cite{Yacoby95}.  \textcite{Golosov06} suggested that this system could display a quantum phase transition between two different  relative occupancies as level energies were varied. This issue was investigated using the CT-HYB method by \textcite{Wang10}.  In the general case of the model presented by Gefen the imaginary part of the hybridization function (giving decay of the dot electrons into the leads) does not commute with the combination of the level Hamiltonian and the real part of the hybridization function (giving the renormalization of the dot energies). This causes a severe sign problem, which prevented any useful simulations in the general case.  \textcite{Wang10} argued that the universal behavior at a quantum critical point (if one existed) could be described by a sign problem-free model  (essentially because at the critical point the combination of  the dot Hamiltonian and real part of  the hybridization function becomes the unit matrix). 

\begin{figure}[tbp]
\begin{center}
\includegraphics[angle=-90,width=0.8\columnwidth]{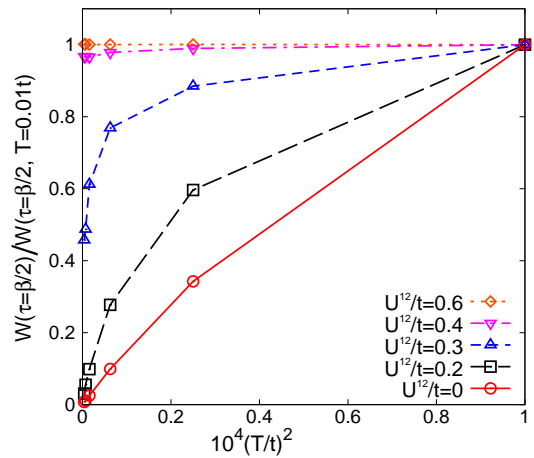}
\end{center}
\caption{
Imaginary time density-density correlation function $W$ of two-level, two-lead model  evaluated using CT-HYB   at midpoint of imaginary time interval and normalized to value at $T=0.01t$,  as function of interaction strength with level energies tuned to be equal.  Weak $T$ and $\tau$ dependence is seen in  non-Fermi-liquid phase ($U=0.4,0.6$) and strong $T$ and $\tau$ dependence in Fermi liquid phase $(U=0,0.2)$. For details see Ref.~\cite{Wang10}.
\label{Wang10fig}
}
\end{figure}
To investigate the criticality, \textcite{Wang10} considered the imaginary-time dependence  of correlation functions  of variables defined on the quantum dot.
Fig.~\ref{Wang10fig} shows three different behaviors at times $\sim \beta/2$: a $T^2$ dependence 
expected for a Fermi liquid for small $U/t$, a power law at the critical point, 
and a constant long time behavior for large $U$ in the non-Fermi liquid phase. 
However, impurity problems may be characterized by exponentially small scales such as the Kondo effect. Distinguishing a very small scale from a true phase transition is numerically challenging. The ability of CT-HYB to access very low temperatures $\sim 10^{-3}t$ provides reasonable evidence of a critical point. However, for problems such as this where the key question concerns the asymptotic low energy behavior, quasi-analytical functional renormalization group methods \cite{Meden06} and NRG approaches \cite{Karrasch07,Bulla08b} may be more powerful.

\subsection{Metal atom clusters on surfaces}
An active area of nanoscience research concerns the properties of one or more transition metal ions on a metal surface. Of particular interest is the density of states, which may be compared to scanning-probe microscopy data.  Savkin and coworkers in \cite{Savkin05} applied the CT-INT scheme to a model of three interacting Kondo impurities on a metallic surface. The ability of the CT-QMC methods to treat realistic interactions allowed an accurate investigation of the interplay of cluster geometry, inter-adatom hopping, local Coulomb interactions and the Heisenberg exchange interactions between magnetic impurities. \textcite{Savkin05} showed that a rotationally invariant antiferromagnetic exchange interaction is almost twice as efficient in suppression of the single site Kondo effect as is the Ising like interaction which was all that could be treated by previous methods.  

The possibility to make quantitative comparisons to experiment highlights the need to incorporate as much material specificity as possible into the calculation. \textcite{Gorelov07} performed a realistic study of $Co$ atoms in the bulk or at the surface of a $Cu$ host. They found that a complete treatment of the problem, including all inequivalent terms of the Coulomb interaction, was essential for obtaining physically relevant results. Inclusion of all of the interaction terms however produces a severe sign problem. While the sign problem can be mitigated to some extent by an appropriate choice of basis, it severely limits the range of temperatures over which results can be obtained.  These calculations represent the current state of the art: they push the CT-INT technique to its limits and demonstrate the need for further algorithmic developments.

To set up the problem, density functional band theory techniques were applied to appropriately chosen supercell geometries. From these calculations wave functions $d_i(r)$ for the $Co$ d-states and itinerant electron wave functions $\Psi_{nk}(r)$ were extracted. The bare local Green's function is then obtained as
\begin{align}
\mathcal{G}^0_{ij}(i\omega_n) = \sum_{nk}\frac{\langle d_i|\Psi_{nk}\rangle\langle \Psi_{nk}|d_j\rangle}{i\omega_n + \mu - \varepsilon_{nk}}
\end{align}
while the Coulomb interaction $H_\text{int} =\frac{1}{2} \sum_{ijkl\sigma\sigma' }$ $U_{ijkl}$ $c_{i\sigma}^\dagger c^\dagger_{j\sigma'}c_{k\sigma'}c_{l\sigma}$ involves matrix elements of the form $U_{ijkl} = \langle d_i(r_1)d_j(r_2) \frac{e^2}{\varepsilon |r_1-r_2|}d_k(r_2)d_l(r_1)\rangle$. The number of interaction terms which must be considered is large, and depends on the choice of basis in the $d$-sector.

Use of symmetries to rearrange the interaction and eliminate redundant terms was also found to be important. Implementing all the symmetries and making an optimal basis choice led Gorleov et al to an expression for the partition function as an expansion in $129$ independent interaction parameters:
\begin{align}\label{Zexp2}
\frac{Z}{Z_{0}}=\sum_{n}\frac{(-1)^{n}}{n!2^{n}}\sum_{\{ijkl\sigma \sigma
^{\prime }\}}\int_{0}^{\beta }d\tau _{1}...\int_{0}^{\beta }d\tau_n \\ \nonumber
U_{i_{1}j_{1}k_{1}l_{1}}...U_{i_{n}j_{n}k_{n}l_{n}}\det {\mathcal{G}}%
^{2n\times 2n}.
\end{align}

The expansion exhibits a ``trivial'' sign problem which may be mitigated by appropriate choice of $\alpha$ parameters as discussed in Sec. \ref{weakcoupling} although the multiplicity of interactions requires a multiplicity of $\alpha$ parameters; for further details see Refs.~\cite{Gorelov07,Gorelov09}. The expansion also suffers from an 'intrinsic' sign problem (not curable by choice of $\alpha$) whose  severity was found to depend on the basis choice. ``Three orbital'' terms $U_{ikkl}$ with $l\neq i$ were found to produce a severe sign problem but do not occur if a spherical harmonic basis is used.  However, ``non-diagonal'' terms $U_{ijkl}$ with $(i\neq j, k\neq l)$ cannot be eliminated by transformations, make important contributions to the physics, and give rise to a sign problem, of a severity which depends on other features such as the Green's functions. 

\begin{figure}[tbp]
\begin{center}
\includegraphics[width=0.5\textwidth]{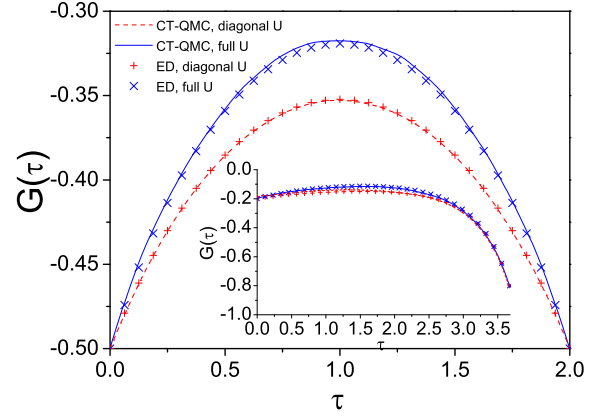}
\end{center}
\caption{(color online) Comparison of CT-INT expansion of Eq.~(\ref{Zexp2}) with exact diagonalization for an isolated $Co$ atom with  $U=1$
eV, $J=0.4$ eV, at inverse temperature $\beta =2$ eV$^{-1}$, for $5$ electrons (main panel) and $8$ electrons (inset). Solid lines and crosses: Full Hamiltonian. Dashed lines and plus symbols: model specified by diagonal terms only.
From Ref.~\cite{Gorelov09}.} \label{TBvsED}
\end{figure}

To test both the expansion and the importance of the non-diagonal terms, Gorelov {\it et al.} determined the Green's function for one orbital of the isolated atom (i.e. with no hybridization function) using both CT-INT based on Eq.~(\ref{Zexp2}) (with the 129 interaction parameters) and by exactly diagonalizing the problem. Fig.~\ref{TBvsED} shows that the CT-INT expansion reproduces the exact result and that the `non-diagonal' terms in the interaction are important. 

\begin{figure}[tbp]
\begin{center}
\includegraphics[width=0.49\columnwidth]{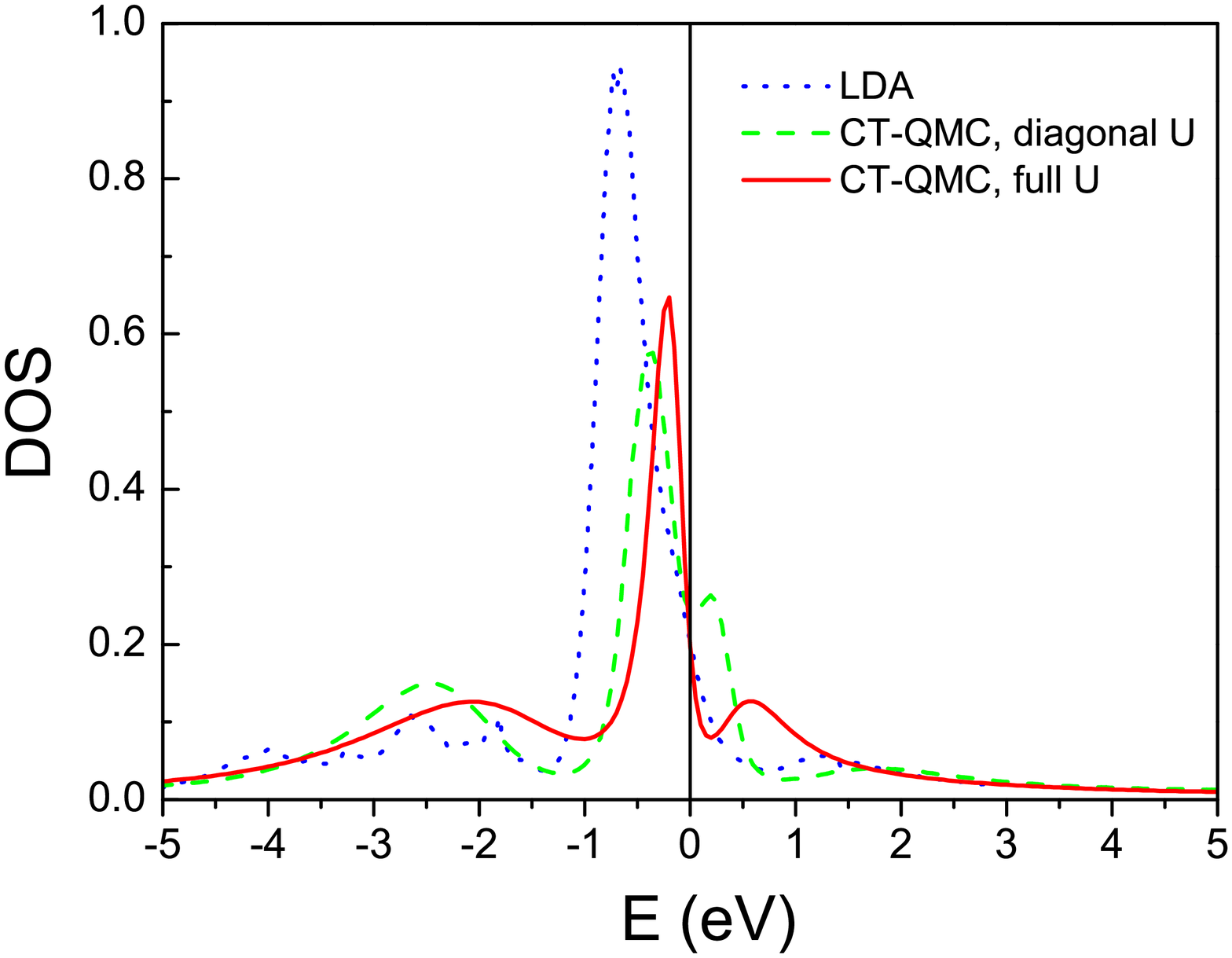}
\includegraphics[width=0.49\columnwidth]{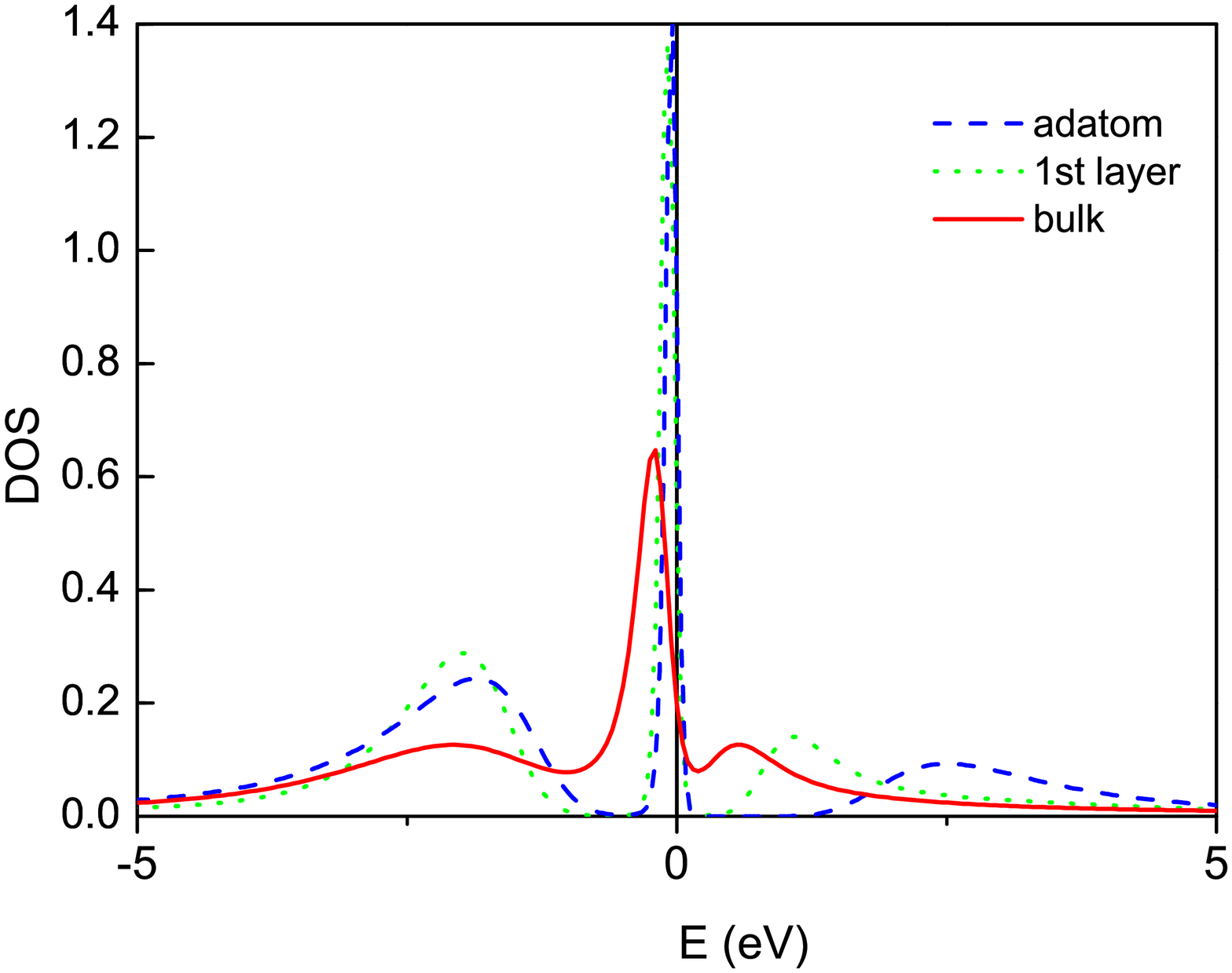}
\end{center}
\caption{(color online) Left panel: Total DOS of 3d orbital of $Co$ atom embedded in $Cu$ matrix.
Model parameters: $U=4$ eV, $J=0.7$ eV, $\beta =10$ eV$^{-1}$ for
5-orbital impurity with 7 electrons.  
Right panel: Total DOS of 3d orbital of $Co$ atom embedded in the bulk
of $Cu$, into 1-st layer and $Co$-adatom on the $Cu$(111) surface. Model
parameters: $U =4$ eV, $J =0.7$ eV, $\protect\beta =10$ eV$^{-1}$ for
5-orbital impurity with 7 electrons. From Ref.~\cite{Gorelov09}.}\label{doses5b_3DOS}
\end{figure}
Gorelov {\it et al.} then computed the local density of states of a $Co$ atom in bulk $Cu$ (far from a surface). For this case the  sign problem, while present, is not severe for the temperatures studied ($T \approx 1200K$). Results are shown in the left panel of Fig.~\ref{doses5b_3DOS}. The non-diagonal interactions have an important effect on the line shape.  (Note that  $T \approx 1200K$ is well above the Kondo temperature, so no peak is evident at the Fermi surface). 

\begin{figure}[tbp]
\begin{center}
\end{center}
\end{figure}
Finally, Gorelov {\it et al.} consider a $Co$ impurity at a surface (right panel of Fig.~\ref{doses5b_3DOS}). Here the relatively large non-diagonal elements of the bath Green's function lead to a serious sign problem. To make a simulation on the surface feasible, Gorelov {\it et al.} in effect restricted the sampling to a constant-sign subset of configuration space, by only allowing updates that did not change the fermionic sign. See Ref.~\cite{Gorelov09} for further details. 

%% file: applicationsnonequilibrium.tex
\section{Applications IV: non-equilibrium impurity models and nanoscale transport}

\subsection{Overview}
CT-QMC methods have been used to study nonequilibrium problems defined on the ``Keldysh'' two-time contour.
These studies are still in their early stages and we present here a few representative preliminary results concerning the current-voltage characteristics of interacting quantum dots, as well as simulations inspired by the newly developing capabilities of performing pump-probe experiments on correlated electron compounds and ``quantum quench'' experiments on cold atom systems.  

\subsection{Results: Current-voltage characteristics \label{interactionquench}}
\subsubsection{Real-time CT-HYB}
\begin{figure}[t]
\begin{center}
\includegraphics[width=0.8\columnwidth]{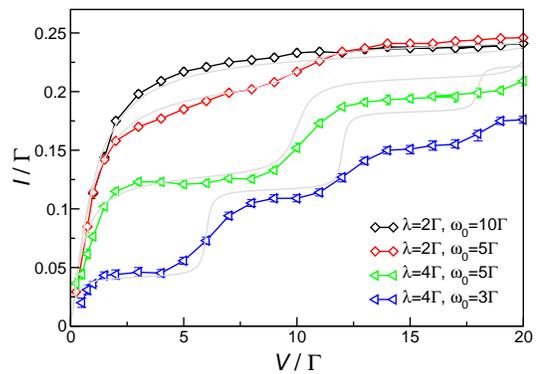}\\\vspace{5mm}
\caption{Points: total current $I$ computed using nonequilibrium CT-HYB methods as a function of the bias voltage for a half filled, spinless, phonon-coupled quantum dot (Eq.~(\ref{H_loc_holstein}) with U=0) with  voltage bias applied symmetrically, $\varepsilon_{d}=0$, $\mu_{L,R} = \pm \frac{V}{2}$ at temperature $T = \frac{\Gamma}{5}$ at electron-phonon coupling strengths $\lambda$ and oscillator frequencies $\omega_0$ indicated. Lines: results of an approximate analytical calculation \cite{Flensberg03}. From Ref.~\cite{Muehlbacher08}. 
}
\label{MuhlbacherSchiroFigure}
\end{center}
\end{figure}
The first nonequilibrium applications of the CT-QMC technique were to the current-voltage characteristics of a quantum dot under a bias voltage. In their pioneering paper, \textcite{Muehlbacher08} showed that the hybridization expansion method could be directly applied on the Keldysh contour and that long enough times could be reached to permit measurements of steady state behavior. They studied a non-interacting dot coupled to phonons (essentially  the Holstein-Hubbard model, Eq.~(\ref{H_loc_holstein}),  with spin neglected and $U=0$); representative results giving the dependence of the  currerent-voltage characteristics on the oscillator frequency and coupling strength  are presented in  Figure~\ref{MuhlbacherSchiroFigure}. 

These calculations start from an initial state in which the dot is decoupled from the leads and the calculation must build in appropriate dot-lead entanglement. This requires a coherence times which depends on the physics. In the calculations of Ref.~\cite{Muehlbacher08} convergence was facilitated by decoherence arising both from the phonons and from  the relatively high $T$ which was studied. Results for the interacting Anderson model (without phonons)  have been published in \cite{Schmidt08, Werner09, Schiro09}. Because the expansion must in this case be performed for both spin flavors and the decohering effect of phonons is not included, reaching a steady state becomes challenging.  Attempts to optimize the algorithm by considering initial states with dot-lead entanglement \cite{Schiro10} have not led to dramatic improvements.

\subsubsection{Real-time CT-AUX}

In the weak coupling methods, for example the CT-AUX algorithm explained in Sec.~\ref{noneqintro}, one may use 'interaction quench' methods in which  the real-time simulation starts from a $U=0$ state with dot-lead entanglement. Temperature enters only as a parameter in the lead correlators, making  it possible to treat arbitrary temperatures, including $T=0$. While the presence of interactions of course modifies this entanglement, it seems that up to interaction strengths of $U\approx10\Gamma$, relatively few perturbation orders are required to reach steady state. The situation is particularly favorable for particle-hole symmetric models with symmetrically applied bias, where odd orders of perturbation theory can be suppressed.    As illustration we present interaction-quench results  for the time-dependence of the current and the current-voltage characteristics of half-filled quantum dots with symmetrically applied voltage bias ($\mu_L=-\mu_R=V/2$).

\begin{figure}[t]
\begin{center}
\includegraphics[angle=-90, width=0.45\columnwidth]{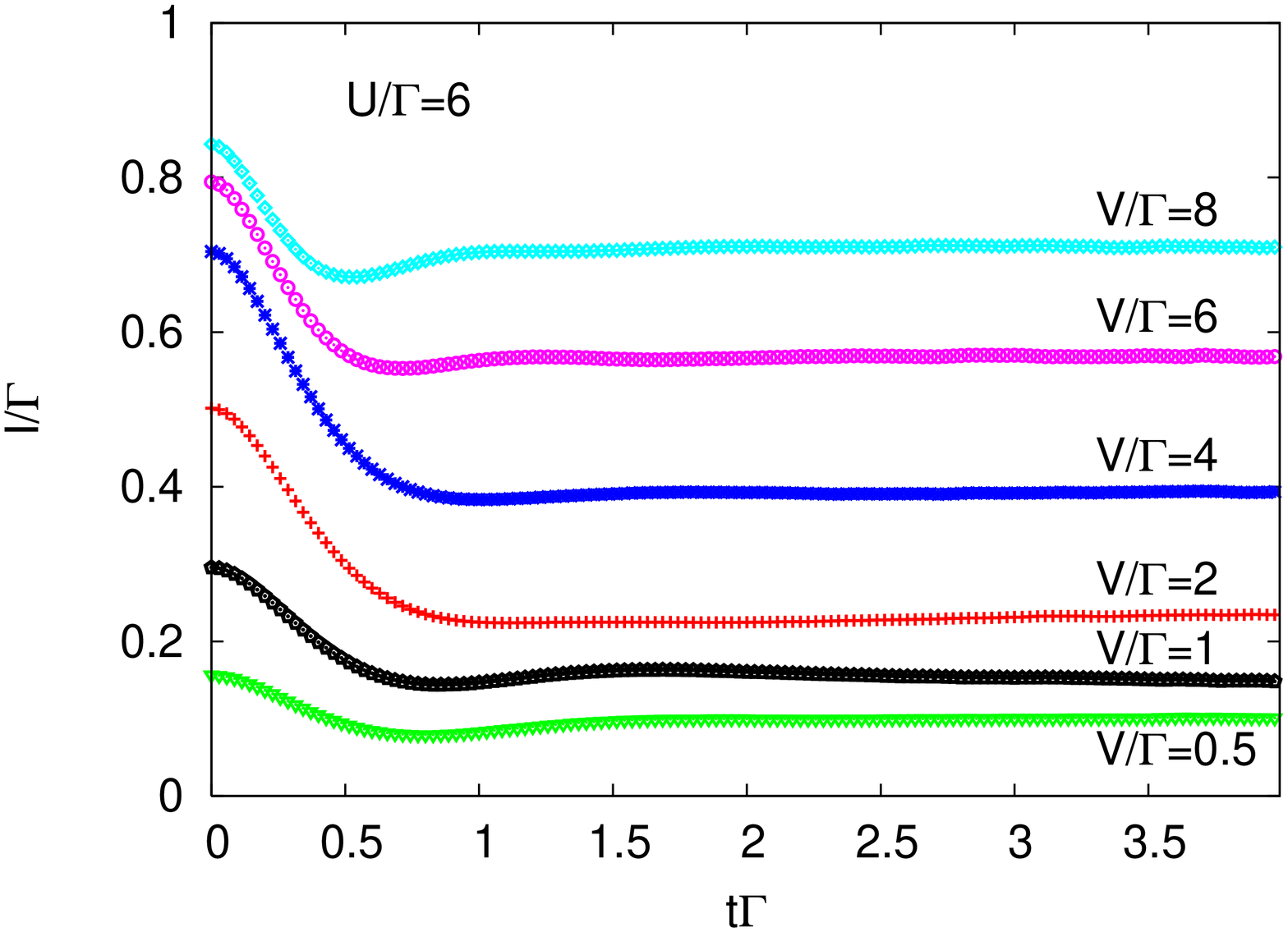}
\includegraphics[angle=-90, width=0.45\columnwidth]{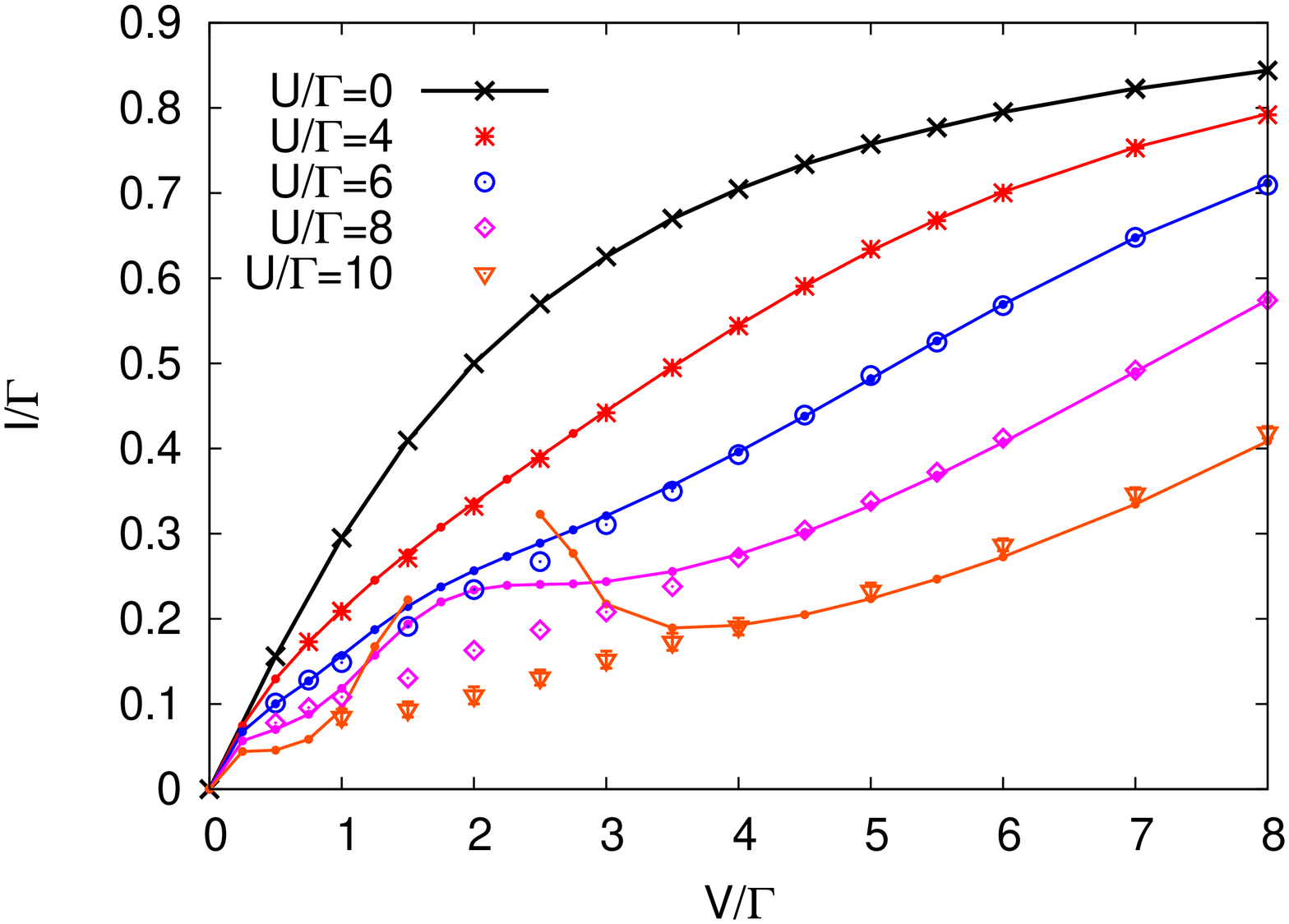}
\caption{Left panel: time evolution of the current for different voltage biases ($U/\Gamma=6,T=0$). Right panel: current-voltage characteristics obtained from long-time limit of the calculated currents. The symbols show Monte Carlo data for $U/\Gamma=4$, 6, 8 and 10, while the lines correspond to fourth order perturbation theory. In the initial state, the current is given by the steady state current through the non-interacting dot. At time $t=0$, the interaction is turned on. From Ref.~\cite{Werner10}.}
\label{i_t_u6}
\end{center}
\end{figure}

At time $t=0$, the system is non-interacting but subject to an applied bias $V$, so a current $I_0(V)$ appropriate to the non-interacting model is flowing through the dot. At $t=0_+$ the interaction is turned on and the system relaxes into the steady-state configuration appropriate to the interacting model.  The left panel of Fig.~\ref{i_t_u6} plots the time evolution of the current for fixed $U/\Gamma=6$ and several voltage biases.  For voltages $V/\Gamma \gtrsim 2$, even though the transient behavior is clearly voltage-dependent, the current settles into the new steady state after a time $t\Gamma \approx 2$. However, as the voltage is decreased below $V/\Gamma\approx 2$ the transient time increases. At $V=\Gamma$ the long time limit is attained only for $t\Gamma \gtrsim 3$ and as $V$ is further decreased, the interaction-quench method cannot reach the steady state.  As discussed in \cite{Werner10}, in the small-$V$ regime, voltage quench simulations are a possible alternative to the interaction quench. In the voltage-quench calculations, the time-evolution starts from the interacting equilibrium state, which is a good starting point for small $V$. However, because the imaginary branch of the L-shaped contour must be explicitly treated, this approach is restricted to non-zero temperatures, and is only advantageous for very small voltage. 

The right panel of Fig~\ref{i_t_u6} presents  $T=0$ results for the voltage dependence of the  steady state current  as well as analytic results obtained from fourth order perturbation theory order. The interacting current initially rises with the same slope as the non-interacting current, and reaches the non-interacting value also in the large-voltage limit.  At intermediate values of $V$ the effect of interactions is to suppress the current (Coulomb blockade). The results show clearly that at intermediate voltages the method can access interaction regimes beyond the scope of analytical perturbation theory. In agreement with conclusions reached in \cite{Werner09} on the basis of (less accurate) hybridization expansion results and also with recent nonequilibrium functional renormalization group calculations \cite{Jakobs10} we see that this model does not display a region of negative differential resistance. 

\subsubsection{Nonequilibrium DMFT}

The real-time CT-QMC methods can also serve as impurity solvers in nonequilibrium-DMFT simulations of bulk systems. The impurity Hamiltonian of Section~\ref{noneqintro} is the impurity problem relevant for the solution of the one-band Hubbard model. In the DMFT context, however, $H_\text{bath}$ (Eq.~(\ref{H_bath})) is a single bath, whose parameters are fixed by a self-consistency equation which in the non-equilibrium context is time dependent \cite{Schmidt02, Freericks06}. 

\begin{figure}
    {\includegraphics[width=0.493\columnwidth]{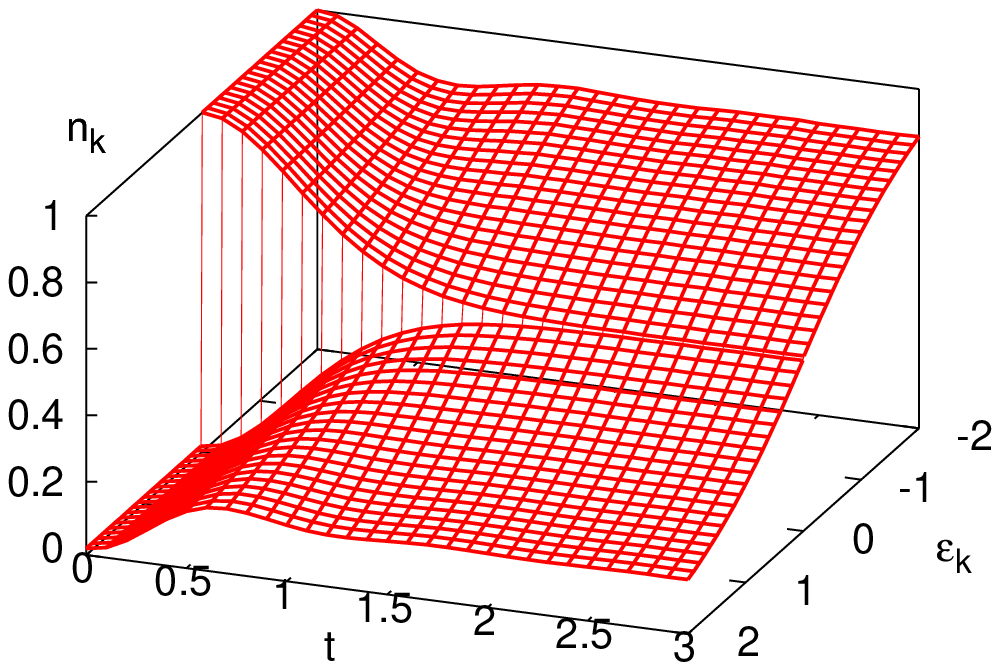}}
    {\includegraphics[width=0.493\columnwidth]{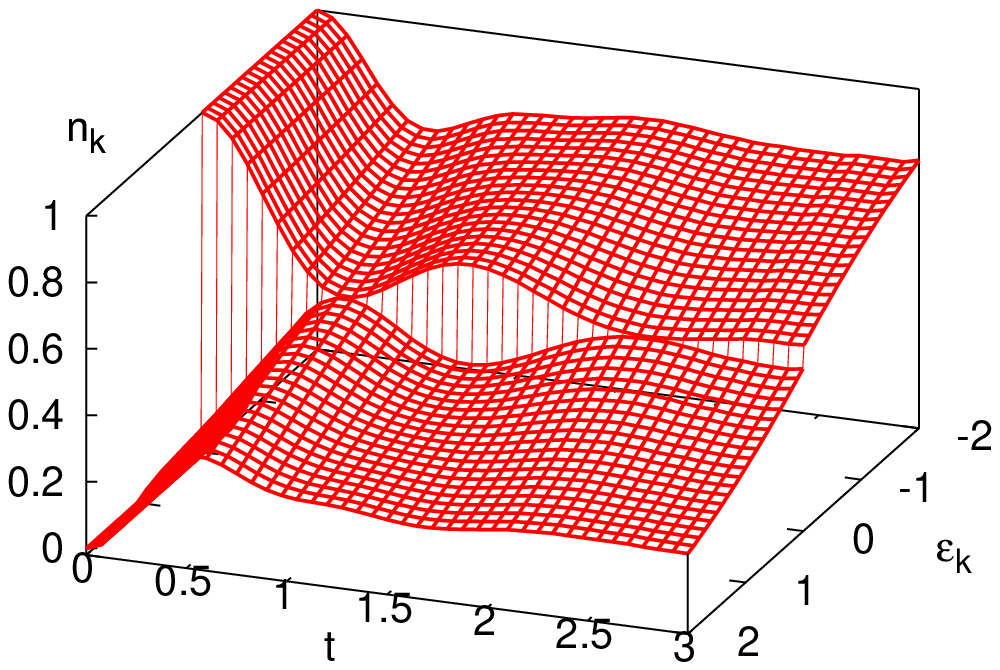}}
     \caption{Momentum distribution $n(\varepsilon_k,t)$ for
      quenches from $U/W$ $=$ $0$ to $U/W$ $=$ $0.75$ (left panel) and $U/W=1.25$
      (right panel), for Hubbard model with semicircular density of states and bandwidth $W$. $t$ is measured in units of the quarter-bandwidth $W/4$. From Ref.~\cite{Eckstein10}. 
    \label{fig:nku5}}
  \end{figure}

CT-QMC have been used by \textcite{Eckstein09, Eckstein10} to study the relaxation dynamics of the half-filled Hubbard model after a sudden switching-on of the electron repulsion $U$. The initial state was the non-interacting equilibrium state at temperature $T=0$, and the DMFT selfconsistency assumed a semi-circular density of states of bandwidth $W=4$. The calculation produces among other observables the time-evolution of the momentum distribution function $n(\varepsilon_k, t)$ which is plotted in Fig.~\ref{fig:nku5} for quenches to $U=3W/4$ and $U=5W/4$. Qualitative differences in the relaxation dynamics appear as the value of the interaction strength is changed. 
  
It was demonstrated in \cite{Eckstein09, Eckstein10} that for a quench to $U=0.8W$ the momentum distribution function and double occupancy relax very fast (within a time $t=6.4/W$) to the thermal equilibrium result compatible with energy conservation. The relevant time-scale is in this case easily accessible with real-time CT-AUX. Away from this ``critical interaction strength", i. e. for quenches to $U\ne 0.8W$, the system is initially trapped in a non-thermal quasi-stationary state, and equilibration occurs on much longer time scales. In this case the accessible times are not long enough to observe the expected thermalization (see right panel of Fig.~\ref{fig:nku5}). Only the initial relaxation into the quasi-stationary state and (for $U<0.8W$) the initial part of the slow crossover towards the thermal equilibrium state are computationally accessible with present techniques. 

%% file: prospects.tex
\section{Prospects and open issues}
Over the last few years, continuous-time quantum Monte Carlo methods for fermionic impurity problems have been developed to a high degree. Because the methods are based on a diagrammatic expansion, they can handle many physically relevant interactions, which were not easily treatable by other methods. Also, by construction  they are  free from  the time-discretization errors associated with methods previously used for fermions, based on the Suzuki-Trotter decomposition \cite{Suzuki76}. Third, a continuous-time formulation 
provides, in a sense, 
a many-body adaptive grid method for the time evolution.  These advantages of the continuous-time formulation  enable a decrease, typically by several orders of magnitude, in the computational effort required to solve a problem of a given complexity, making previously intractable problems tractable and creating new opportunities for physics by allowing rapid and routine investigation of problems which had previously required access to supercomputer facilities.

The methods have become very important to the field of  correlated electron physics  (via the connection to single \cite{Georges96} and cluster \cite{Maier05} dynamical mean field theory) and   are having an increasing impact on nanoscience.  However, the first generation of results has only begun to explore what is possible.   We expect that over the next few years the methods will be  increasingly widely used in dynamical mean field computations of correlated electron materials and condensed atomic gases,  and in studies of  the equilibrium and nonequilibrium phenomena arising in the impurity models relevant to nanoscience. The nonequilibrium applications in particular represent an entirely new field with many exciting possibilities. Methodological improvements, including combinations of hybridization and coupling constant expansions and the further study of projected Hilbert space methods such as CT-J are likely to be fruitful.  We hope that an increasing number of scientists will take advantage of the opportunities, by applying the methods to yet wider classes of problems and by developing them further.   

\begin{table*}[htdp]
\begin{center}\begin{tabular}{|c||c|c|c|c|}\hline
Scaling / Algorithm & CT-INT & CT-AUX &CT-HYB (segment) & CT-HYB (matrix)\\
\hline\hline
diagonal hybridization & $N (\beta U)^3       $&$N (\beta U)^3$& $N\beta^3  $&$a e^N\beta^2+ b N\beta^3, a \gg b$\\
non-diagonal hyb.      & $(N \beta U)^3,\ \sp  $&$(N\beta U)^3,\ \sp$& $(N\beta)^3, \sp $&$a e^N\beta^2+ b (N\beta)^3, a \gg b,\ \sp $\\
diagonal interaction   & $(N \beta U)^3,\ \sp  $&$(N\beta U)^3,\ \sp$ & $(N\beta)^3, \sp $&$a e^N\beta^2+ b (N\beta)^3, a \gg b,\ \sp $\\
general $U_{ijkl}$     & $(N^2 \beta U)^3,\ \sp $& n/a           & n/a       &$a e^N\beta^2+ b (N\beta)^3, a \gg b,\ \sp $ \\
\hline
\end{tabular}
\end{center}
\caption{Summary of scaling and sign metrics in the equilibrium case for most widely studied continuous-time quantum Monte Carlo methods. CT-INT, CT-AUX, CT-HYB refer to the interaction expansion (Sec.~\ref{weak_chapter}), auxiliary field (Sec.~\ref{ctaux_chapter}) and hybridization (Sec.~\ref{hyb_seg_chapter}) expansion algorithms respectively. ``Segment'' refers to the case of the hybridization interaction where the hybridization function, local Hamiltonian and interaction are all diagonal in the same basis (\ref{segment_section}), while ``matrix'' refers to the general implementation in Sec.~\ref{hyb_matrix_section}. We distinguish Green functions which can be diagonalized by one single canonical transformation from general Green functions where the hybridization function, local Hamiltonian and self energy do not all commute, and we distinguish interactions such as the Hubbard $U$ which are diagonal in an appropriate single-particle occupation number basis from those such as spin exchange and ``pair hopping'' which cannot be diagonalized. $\sp$ indicates the possibility of the presence of a fermionic sign problem. }
\label{scalingtable}
\end{table*}

We conclude our discussion by summarizing what we perceive to be the strengths and weaknesses of  the different CT-QMC methods, and suggesting some issues that may warrant further attention. The fundamental issues for any algorithm are the scaling with temperature, interaction strength, and system size. In addition, for fermions, one must consider the sign problem.  Table [\ref{scalingtable}]  summarizes what we know about these scalings. 

The hybridization expansion algorithm, CT-HYB, diagonalizes the local Hamiltonian and expands in the impurity-bath hybridization. The principal advantage of this approach is that  instantaneous (Hamiltonian)  interactions of essentially arbitrary strength and functional form can be handled (retarded interactions can be conveniently treated only in special, but physically relevant cases such as the screened density-density interaction). The hybridization expansion appears to suffer from a severe sign problem if the hybridization function does not commute with the one-body part of the local Hamiltonian, and this limits its use in the most general contexts. It appears to be most useful for the  single-site dynamical mean field theory of materials with partly filled $d$ and $f$ shells, where its ability to treat the full complexity of general multiplet interactions is unmatched and the point symmetry ensures that the local Hamiltonian and hybridization functions commute.  

The fundamental computational bottlenecks of the hybridization method are the need to manipulate matrices whose size is set by the dimension of the full fermionic  Hilbert space of the impurity Hamiltonian and the need to compute determinants of hybridization matrices of a size linearly growing with $\beta$. The computational burden grows exponentially with the size of the fermionic problem and as the cube of the inverse temperature, and the system-size constraint is therefore more severe. For a model of $N$ spin degenerate orbitals the Hilbert space size is $4^N$. At present, $5$ orbital models are accessible with large scale computing resources. For larger systems a straightforward approach is not feasible yet without  truncation. The accuracy of truncation schemes is not yet established. Of course, in special cases block diagonalization is possible so the full Hilbert space need not be treated. In the most favorable case, the local Green function, hybridization matrix and interaction may all be diagonalized in the same single-particle occupation number basis (this occurs in the $N$-orbital impurity model with density-density interactions, if each orbital hybridizes with a different bath) and the segment representation of the hybridization expansion may be used. In this case there is no sign problem and the cost is linear in the number of orbitals and cubic in the inverse temperature. Thus if a segment representation exists, it should be used. 
Unfortunately, in most problems of physical interest either hybridizations or interactions entangle the different single-particle basis states, and a general matrix formulation is required. In this case the exponential scaling associated with the Hilbert space size is the crucial constraint, and an important open problem concerns the degree to which the Hilbert space can be block diagonalized or truncated. Haule pioneered the use of symmetry-based  block diagonalization and of truncation \cite{Haule07}. An alternative approach based on sparse matrix-vector instead of dense matrix-matrix multiplication is the Krylov technique discussed in Sec.~\ref{Krylov}. Further research along these and related lines appears to be worthwhile. 

The interaction expansions CT-INT and CT-AUX   are based on an expansion about the free-fermion limit. The computational burden therefore increases with the interaction strength, as well as with inverse temperature and the system size, making it difficult to access the very strong coupling regime. However, the scaling with system size is power-law,  rather than exponential,  so that these methods are the only ones feasible when many orbitals or many sites are important to the physics. At present, a lack of good auxiliary field decompositions means that the CT-AUX method can only be used for models with density-density interactions. Its main application has been to cluster dynamical mean field studies of the Hubbard model. A natural subject for further investigations is the application of the method to wider classes of models, including more general (but still density-density) interactions. The CT-INT method is equivalent to the CT-AUX method for Hubbard like interactions (although the present CT-AUX implementations appear to be more efficient), and is applicable to models with general (non density-density) interactions. However, sign problems occur and grow in severity as the complexity of the interaction increases. 

Our experience in the nonequilibrium context is that in a particle-hole symmetric model, the CT-INT and CT-AUX methods are to be preferred over the hybridization methods because odd perturbation orders can be suppressed \cite{Werner10}, resulting in a less severe sign problem and longer accessible times. 

There are two limitations associated with the CT-INT and CT-AUX methods. 
One issue for more realistic models with more complicated interactions is the need to make a multiple expansion in all components of the interaction. This is not a serious issue as long as no sign problem is encountered. 
The more fundamental limitation is the sign problem, which 
can arise in cluster dynamical mean field calculations from the presence of physical (real-space) fermionic loops or more generally from non-commutativity of operators appearing in the impurity model, due for example to exchange interactions or to hybridization functions which cannot be diagonalized by a single (time-independent) basis change. Sign problems are in general dependent on the choice of basis and further exploration of different representations of the interaction and the Green function, especially in the case of non-diagonal interaction, may be worthwhile.

\acknowledgments We are greatly indebted to our many
collaborators and colleagues,
Fakher Assaad, 
Evgeny Burovski, 
Armin Comanac, 
Martin Eckstein, 
Michel Ferrero, 
Sebastian Fuchs, 
Antoine Georges, 
Hartmut Hafermann, 
Kristjan Haule, 
Mark Jarrell, 
Mikhail Katsnelson, 
Marcus Kollar, 
Andrei Komnik,
Gabriel Kotliar, 
Evgeny Kozik, 
Jan Kunes, 
Hiroaki Kusunose, 
Andreas L\"auchli, 
Gang Li, 
Chungwei Lin, 
Nan Lin, 
Lothar M\"uhlbacher, 
Thomas Maier, 
Luca de Medici,
Karlis Mikelsons, 
Hartmut Monien, 
Takashi Oka, 
Junya Otsuki, 
Olivier Parcollet, 
Lode Pollet, 
Nikolay Prokof'ev, 
Thomas Pruschke,
Stefan Rombouts, 
Vladimir Savkin, 
Thomas Schmidt, 
Michael Sentef, 
Boris Svistunov, 
Dieter Vollhardt, 
and Xin Wang,
without whose input such rapid development would not have been
possible. The work of E.~G. and A.~J.~M. was  supported by NSF under Grant No.
DMR-0705847, DMR-1006282 and the US Department of Energy under grant ER-46169, P.~W. and M.~T. by the Swiss National Science Foundation, A.~L. and A.~R. by RFFI-DFG grant 436 RUS
113/938/0 -- 08-02-91953, SFB 668, the Kurchatov institute, and the Cluster of Excellence ``Nanospintronics''. Preparation of this manuscript was supported in part by the National Science Foundation under Grant No. PHY05-51164.

{\it Open-source implementations of some of the algorithms we describe in this review have been made available for download. Codes for the interaction expansion algorithm implemented by Rubtsov {\it et al.} are  available at \cite{Rubtsovcodes}. A hybridization expansion code for density density interactions, corresponding to the description in Sec.~\ref{segment_section}, as well as an interaction expansion implementation are also available as part of the ALPS project \cite{ALPS20} and have been published in Ref.~\cite{ALPS_DMFT}.}